\documentclass[a4paper,11pt]{article}
\usepackage{jcappub}

\usepackage{bm}
\usepackage{subcaption}
\usepackage{graphicx}
\usepackage{multirow}
\usepackage{amsmath}
\usepackage{mathrsfs}
\usepackage{amssymb}
\usepackage{hyperref}
\usepackage{yfonts}
\usepackage{color}
\usepackage{xspace}
\usepackage{verbatim}
\usepackage{mathtools}
\usepackage{booktabs}
\usepackage{tabularx}
\usepackage[utf8x]{inputenc} 
\usepackage[dvipsnames,table]{xcolor}
\hypersetup{urlcolor=Mulberry,
	    citecolor=OliveGreen,
	    linkcolor=Blue, 
	    colorlinks=true}
\definecolor{lightgray}{rgb}{0.9,0.9,0.9}	    
\definecolor{green}{rgb}{0,0.5,0}
\definecolor{red}{rgb}{0.5,0,0}
\definecolor{blue}{rgb}{0,0,0.5}

\newcommand{\epsa}{\hat{\boldsymbol{\epsilon}}_1}
\newcommand{\epsb}{\hat{\boldsymbol{\epsilon}}_2}
\newcommand{\vstr}{\mathbf{v}_\mathrm{str}}
\newcommand{\vlab}{\mathbf{v}_\mathrm{lab}}
\newcommand{\lab}{\mathrm{lab}}

\newcommand{\kms}{\textrm{ km s}^{-1}}

\newcommand{\dbd}[2]{\ifmmode \frac{\textrm{d}#1}{\textrm{d}#2}\else $\textrm{d}#1/\textrm{d}#2$\fi}
\newcommand{\pbp}[2]{\ifmmode \frac{\partial#1}{\partial#2}\else $\partial#1/\partial#2$\fi}

\newcommand{\ra}[1]{\renewcommand{\arraystretch}{#1}}

\newcommand{\gl}{\mathpzc{g}_\ell}
\newcommand{\gq}{\mathpzc{g}_q}
\DeclareMathAlphabet{\mathpzc}{OT1}{pzc}{m}{it}

\newcommand{\be}{\begin{equation}}
\newcommand{\ee}{\end{equation}}
\newcommand{\bea}{\begin{eqnarray}}
\newcommand{\eea}{\end{eqnarray}}
\newcommand{\vv}[2]{ \begin{pmatrix}   #1 \\ #2 \end{pmatrix}}

\renewcommand\vec[1]{{\bf #1}}
\newcommand\vp{{\bf p}}
\newcommand\psin[1]{\int\frac{{\rm d}^3{\bf #1}}{(2\pi)^3}}
\renewcommand\({\left(}
\renewcommand\){\right)}

\renewcommand\]{\right]}

\title{Directional axion detection}
\subheader{\hfill MPP-2018-56}
\author[a]{Stefan Knirck,}
\author[a]{Alexander~J.~Millar,}
\author[b]{Ciaran A. J. O'Hare,}
\author[a,b]{Javier~Redondo,}
\author[a]{Frank~D.~Steffen}
\affiliation[a]{Max-Planck-Institut f\"ur Physik (Werner-Heisenberg-Institut),
F\"ohringer Ring 6,\\ 80805 M\"unchen, Germany}

\affiliation[b]{Universidad de Zaragoza, P.\ Cerbuna 12, 50009 Zaragoza, Espa\~{n}a}\emailAdd{cohare@unizar.es}
\emailAdd{knirck@mpp.mpg.de}
\emailAdd{millar@mpp.mpg.de}
\emailAdd{jredondo@unizar.es}
\emailAdd{steffen@mpp.mpg.de}

\abstract{We develop a formalism to describe extensions of existing axion haloscope designs to those that possess directional sensitivity to incoming dark matter axion velocities. The effects are measurable if experiments are designed to have dimensions that approach the typical coherence length for the local axion field. With directional sensitivity, axion detection experiments would have a greatly enhanced potential to probe the local dark matter velocity distribution. We develop our formalism generally, but apply it to specific experimental designs, namely resonant cavities and dielectric disk haloscopes. We demonstrate that these experiments are capable of measuring the daily modulation of the dark matter signal and using it to reconstruct the three-dimensional velocity distribution. This allows one to measure the Solar peculiar velocity, probe the anisotropy of the dark matter velocity ellipsoid and identify cold substructures such as the recently discovered streams near to Earth. Directional experiments can also identify features over much shorter timescales, potentially facilitating the mapping of debris from axion miniclusters.}

\begin{document}
\maketitle
\flushbottom

\section{Introduction}
The axion is a very light pseudoscalar particle that appears as a consequence of the solution of Peccei and Quinn~\cite{Peccei:1977hh, Kim:2008hd} to the strong CP problem of quantum chromodynamics (QCD). The axion has long been an alluring particle candidate to explain the dark matter (DM) that seems to dominate the mass content of the Universe. But now in recent years, with the persistent lack of unambiguous positive signals for any weakly interacting massive particles (WIMPs) from direct and indirect probes, the axion has been enjoying growing popularity. 

Through a variety of mechanisms, axions can sizably contribute to the abundance of dark matter. The subject of axion cosmology is reviewed comprehensively in ref.~\cite{Marsh:2015xka}. Cold dark matter can be produced via the oscillations of the axion field associated with the vacuum realignment mechanism~\cite{Preskill:1982cy,Dine:1982ah,Abbott:1982af,Wantz:2009it}.  In the scenario in which the Peccei--Quinn symmetry is broken before inflation and not restored thereafter, this contribution depends on the one initial misalignment angle $\theta_\mathrm{I}$ in our observable patch of the Universe, with any contributions from topological defects diluted away by inflation. In contrast, in the scenario with post-inflationary Peccei-Quinn symmetry breaking, many different $\theta_\mathrm{I}$ values occur and effects associated with topological defects (domain walls and cosmic strings) have to be taken into account~\cite{Davis:1986xc,Hiramatsu:2012gg,Hiramatsu:2012sc,Kawasaki:2014sqa,Gorghetto:2018myk}. Furthermore, in this scenario sufficiently overdense regions of the axion field that enter matter-radiation equality earlier than their surroundings will have their axions gravitationally bound faster than the surrounding Hubble expansion. The collapse of the mass inside the horizon at this time leaves behind stable clumps of axions called `miniclusters'~\cite{Hogan:1988mp,Kolb:1993zz,Kolb:1993hw,Kolb:1994fi,Kolb:1995bu,Berezinsky:2013fxa,Enander:2017ogx}. These miniclusters may also host solitonic oscillating configurations of the axion field, variously called oscillatons, axitons~\cite{Kolb:1993hw}, axion stars~\cite{Braaten:2015eeu,Eby:2015hyx,Chavanis:2017loo,Visinelli:2017ooc}, bose stars~\cite{Colpi:1986ye,Levkov:2016rkk} or drops~\cite{Davidson:2016uok}. If any of these objects are abundant enough (and indeed stable enough to survive the formation of galactic halos) there may be prospects for their direct~\cite{Tinyakov:2015cgg,Dokuchaev:2017psd}, or indirect detection~\cite{Iwazaki:2014wka,Tkachev:2014dpa,Iwazaki:2014wta,Raby:2016deh,Munoz:2016tmg,Iwazaki:2017rtb,Pshirkov:2016bjr,Fairbairn:2017dmf,Fairbairn:2017sil} today.

Laboratory searches for axions, and their phenomenological generalisation, the axion-like particle (ALP), predominantly rely on their coupling to photons $g_{a\gamma}$. This coupling conveniently allows for the mixing of axions to photons inside magnetic fields. Hence if such particles exist there is the possibility for a measurable flux of ALPs emitted by the Sun (potentially to be observed by the helioscope CAST~\cite{Zioutas:2004hi} and in the future by IAXO~\cite{Armengaud:2014gea}). Moreover, ALPs could  be produced and detected in a purely laboratory setup (such as in the `light-shining-through-a-wall' experiment~\cite{VanBibber:1987rq} ALPS~\cite{Bahre:2013ywa}). However if axions comprise a significant fraction of galactic DM then the value of the axion field should be perpetually oscillating around us at the frequency of the axion mass. So if an experiment can proffer our local DM population a strong enough magnetic field in which to convert, and we are able to precisely measure the subsequent electromagnetic (EM) response, then we will find the axion. Of course the axion mass is unknown, and existing constraints on the ALP-photon coupling tell us that a signal, if present, must be terribly small. So in looking for the axion an experiment must be able to cover a range of frequencies as well as somehow enhance the signal to something measurable. 

Historically the most popular way to enhance a potential axion signal experimentally is to couple it to the resonant mode of a cavity. The ADMX collaboration~\cite{Asztalos:2009yp} found great success with this method and have recently achieved sufficient sensitivity to probe the DFSZ QCD axion model for the first time in a dark matter search~\cite{Du:2018uak}. ADMX are now followed by fervent activity from bright-eyed resonant cavity enthusiasts such as HAYSTAC~\cite{Brubaker:2016ktl,Rapidis:2017ytq,Brubaker:2017rna,Zhong:2017fot,Brubaker:2018ebj}, CULTASK~\cite{Chung:2016ysi,Lee:2017mff,Chung:2017ibl}, Orpheus~\cite{Rybka:2014cya}, ORGAN~\cite{McAllister:2017lkb,McAllister:2017ern} and RADES~\cite{Melcon:2018dba}. The resonant cavity can, and indeed has, set extremely stringent constraints on the axion-photon coupling thanks to rapid development in highly sensitive receiver and amplification technology with noise temperatures nearing the quantum limit. However there are substantial difficulties to be encountered in designing cavities for higher $m_a$ since higher resonant frequencies generally require smaller volumes. Such smaller experiments would suffer in signal strength and therefore sensitivity, unless novel modifications and complex structures are employed, as envisioned in the recent RADES proposal~\cite{Melcon:2018dba}. 

Cavities are well suited to cover axion masses in the range 1--40 $\mu$eV. The search towards higher masses however might be better handled by entirely different designs. For instance some are considering measuring DM axion-induced photon emission from magnetised surfaces. This property is to be exploited by MADMAX~\cite{TheMADMAXWorkingGroup:2016hpc,MADMAXinterestGroup:2017bgn} which is designed to coherently enhance the emitted photons with a series of dielectric disks (see e.g.~ref.~\cite{Millar:2016cjp}). Similarly BRASS~\cite{Brass} is planned to measure this effect as well, but inside a dish antenna configuration, thus achieving a huge effective volume (see also refs.~\cite{Horns:2012jf,Suzuki:2015sza,Jaeckel:2015kea,Knirck:2018ojz}). These experiments will be free from the volume-frequency restriction of the resonator, so are the natural choice to probe larger values of $m_a$. 

The low mass window, below the reach of ADMX, still waits to be explored as well. The vanguard of this region are experiments that persuade the axion field to generate a secondary magnetic flux by circulating the primary axion-induced electric field~\cite{Sikivie:2013laa,Kahn:2016aff}. The ABRACADABRA~\cite{Kahn:2016aff} and DM-Radio~\cite{Silva-Feaver:2016qhh} groups are making progress with this approach, as well as BEAST~\cite{McAllister:2018ndu} which looks to measure the axion-induced electric field directly. 

In this paper we neglect the discussion of the detection of axion couplings to fermions, suffice to say that there are experiments in the planning such as CASPEr~\cite{Budker:2013hfa,Graham:2013gfa} and QUAX~\cite{Barbieri:2016vwg,Crescini:2016lwj,Ruoso:2015ytk,Crescini:2018qrz} to look for them. For an up to date review of all past, present and future experimental searches for axions see ref.~\cite{Irastorza:2018dyq}. 

The primary calling of an axion search experiment is, one will be surprised to hear, to find the axion. However there is good motivation for asking what such a fortunate experiment might be able to offer particle and astrophysics, beyond the initial identification of the axion mass. One possible avenue that has recently been spotted beyond the horizon is the possibility of haloscopes fulfilling their namesake and becoming devices for doing astronomy. Although usually unimportant when exploring over a relatively large range of masses, the thermal distribution of DM velocities would cause a very small spread in the frequency of emitted photons with a width roughly given by the virial velocity dispersion of the DM halo~\cite{Krauss:1985ub}. Past axion searches with ADMX have incorporated some of these astrophysical uncertainties, for example by searching for discrete flows of axions~\cite{Duffy:2006aa,Hoskins:2016svf,Hoskins:2011iv} or applying constraints to different halo models~\cite{Sloan:2016aub,Vergados:2016rlh}. Furthermore there would also be an order 1\% modulation of this lineshape in time due to the relative velocity of the Earth and Sun with respect to the DM halo `wind'~\cite{Turner:1990qx,Ling:2004aj,Vergados:2016rlh}. These are the signals that we can be confident must be present in any successful axion detection and would be essential cross-checks for confirming the discovery of DM. However irregularities in the shape of the axion spectrum and its time evolution would naturally be expected in a halo formed from the hierarchical merger and accretion of subhalos. These irregularities are of additional interest for the study of the history of the Milky Way (MW), galaxy formation in general, as well as improving our understanding of the mechanisms of axion DM production mentioned earlier. More fundamentally the phase space structure of the DM halo on solar system scales ($<$mpc) can only be explored by a terrestrial DM experiment. This epistemology, `axion astronomy', was introduced and studied in detail recently in refs.~\cite{OHare:2017yze,Foster:2017hbq}.

\begin{figure}[t]
\centering
\includegraphics[width=\textwidth]{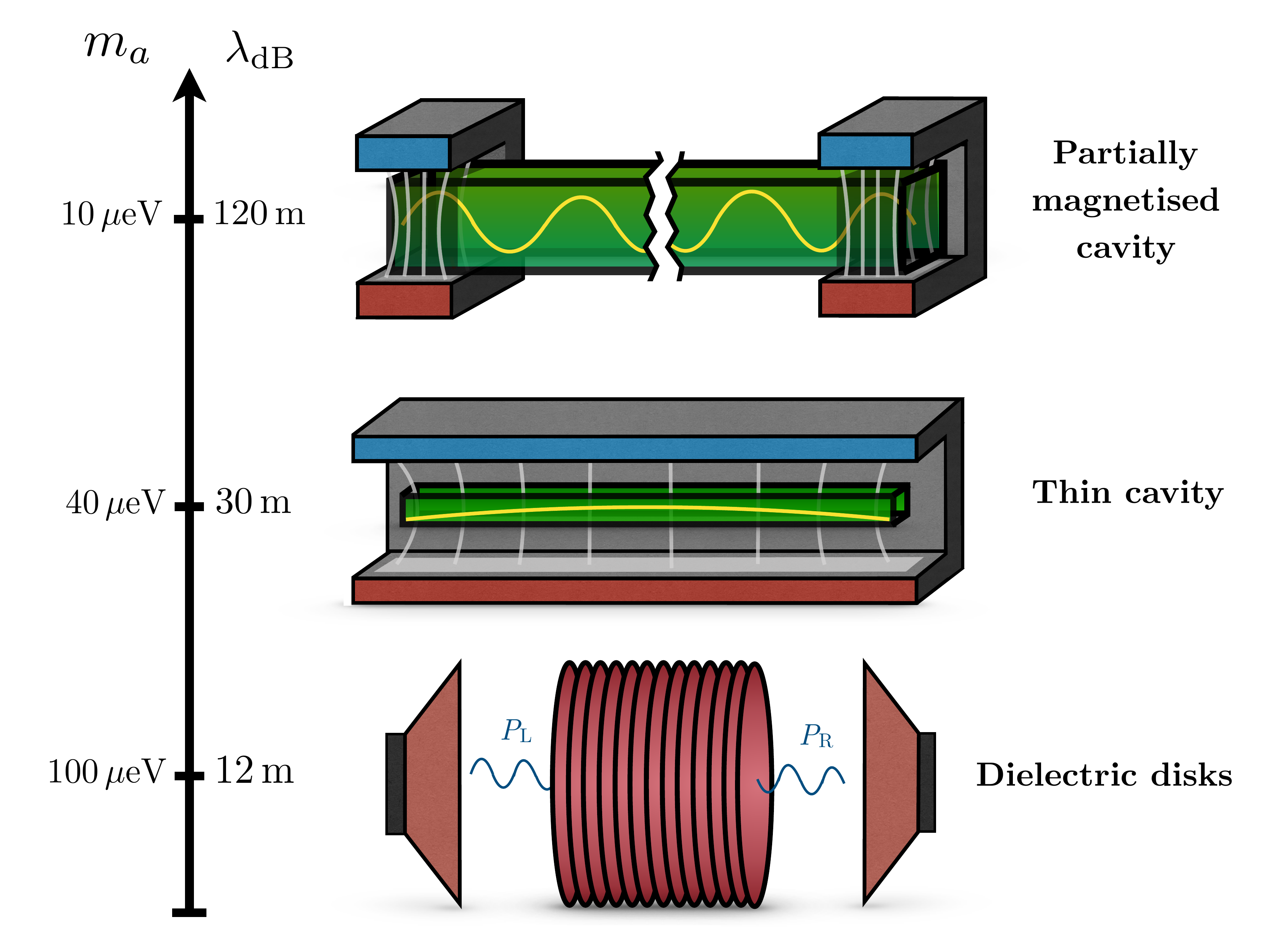} 
\caption{Illustration of our three example directional haloscope designs with their corresponding axion mass and de Broglie wavelength. The experimental parameters needed to achieve directional sensitivity are summarised in table~\ref{tab:benchmarks}.}
\label{fig:haloscope_designs}
\end{figure}
In this paper we aim to enhance the prospects of axion astronomy in future haloscopes by introducing directional effects, first suggested in ref.~\cite{Irastorza:2012jq} in the context of cavities. The advantages offered by \emph{directional} DM detection are well known in the WIMP community~\cite{Mayet:2016zxu}, especially with regard to DM astronomy~\cite{Lee:2012pf,Kavanagh:2016xfi}. We demonstrate here the prospects for the case for axions, which in many cases (as with the findings of the aforementioned non-directional studies) greatly exceed the prospects for WIMPs. The most striking effect when considering directionality in axion experiments is the extremely prominent $\mathcal{O}(1)$ daily modulation present when an experiment has an elongated axis. For comparison the daily modulation in a non-directional experiment is at the $0.2\%$ level. We suggest that one might be able to construct some manner of axion observatory, if multiple experiments are placed adjacent to one another, pointing along orthogonal axes. Although the axion velocity effects can be written in a unified framework, we highlight the technical restrictions on doing astronomy in three example haloscope designs --- two using a resonant cavity setup and one using layered dielectric disks --- covering axion masses between 10 and 100~$\mu$eV. We illustrate these designs in figure~\ref{fig:haloscope_designs}. See table~\ref{tab:benchmarks} and section~\ref{sec:benchmarks} for further numerical details on the required experimental parameters for each. 

We structure this paper as follows. To begin in section~\ref{sec:axiondm} we sketch a description of the behaviour of the axion DM field at ultralocal scales, this will inform our input to the calculation of the expected signal and will allow us to connect a detected signal with the astrophysical velocity distribution for DM, which we also review briefly in this section. Then in section~\ref{sec:directional} we develop our formalism for describing directional effects in axion experiments. We show a general description at first before detailing how one would apply this formalism in practice. In section~\ref{sec:stats} we outline the statistical analysis methodology we will adopt in order to give analytic estimates to the experimental requirements for axion astronomy. We present these results in section~\ref{sec:results}, before concluding in section~\ref{sec:summary}.

\section{Axions and dark matter}\label{sec:axiondm}
\subsection{The local axion field}
The axion DM field $a(\mathbf{x},\,t)$ is born as a coherent state that retains a very large occupation number until today. It is appropriate then to describe it as a classical field. We consider a large box of volume $V_\odot$ centered around the Solar System and describe the axion field as a superposition of plane waves of momentum $\mathbf{p}$,
\begin{equation}\label{modesofaxionfield}
a(\mathbf{x},t)= \sqrt{V_\odot} \int \frac{\textrm{d}^3 \mathbf{p}}{(2\pi)^3} 
\frac{1}{2}\left[a(\mathbf{p}) e^{i(\mathbf{p}\cdot\mathbf{x} - \omega_\vp t)}+
a^*(\mathbf{p}) e^{-i(\mathbf{p}\cdot\mathbf{x} - \omega_\vp t)}
\right]\, , 
\end{equation}
where $\omega_\vp$ is given implicitly by the dispersion relation\footnote{In the gravitational field of the Galaxy the dispersion relation is modified by the gravitational potential, $\Phi$, by $m_a^2\to m_a^2(1+2\Phi(\mathbf{x}))$ at first order. 
The overall effect of the Galaxy can be reabsorbed in a redefinition of time while the spatial variations due to local inhomogeneities in our volume will be neglected.} $\omega_\vp^2=|\mathbf{p}|^2+m_a^2$. 
The average energy density is,   
\be
\label{avdens}
\bar{\rho}_a =   \frac{1}{V_\odot}\int_{V_\odot} {\rm d}^3\mathbf{x}\, \rho_a(\mathbf{x}) = \int \frac{\textrm{d}^3 \mathbf{p}}{(2\pi)^3} \frac{1}{2}\omega_\vp^2 |a(\mathbf{p})|^2 , 
\ee
which must be consistent with local determinations of the dark matter density inferred astronomically at relatively large scales, $\sim\mathcal{O}(100\,{\rm pc}$ -- ${\rm kpc})$. We have densities $\bar{\rho}_a \simeq \rho_0 \simeq 0.4$~GeV~cm$^{-3}$ locally, where $\rho_a$ is the density of axions and $\rho_0$ is the astronomically measured value.
The group velocity of axion waves is $\mathbf{v}= d \omega/ d\mathbf{p} = \mathbf{p}/\omega$. 
A change of variables in eq.~\eqref{avdens} allows us to identify the DM velocity distribution with the Fourier decomposition, 
\be
\label{veldist}
\bar \rho_a \equiv \bar \rho_a \int \textrm{d}^3 \,\mathbf{v} f(\mathbf{v})  \quad , \quad
f(\mathbf{v})  \simeq   \frac{1}{\bar \rho_a}\frac{m_a^3}{(2\pi)^3} \frac{1}{2}m_a^2|a(\mathbf{p})|^2 \, ,
\ee
where we have used $\omega\sim m_a$ in the multiplicative factors. Since DM velocities are of the order of $10^{-3}$ they amount to corrections of order $10^{-6}$ in the formula\footnote{We use natural units $\hbar = c = 1$ throughout except at certain points when we reintroduce $c$ for clarity.}. 

We have a relatively clear idea of the distribution of DM on $\lesssim$ kpc scales, both from observations as well as from N-body and hydrodynamic simulations: the density ought to be essentially homogenous and the velocity distribution will be something resembling a Maxwellian,
\begin{equation}
\label{velocitydist}
f(\vec v)\sim \exp\left(-\frac{|\vec v|^2}{2\sigma_v^2}\right) \, .
\end{equation}
The precise description of this is dealt with in section~\ref{sec:fv}. We must admit however a degree of ignorance when we discuss the DM distribution on the much smaller scales we can probe in an experimental campaign. In 10 years of observation, our laboratories together with the Sun sample only $\sim 2$ mpc of the MW halo. At these scales we have no direct handle of the distribution of DM in simulations or through observation, so we must rely on methods of extrapolation. In particular the question of the ultrafine homogeneity of the MW halo is such a critical one for any successful direct detection of DM, that many attempts have already been made to address it. The possibility of a distribution too clumpy to realistically observe from Earth is a grave one. To soothe one's anxiety, take note of the result of Vogelsberger \& White~\cite{Vogelsberger:2010gd}. In this study the authors follow particle trajectories placed inside an N-body distribution, to trace the subgrid evolution of accreted structure. They find that the typical DM distribution we would sample at Earth is the sum of many $\sim 10^{14}$ ancient streams, with half of all particles contained in streams with densities less than $10^{-7}\rho_0$ today.  With these claims --- supported by other analyses using a range of alternative approaches to the same problem~\cite{Schneider:2010jr,Hofmann:2001bi,Green:2003un,Green:2005fa,Diemand:2005vz} --- we notice that the general opinion tends towards the conclusion of relative homogeneity on the relevant mpc scales. Nevertheless, we must keep in mind the possibility of any non-gravitational `beyond-CDM' interactions that would not be accounted for in these particle-agnostic studies. Even the case of axionic DM alone would warrant a devoted analysis, but this is beyond the scope of our paper. Instead we simply adopt the \emph{assumption} of homogeneity (as suggested by the aforementioned simulations). This is far from a new argument --- almost every theoretical study of direct DM detection works from this assumption --- but in the case of axions there are unexpected consequences. So we should identify the impact of this assumption on our analysis.

The assumption of homogeneity is usually done in an statistical way. The axion density at a point can be expanded into the modes of the field (see ref.~\cite{Enander:2017ogx} for a similar treatment in the context of miniclusters), 
\be
\label{rhox}
\rho_a(\vec x) = \psin{p}\psin{q} \frac{m_a^2}{2}\[e^{i (\vec p-\vec q)\cdot \vec x} 
\cos\big( (\omega_{\vec q}-\omega_{\vec p} ) t\big)a(\vec p)a^*(\vec q)  + ... \]
\ee
where the ellipsis stands for factors of order ${\cal O}( a^2 p^2/m_a^2)$ and therefore negligible. At $\vec x~=~0$, $t = 0$, the density is given by the square of the integral over the complex amplitudes of the modes, $m_a\int d^3\vp\, a(\vp)$. Assuming the distribution of amplitudes with momentum $\mathbf{p}$ is Gaussian, the integral is also a Gaussian. This means that the modulus squared (i.e. the energy density) will be distributed according to an exponential distribution,
\begin{equation}
\label{fkuxturandom}
\frac{d{\cal P}}{d\rho_a} = \frac{1}{\bar \rho_a} \exp\(-\frac{\rho_a}{\bar \rho_a}\). 
\end{equation}
We can use this distribution for any point since $\vec x=\vec 0$ should not be particularly special, it only follows from the randomness of the amplitude coefficients. In our local volume however, $a(\vp)$ is not in fact a statistical variable at all, it is just one fixed complex number (that we would like to eventually measure). But when we sum these complex numbers to measure $\rho_a(\vec x)$ over a volume swept out during an observation, we must account for the phase factor $e^{i\vp\cdot \vec x}$ and the oscillatory cosine which are not constant. Therefore, beyond a certain length and time, the phases at one end of the integral will be uncorrelated with the ones at $\vec x=\vec 0$ and the density we observe will be drawn again from the exponential distribution. The length and time of coherence can be read from the distribution of modes, noting that $|a(\vp)|^2$ are exponentially suppressed above $p_c \sim m_a \sigma_v$.  So $\vp\cdot\vec x \ll 1$ is only true for length scales $|\vec x|\ll 1/p_c\equiv L_c$ and timescales  $t \ll 1/(p_c^2/2m_a)\equiv t_c$. With these coherence scales in mind, consider making repeated observations that sweep out a large enough volume where $V\gg L_c^3$. The fluctuations in the measured density between each of these volumes will now be smaller than suggested by the exponential distribution. Because in each $V$ we have $N = V/L_c^3$ coherence volumes, meaning the integral encodes a random walk over many uncorrelated phases. Eventually the standard deviation of eq.~\eqref{fkuxturandom} will get suppressed by $1/\sqrt{N}\propto \sqrt{L_c^3/V}$.

Importantly for us, this argument will also apply to the fraction of energy associated to axions with frequencies between $\omega$ and $\omega+{\rm d}\omega$, ${\rm d}\rho(\omega)$ and hence a measurement of the velocity distribution. As long as the phases of the integrals in eq.~\eqref{rhox} really are random, the statistics of eq.~\eqref{fkuxturandom} and its suppression as we sum over many coherence volumes will follow,
\be
{\rm d}\bar\rho_a\propto |a(\vec p)|^2 {\rm d}\omega \, .
\ee
where the proportionality factor can be read from eq.~\eqref{avdens}. In any case, the coherence time for modes in a small bin of frequencies is much longer $t_c\sim 1/{\rm d}\omega$ and thus much longer observations are required for the measured density to be drawn again from the distribution. The fundamental statistical nature of the measurement of an axion DM signal was identified only recently by ref.~\cite{Foster:2017hbq} since it was missed in previous in work. The argument sketched here agrees in the final statistical distribution of the signal but is derived in a different way.

A word of warning is in order with respect to the randomness of the Fourier coefficients, their phases in particular. Even if we do believe that the assumption of homogeneity may adequately reflect axion DM produced in the pre-inflationary scenario, in the post-inflationary scenario there is the issue of miniclusters. They have been shown to form in simulations of the axion field at early cosmological times from density perturbations collapsing and decoupling from the Hubble flow (see e.g. refs.~\cite{Hogan:1988mp,Kolb:1993zz,Kolb:1993hw,Kolb:1994fi,Kolb:1995bu}). The characteristic mass of a minicluster is set by the horizon size at formation, typically around the mass of a large asteroid, $M_{\rm mc} \sim 10^{-12}\,M_\odot$. The abundance of miniclusters at formation can be quite high, potentially constituting the leading fraction of the DM~\cite{Kolb:1995bu}. Sadly, it is highly unlikely that we will pass through one in our lifetime\footnote{Though a prediction like this depends on the mass function, density profile, spatial extent and overall abundance of a minicluster population, all of which are being actively investigated~\cite{Enander:2017ogx,Fairbairn:2017sil}.}. Even if the entirety of the dark matter were in the form of miniclusters and there were on the order of $\sim 10^{19}~\textrm{kpc}^{-3}$ locally, a direct encounter would occur less than once every $10^5$ years. 

With the assumption of completely random coefficients, large upwards fluctuations of the density are relatively rare. For instance, for an axion mass of $10 \,\mu\rm eV$, with $L_c\sim 1/m_a\sigma_v\sim 20$ m, we expect only $\sim 1$ volume $L_c^3$ in the entire local kpc$^3$ that would have the phases and amplitudes arranged in such a way to give a measurement of an overdensity $\sim 100 \rho_0$. On the other hand, the typical minicluster can easily reach an overdensity many orders of magnitude larger than this, even though the large scale averaged velocity distribution for miniclusters and a homogenous axion field should be the same. So how can it be that the same distribution of Fourier amplitudes $|a(\vp)|$ can describe both a consistently observable smooth population of dark matter, and an almost unobservable sparse distribution of miniclusters? The information of such extreme clumpiness can only be encoded in the \emph{correlations} of the Fourier phases. For miniclusters the phases are such that only around one particular fine tuned place do they add coherently. For an example, consider the following model for a Gaussian minicluster of radius $R$. Taking the Fourier transform of this lump of axions we have, 
\be
a \sim a_0 e^{-\frac{|\mathbf{x}-\mathbf{x}_0|^2}{2 R^2} }\cos(m_a t)  \quad \to \quad 
a(\mathbf{p}) \propto a_0 e^{i m_a t} e^{-\frac{|\mathbf{p}|^2 R^2}{2}} e^{i \mathbf{x}_0\cdot \mathbf{p}} 
\ee
revealing $|a(\mathbf{p})|\propto a_0$ and ${\rm arg}\left[a(\vp)\right] = \mathbf{x}_0\cdot \mathbf{p}$. The Gaussian envelope retains no information about the position of our lump, but the phase does. It is clear from the spatial representation that for $|\mathbf{x}_0|\gg R$ our detectors will not see the axion DM lump. In Fourier space this is encoded in the \emph{correlated} but extremely quickly varying phase if $|\mathbf{x}_0|\gg 1/p_m$ where $p_m$ is the characteristic momentum of the minicluster distribution. So in a sense, if one is far outside of the lump then the phase is oscillating so wildly between momenta that each `step' in the random walk is cancelling the previous one. On the other hand, inside the lump the phase can vary slowly and allow the measured density to build up to a very large value.

In this paper we will assume that a smooth distribution of axion DM at kpc scales is still valid at the mpc scales relevant for experiments and axions are not overwhelmingly bound up in miniclusters. In any case, our study begins from the hypothesis that the axion has already been found in an experiment, so the argument is at the very least self-consistent. 

\subsection{Detecting axions}
We explore directional effects in haloscope experiments, i.e. those that exploit the axion coupling to the photon $g_{a\gamma}$ allowing a mixing between axion and EM fields inside static magnetic fields. The QCD axion-photon coupling is related to the axion mass via the relation,
\begin{equation}\label{eq:axioncoupling}
\frac{g_{a\gamma}}{{\rm GeV}^{-1}} = 2.0\times10^{-16} C_{a\gamma}\frac{m_a}{\mu {\rm eV}} \, .
\end{equation} 
Where the $\mathcal{O}(1)$ number $C_{a\gamma}$ is model dependent (see ref.~\cite{Irastorza:2018dyq} for a discussion). Throughout we make the supposition that the discovered axion turned out to be from the KSVZ model~\cite{Kim:1979if,Shifman:1979nz} so $|C_{a\gamma}| = 1.92$. Since signals in haloscope experiments depend on the coupling as $g_{a\gamma}^2$, one should use this fact to rescale our results to match any alternative QCD axion (or indeed ALP) model at the quoted masses\footnote{The same is true for the axion density and fraction of axionic dark matter which would scale the signal linearly; though throughout we assume $\bar\rho_a = \rho_0 = 0.4\, {\rm GeV\,cm}^{-3}$.}. The derivation of the effects in question begin with the axion-modified Maxwell's equations for magnetic and electric fields $\mathbf{B}$ and $\mathbf{E}$,
\begin{eqnarray}
\nabla \cdot \mathbf{E} &=& \rho_q - g_{a\gamma} \mathbf{B} \cdot \nabla a \\
\nabla \times \mathbf{B} - \dot{\mathbf{E}} &=& \mathbf{J} + g_{a\gamma}(\mathbf{B}\,\dot{a} - \mathbf{E} \times \nabla a) \\
\nabla \cdot \mathbf{B} &=& 0 \\
\nabla \times \mathbf{E} + \dot{\mathbf{B}} &=& 0 \\
(\Box + m^2_a)a &=& g_{a\gamma} \mathbf{E}\cdot \mathbf{B}\, ,
\end{eqnarray}
where $\rho_q$ and $\mathbf{J}$ are the electric charge density and current. In the following we will assume that a static magnetic field $\mathbf{B}_e$ is applied over some experimental volume $V$, and the resulting axion-photon oscillations are enhanced through a coupling to a resonant mode, or through the correct spacing of a series of dielectric disks. The dependence on $f(\mathbf{v})$ appears when one offers these equations an axion plane wave $a \sim a_0 \exp{\left[i(\mathbf{p}\cdot\mathbf{x} - \omega t)\right]}$. The plane wave will have some frequency and momentum selected from the DM velocity distribution that will be reselected over the characteristic coherence length and time. For a typical speed of $300 \kms$ these are,
\begin{equation}
t_c = \frac{2\pi}{m_a v^2} \,  = 40 \, \mu{\rm s} \left( \frac{100 \, \mu{\rm eV}}{m_a} \right) \, ,
\end{equation}
\begin{equation}\label{eq:coherence}
L_c = \frac{\pi}{m_a v} = 6.2 \, {\rm m} \left( \frac{100 \, \mu{\rm eV}}{m_a} \right) \, .
\end{equation}
The characteristic time of coherence is considerably shorter by many orders of magnitude than the typical integration times of most experiments (even for lower masses than the benchmark used here). So the Fourier transform of the signal collected over many thousands of these durations will approach the \emph{speed} distribution $f(v)$ up to some exponentially distributed coefficient at each speed/frequency bin coming from the uncorrelated nature of the phases as described earlier. This type of measurement is the focus of refs.~\cite{OHare:2017yze, Foster:2017hbq}. Here we account for an additional effect; if the linear scale of $V$ is larger than the typical $L_c$ then the axion will oscillate with a slightly different phase across the dimensions of the experiment. So any measured signal will be modified slightly by how out of phase the oscillation is at one end of the device compared with the other. The size of this effect at some instant will be given by the angle between the axion direction and the preferred axis for the experiment. Accounting for this effect on a power spectrum measurement over some finite time essentially constitutes a correction from a weighted integral of the \emph{velocity} distribution $f(\mathbf{v})$. This effect was introduced in ref.~\cite{Irastorza:2012jq} but how one can exploit it to make a measurement of $f(\mathbf{v})$ in 3D has not been studied in detail before.

\subsection{The velocity distribution}\label{sec:fv}
Most dark matter detection analyses are performed under a simple assumption for the MW halo known as the standard halo model (SHM)~\cite{Green:2011bv}. This is a spherically symmetric isothermal halo model. Its $1/r^2$ density profile yields a Maxwell-Boltzmann velocity distribution with peak speed $v_0 $ and dispersion $\sigma_v = v_0/\sqrt{2}$. The distribution ought to be truncated at the escape speed of the Galaxy~\cite{Piffl:2013mla}, but given the exponential suppression of fast $v$, this has an extremely marginal effect for most axion direct detection signals. The velocity distribution in the galactic frame is given by:
\begin{equation}\label{eq:shm}
f(\mathbf{v}) = \frac{1}{(2\pi \sigma_v^2)^{3/2}} \, \exp \left( - \frac{|\mathbf{v}|^2}{2\sigma_v^2}\right) \,.
\end{equation}
We may also allow for the velocity distribution to be anisotropic in the galactic frame. We discuss this possibility and prospects for detection in section~\ref{sec:anisotropy}.

There have been long-standing concerns raised by the results of DM-only N-body simulations that the SHM may be a poor reflection of the real MW halo~\cite{Vogelsberger:2008qb,Maciejewski:2010gz,Mao:2012hf}.  Interestingly however, more recent analyses of hydrodynamic simulations have found that the simple Maxwellian distribution of the SHM may, at least functionally, be sufficient to describe the local velocity distribution for the purposes of direct detection~\cite{Bozorgnia:2016ogo,Sloane:2016kyi,Kelso:2016qqj,Lentz:2017aay}. However there are quantitative disagreements about whether the local $f(v)$ should be shifted higher or lower peak speeds from the SHM value of $220 \kms$. The solution suggested by ref.~\cite{Kelso:2016qqj} is the correlation between the circular rotation speed (which is related to the peak speed) and the stellar mass of the halo. Despite these quantitative discrepancies, they do agree that the addition of baryons improves the fit to the Maxwellian locally.
 
In the absence of a detection, a narrower speed distribution strengthens constraints on axions since a narrower line shows up more strongly over thermal noise. In the case of a detection (which is the focus of our work) the issue is immaterial since we simply measure the peak and width of the distribution directly. Indeed the comparison between this direct measurement and the aforementioned simulations will be an excellent way to refine the mass model and evolution history of the galaxy, in particular the relationship between the stellar and dark matter halos. Ultimately though a measurement from Earth is the most direct way to learn about the structure of dark matter halos on the scales inaccessible to simulations, and about \emph{our} galaxy in particular. However along these lines we must mention recent work showing that information on a slightly larger scale about our DM velocity distribution could be determined empirically using astrometric survey data. Reference~\cite{Herzog-Arbeitman:2017zbm} showed that the kinematics of metal-poor stars, those which populate the stellar part of the halo, can be used as tracers for the velocity distribution of the virialised dark matter part of the halo. A determination was made applying this method to stars from RAVE and \emph{Gaia} in ref.~\cite{Herzog-Arbeitman:2017fte}. They observe a narrower distribution than the SHM prediction, in agreement with the N-body inspired axion lineshape~\cite{Lentz:2017aay} which is currently used by ADMX~\cite{Du:2018uak}. In the future these three complementary methods --- simulations, astronomy and direct detection --- will comprise a powerful multi-perspective view of the structure and growth of galactic halos on a wide range of scales.

\subsection{Streams}\label{sec:intro_streams}
One of the most interesting questions we can ask of our local population of dark matter is about the presence of substructure. For instance streams are seen generically in simulations of Milky Way-like galaxies as smaller subhalos become absorbed by their larger host. In fact they are an inevitable consequence of the hierarchical growth of structure. The early numerical simulations of ref.~\cite{Stiff:2001} suggested that there was an $\mathcal{O}(1)$ probability for a stream to make up 1--5\% of our local density. We now know of many examples of such substructure in the inner Milky Way~\cite{NEWBERG:2016,Myeong:2017skt,Lancaster:2018,Shipp:2018yce,Myeong:Preprint}. Nearby streams can be identified either by looking for overdensities of individual stars or as phase space structures that have remained kinematically cold. Some have been known for many years, for example the stream from the famous Sagittarius dwarf~\cite{Newberg:2003cu,Yanny:2003zu,Majewski:2003ux,Luque:2016nsz} (a favourite benchmark for direct detection theory papers~\cite{Lee:2012pf,O'Hare:2014oxa,Savage:2006qr,Foster:2017hbq,OHare:2017yze,Kavanagh:2016xfi}). Unfortunately 
after several years of mapping across the sky with multiple stellar tracer populations, the Sagittarius stream is now known to not pass close to the Sun~\cite{Koposov:2012, Belokurov:2014}. 

Nevertheless our local neighbourhood may not be so bereft of streams after all. Thanks to the transformative data set from \emph{Gaia}~\cite{Gaia:2016}, more candidates have been found, including six stream-like or `clumpy' objects which were shown to approach the Solar position~\cite{Myeong:2017skt}. One object in particular denoted `S1' is certainly a stream and with a judicious selection of stars in phase space can be shown to have a mean position consistent with our galactic location~\cite{Myeong:Preprint}. The S1 stream has a galactocentric velocity of around 300 km s$^{-1}$ and is incoming in the same direction as the dark matter wind. Whilst these velocities can be well-measured, there is still some doubt regarding how much one can assume about the dark component of a stream from its stars. S1 is believed to have an infallen over a time of $\gtrsim$~9 Gyr from a progenitor with a total mass of around $10^{10}\,M_\odot$ (around the mass of the largest MW dwarf spheroidal, Fornax), so there is a good case to be made for a sizable dark matter component. Furthermore there may indeed be streams from dark subhalos that never contained stars to begin with. It is expected that around $\sim 100$--$200$ streams will be found in the inner halo of the MW over the next few years with the phase space method~\cite{Lancaster:2018}. For us there is no need to make any assumptions, but these objects are attractive as a first set of benchmarks that are in some way grounded in reality. Again, the mysteries of the dark hearts of streams ought to be unveiled by detecting dark matter! We discuss the detectability of streams more in section~\ref{sec:stream}.

If a stream passed through the solar system it would exist as a distinct component of the local dark matter phase space distribution with speeds tightly concentrated around a single velocity $\vstr$. The velocity distribution of a stream can be written similarly\footnote{In using models like this one should decide whether the stream is an additional contribution to dark matter on top of $\rho_0 \sim 0.4$ GeV cm$^{-3}$ or if it would comprise a fraction of $\rho_0$. The former would be best if the substructure is small enough in extent to not affect local determinations of the dark matter density with stars beyond a few parsecs away, i.e. the stream surrounds the Earth but not nearby stars. On the other hand if the substructure is on the order a few hundreds of parsecs in size or larger (as is expected for streams from dwarfs) then it would contribute to the local gravitational potential and hence determinations of $\rho_0$.},
\begin{equation}
f_\mathrm{str}(\mathbf{v}) = \frac{1}{(2\pi \sigma_\textrm{str}^2)^{3/2}} \, \exp \left( - \frac{(\mathbf{v} - \mathbf{v}_\textrm{str})^2}{2\sigma_\textrm{str}^2}\right) \,,
\end{equation}
where $\sigma_{\rm str}$ would be $\mathcal{O}(10) \kms$.

We (like others before us~\cite{Freese:2003tt,O'Hare:2014oxa,Foster:2017hbq}) will focus on substructure in the form of streams, since the case for their presence nearby is the most compelling. However other creatures have been suggested variously in the literature such as debris flows~\cite{Kuhlen:2008qj,Lisanti:2011as,Kuhlen:2012fz,Vergados:2012xn}, shadow bars~\cite{Petersen:2016xtd,Petersen:2016vck} and dark disks~\cite{Bruch:2008rx,Read:2009iv,Purcell:2009yp,Schaller:2016uot,Schutz:2017tfp}. The latter of these would lead to an enhancement of $f(v)$ at low speeds. Such a situation would be of no great concern for the detection of axions, in fact an enhanced low-speed population of dark matter would only increase the signal strength (the reverse is true for WIMPs~\cite{Billard:2012qu}). In any case the dark disk scenario is believed to be unlikely since they are usually formed after a significant late merger~\cite{Schaller:2016uot} and can be constrained with astrometric data, as in ref.~\cite{Schutz:2017tfp} for example. 

Finally we comment that the bestiary of substructure roaming our local halo may be enriched by the mechanisms involved in the cosmological production of dark matter. As discussed in the previous subsection, for axions produced in the post-inflation scenario, substructure in the form of miniclusters is expected. We mentioned that it is highly unlikely that we will pass through an individual minicluster in our lifetime, but an interesting prospect for direct detection is if this initial population of axion miniclusters are tidally disrupted by stellar interactions inside a galactic halo over many orbits through the disk and bulge~\cite{Tinyakov:2015cgg, Dokuchaev:2017psd}. This could result in a network of streams wrapping the Milky Way each with much smaller radii than tidal streams from the stripping of satellites. A journey through this network would be characterised by temporary enhancements in the axion signal over timescales between a few hours to many days depending on the size of the original minicluster. Clearly if we wish to detect a ministream we need an experiment that can measure signals that tell us its velocity components within this duration.
 
\subsection{Signal modulations} 
We observe the velocity distribution of DM particles in the rest frame of the laboratory, so the $f(\mathbf{v})$ that we use to construct our power spectrum must undergo a Galilean transformation into to the lab rest frame by the time dependent velocity $\vlab(t)$. In section~\ref{sec:directional} we describe how we can build experiments that are most sensitive in a particular direction. So to measure the velocity distribution in 3D it is sensible to arrange three of these experiments orthogonal to one another in a $(\hat{\mathcal{N}},\,\hat{\mathcal{W}},\,\hat{\mathcal{Z}}) = ({\rm North,\,West,\,Zenith})$ coordinate system. We assume that the experiment is located at latitude and longitude $(\lambda_\textrm{lab},\,\phi_\textrm{lab})$. The angle between $\vlab$ and these axes will be diurnally modulated by the rotation of the Earth. We describe the calculation of these three daily modulations in appendix~\ref{sec:labvelocity}. For now we skip to the final result which is the daily modulation of $\vlab(t)$ projected along each axis, 
\begin{eqnarray}\label{eq:costhlabs}
v_{\rm lab}^\mathcal{N}/v_{\rm lab} = \cos{\theta_{\rm lab}^\mathcal{N}}(t) &=& b_0 \cos{\lambda_{\rm lab}} -b_1 \sin{\lambda_{\rm lab}} \cos{\left(\omega_d t + \phi_{\rm lab} + \psi \right)} \, ,\\
v_{\rm lab}^\mathcal{W}/v_{\rm lab} = \cos{\theta_{\rm lab}^\mathcal{W}}(t) &=& b_1 \cos{\left(\omega_d t + \phi_{\rm lab} + \psi-\pi\right)} \, ,\\
v_{\rm lab}^\mathcal{Z}/v_{\rm lab} = \cos{\theta_{\rm lab}^\mathcal{Z}}(t) &=& b_0 \sin{\lambda_{\rm lab}}  + b_1 \cos{\lambda_{\rm lab}} \cos{\left(\omega_d t + \phi_{\rm lab} + \psi\right)} \, ,
\end{eqnarray}
where the frequency is $\omega_d = 2\pi/$(1 siderial day), and the constants $b_1$, $b_2$ and $\psi$ vary slowly over the year, but can be taken as approximately constant over a duration of a couple of days. These constants can be inverted to find the three components of the Solar velocity, $\mathbf{v}_\odot$, see eq.~\eqref{eq:vodot_sol}. The angle between the Earth's equator and $\vlab(t)$ varies between 41$^\circ$ and $54^\circ$ degrees over the year so locations between these latitudes would be optimally placed to have a large daily modulation in all three experiments throughout the whole year\footnote{The locations of CAST, ADMX, HAYSTAC and ABRACADABA, as well as the proposed site for MADMAX already satisfy this condition.}. An experiment could also be placed on a tilt to mimic the effect of being at a different latitude. Every example we use takes the location of the experiment to be Munich with coordinates $(\lambda_{\rm lab},\,\phi_{\rm lab}) = (48^\circ,\,12^\circ)$.

The modulations due to the movement of the laboratory with respect to the DM wind are the only ones we consider since we make the assumption of homogeneity in the smooth component of the axion field on our mpc scales. However we would like to briefly note that we know that there will certainly be inhomogeneities induced even more locally than this due to the gravitational field of the Sun~\cite{Lee:2013wza}. This effect of gravitational focusing was identified as an issue for axion astronomy by the authors of ref.~\cite{Foster:2017hbq} who implement it perturbatively at leading order in $G$ as a correction to the velocity distribution, see ref.~\cite{Buschmann:2017ams}. We are behind the Sun with respect to the DM wind during March so the greatest amount of focusing is observed during Northern Hemisphere spring. The effect is around 1--2\% at the level of the distribution and is largest for small values of $v$. This means that the measurement of signals which modulate with a period of a year or those at low speeds in the distribution will be biased by not taking this effect into consideration. The modification turns out to be at a higher harmonic order than a simple amplitude or phase shift. We neglect gravitational focusing here since the bulk of our analysis involves comparing diurnally modulating signals as well as fast substructure such as streams. For these focusing amounts to an essentially negligible correction that comes with a rather large computational expense. However as demonstrated in ref.~\cite{Foster:2017hbq}, to make an unbiased measurement of the phase and amplitude of \emph{annual} modulation, the focusing effect should be accounted for.

\section{Directional axion haloscopes}\label{sec:directional}
\subsection{General formalism}
\label{generalformalism}
To have a consistent discussion of directionally sensitive experiments, we need a unified framework on to which we can map specific experimental designs. The following subsections~\ref{sec:cavity} and~\ref{sec:dielectrics} will deal with cavity and dielectric experiments respectively. Fortunately, both of these haloscope designs permit an overlap integral formalism. This can be seen either by classical electromagnetic calculations or from the lowest order of perturbation in quantum field theory~\cite{Sikivie:1983ip,Millar:2016cjp,Ioannisian:2017srr}. The latter is useful as we only need to do one calculation to cover both cavities and dielectric haloscopes. The inverse lifetime for a single axion with energy $\omega_a$ to convert to a photon is 
\begin{equation}\label{eq:decayrate}
\Gamma_{a\to\gamma}=2\pi\sum_{\bf k}|{\cal M}|^2\,\delta(\omega_a-\omega_{\bf k})\,.
\end{equation}
Here ${\cal M}=\langle {\rm f}|H_{a\gamma}|{\rm i}\rangle$ is the matrix element of the interaction Hamiltonian between the initial and final state given by
\begin{equation}\label{eq:general-matrix-element}
{\cal M}=\frac{g_{a\gamma}}{2\omega V}\int {\rm d}^3{\bf x}\,
\,{\bf B}_{\rm e}({\bf x})\cdot{\bf E}^*_{\bf k}({\bf x})e^{i{\bf p}\cdot{\bf x}}\, ,
\end{equation}
for $\omega = \omega_a = \omega_\mathbf{k}$, where ${\bf E}^*_{\bf k}$ is the free photon wave function and ${\bf B}_{\rm e}$ is the external magnetic field. In the dielectric haloscope ${\bf E}_{\bf k}$ is given by a Garibian wave function~\cite{Ioannisian:2017srr}. In general ${\bf k}$ denotes some general set of quantum numbers that describe the photon wave functions, for example momenta or mode numbers. Note that the quantum field calculation described above has a limitation: formally one must know the final state to which the axion converts, which to be detectable must be a state that extends outside the cavity. Exactly how the signal leaves the cavity is usually unspecified in the literature when discussing a generic setup. We assume for simplicity that the energy leaves the cavity via photons. The part of the photon wave function outside the cavity will generally be oscillatory so does not contribute to the integral. Thus the only way the external, measurable part of the photon wave function enters the calculation is in the normalisation of ${\bf E}_{\bf k}$. So for a cavity ${\bf E}_{\bf k}$ is given by the resonant mode up to some normalisation from a quality factor, which describes the transition rate of photons inside the system to energy outside of the system. 

Thus the power generated for a given axion momentum $\bf p$ is
\begin{equation}
P_{\bf p}\propto \left|\int {\rm d}^3{\bf x}\, {\bf E_{\bf k}}({\bf x})\cdot {\bf B}_{\rm e}e^{i{\bf p}\cdot{\bf x}}\right|^2\, .
\end{equation}

If we write the number of axions inside the device at a time $t$ with velocity $\bf v$ as $N(\mathbf{v}; t)$ then we can write the corresponding power in a form more familiar to those conversant with cavity experiments,
\begin{equation}
P_{\bf p} = \kappa g^2_{a\gamma} B_{\rm e}^2 C({\bf v})N(\mathbf{v}; t) Q_{\rm eff} \, , \label{eq:power}
\end{equation}
where $\kappa$ is the coupling efficiency, $V$ is the volume of the device, $Q_{\rm eff}$ is some effective ``quality factor" and the form factor $C$ is given by
\be
C({\bf v})=\frac{2\left|\int {\rm d}V {\bf E_{\bf k}}({\bf x})\cdot {\bf B}_{\rm e}e^{i{\bf p}\cdot{\bf x}}\right|^2}{B_{\rm e}^2V\int {\rm d}V\left[\epsilon({\bf x})|{\bf E_{\bf k}}({\bf x})|^2+\mu({\bf x})|{\bf B_{\bf k}}({\bf x})|^2\right]}\,, \label{eq:formfactordef}
\ee
with $\epsilon({\bf x})$ being the relative permittivity and $\mu({\bf x})$ the permeability (which will generally be set to 1). Formally non-resonant devices do not have a quality factor, however an analogous quantity can be defined for dielectric haloscopes~\cite{TheMADMAXWorkingGroup:2016hpc}. In the case of resonant cavities, rather than a general proof, which requires detailed knowledge of the final photon state, we note that the normalisation of the photon wave function is unaffected by the velocity of the axion, and thus must agree with Sikivie's original calculation~\cite{Sikivie:1983ip}. However, one can explicitly show that eq.~\eqref{eq:power} holds for more specific cases where the final state is specified. For example, for open resonators this normalisation was shown in refs.~\cite{Millar:2016cjp,Ioannisian:2017srr}\footnote{The former reference shows that in the zero velocity limit, Sikivie's original calculation agrees with the classical calculation of dielectric haloscopes in a resonant limit, and the latter reference shows that such a calculation is equivalent to a perturbative  quantum field calculation as described here.}.  Such an argument can be applied to a rectangular cavity, under the assumption that the leaked power is due to a non-zero transmissivity in the end caps.

We will make the assumption here that the DM density measured in the experiment agrees with the average local DM density $\bar \rho_a$, so we see that the total power is given by
\begin{equation}\label{eq:totpower}
P=\int {\rm d}^3{\bf v} P_{\bf p}=\kappa g^2_{a\gamma} B_{\rm e}^2 V \frac{\bar \rho_a}{m_a} Q_{\rm eff}\int {\rm d}^3{\bf v} f({\bf v})C({\bf v})\, .
\end{equation}
To see how $C$ depends on $\bf v$, we note that there are only two effects from the velocity of the axion: a change in the frequency, and a change of phase. Only the change in phase of the axion can provide a directional sensitivity but since the velocity is multiplied by the dimensions of the device, such an effect can be very significant. Expanding the axion phase we have
\be
C({\bf v})\propto\left|\int {\rm d}^3{\bf x}\, {\bf E_{\bf k}}({\bf x})\cdot {\bf B}_{\rm e}\left(1+i{\bf p}\cdot{\bf x}-\frac{({\bf p}\cdot{\bf x})^2}{2}+...\right)\right|^2 \, .\label{eq:cexpand}
\ee
Note that if ${\bf E_{\bf k}}$ is a standing wave then it has no spatial phase variation so can be treated as real. Then after taking the modulus squared no linear order terms in the velocity can survive since they enter purely imaginarily. Cavity haloscopes always satisfy this condition, meaning that they never have a linear dependence on the velocity. However, it is possible to design dielectric haloscopes for which the free photon wave function has traveling behaviour~\cite{Millar:2016cjp}. Thus at lowest order the geometry factor will be either linearly ($\ell$) or quadratically ($q$) dependent on the velocity components, allowing us to define
\be
C({\bf v})=C_0\left(1-\mathcal{G}_{\ell,\,q}({\bf v})\right) \, ,
\ee
containing either,
\begin{equation}
\mathcal{G}_{\ell}(\mathbf{v}) =  \sum_{i = 1}^3 \gl^i v_i\, ,
\end{equation}
or,
\begin{equation}\label{eq:quadratic}
\mathcal{G}_{q}(\mathbf{v}) = \sum_{i = 1}^3 \sum_{j = 1}^3 \gq^{ij} v_iv_j\, ,
\end{equation}
where we have pulled out the form factor $C_0$ in the 0-velocity limit. Keep in mind that if ${\bf p}\cdot{\bf x}\sim 1$ then one cannot only consider just the lowest order contributions. To gain directional sensitivity, we enforce the primary effect on the geometry factor to come from a single direction, corresponding to an elongated dimension. While in general eq.~\eqref{eq:quadratic} could contain cross terms, we will see that in our examples there are only factors proportional to $v_i^2$, so we will drop one of the superscripts and just give $\gq^i$. In these instances there is the unfortunate aspect that the geometry factor is insensitive to the sign of $v_i$.

\begin{figure}[t]
\centering
\includegraphics[width=0.78\textwidth]{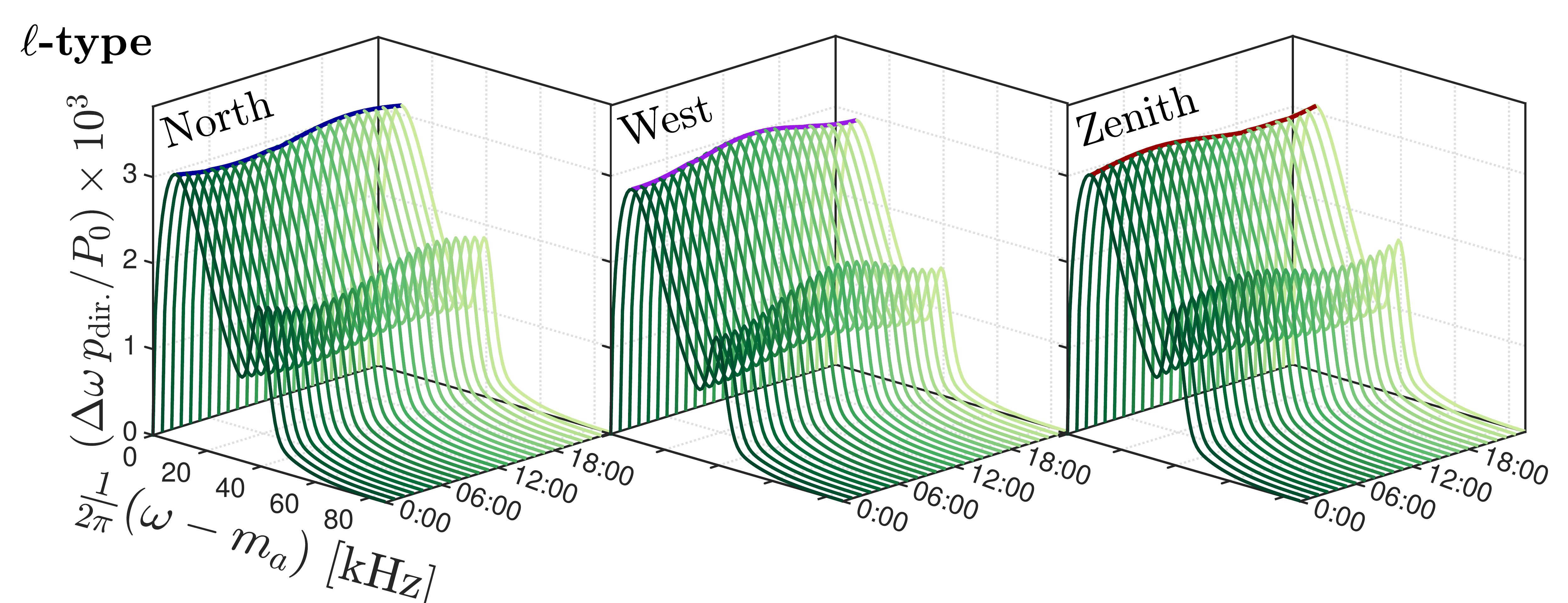} 
\includegraphics[width=0.78\textwidth]{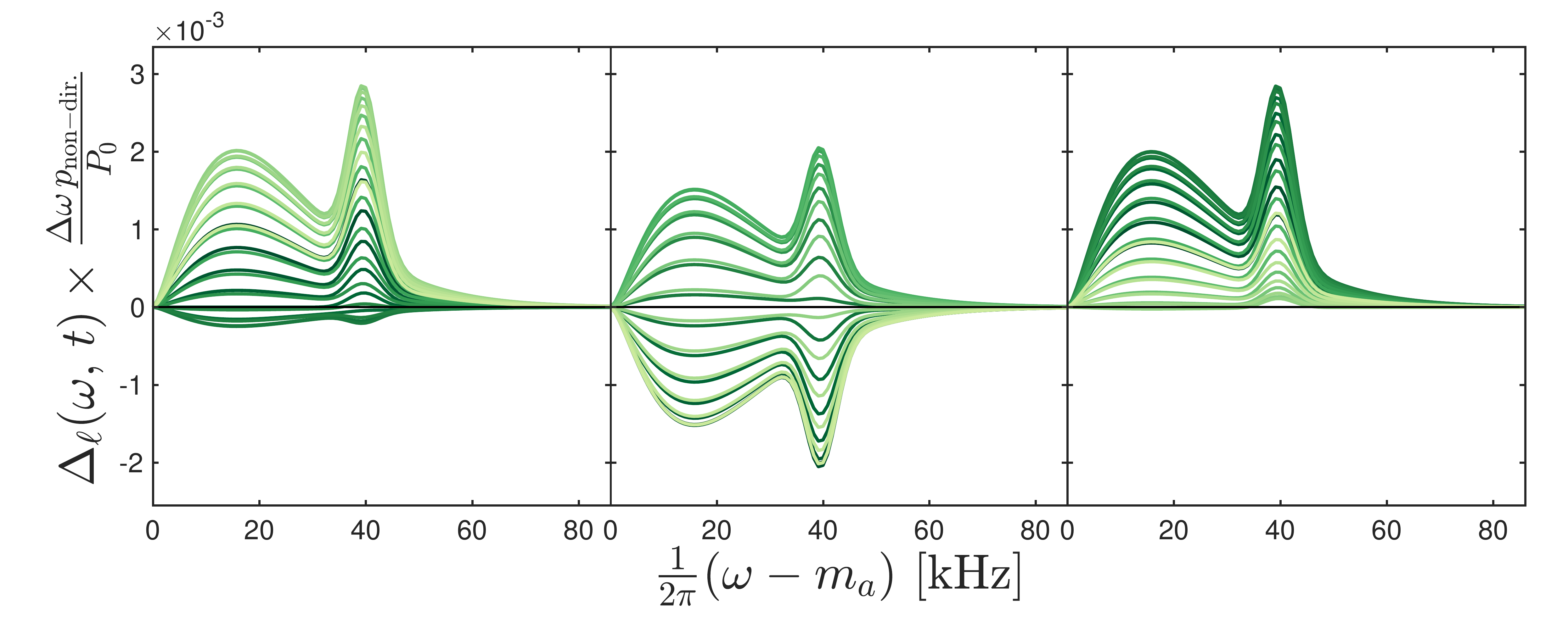} 
\includegraphics[width=0.78\textwidth]{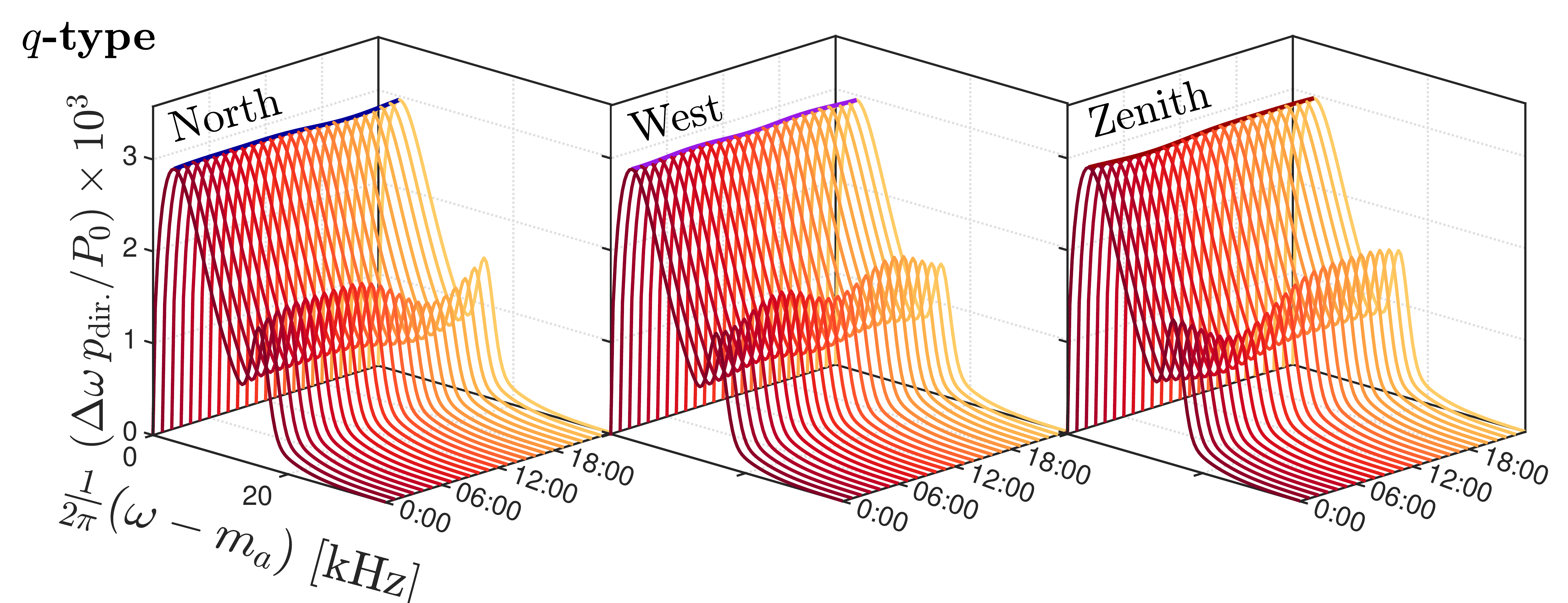} 
\includegraphics[width=0.78\textwidth]{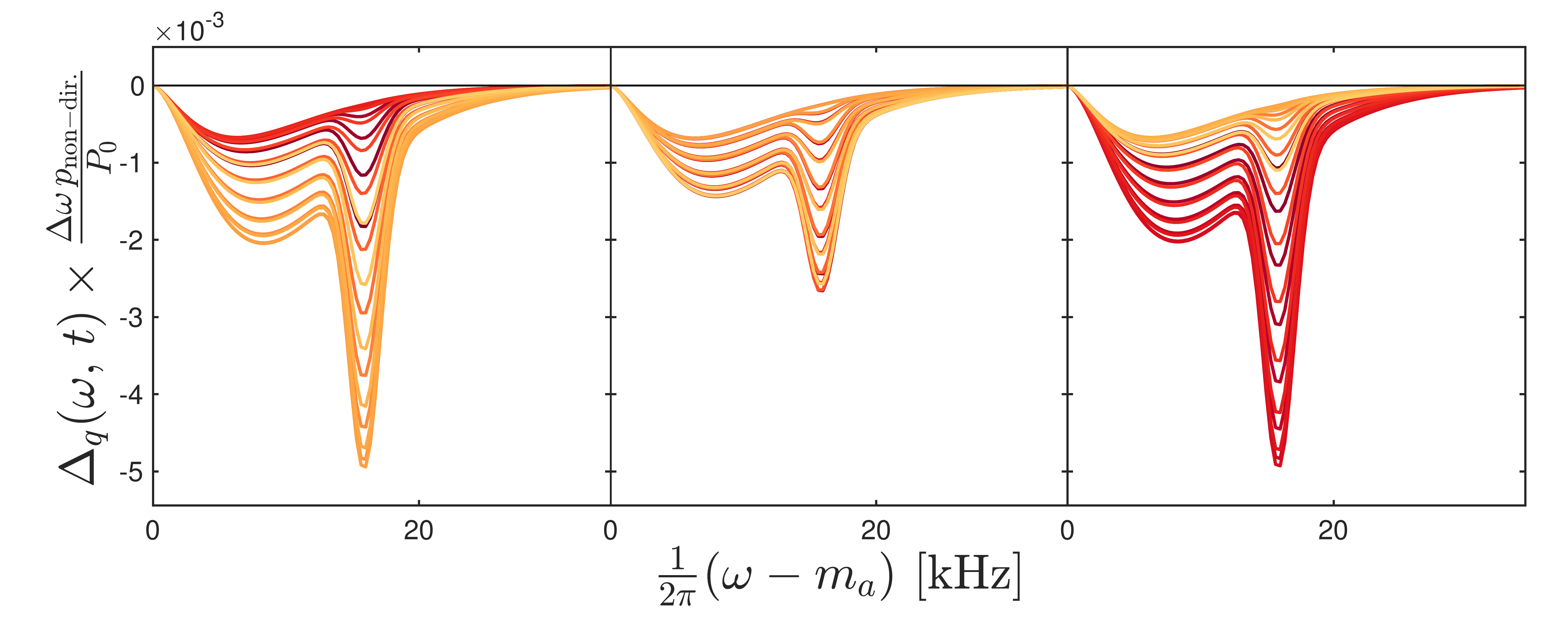} 
\caption{Directionally sensitive power spectra for a smooth Maxwellian distribution of axions with a 10\% contribution from the S1 stream. We show the expected signals for both linear (``$\ell$-type'': top, red) and quadratic (``$q$-type'': bottom, green) experiments. The upper panels in each case are the binned values of power relative to the total power $P_0$ whereas the lower panels isolate the directional effect.}
\label{fig:powerspectra}
\end{figure}
The type of signal we want to use to measure the velocity distribution is a power spectrum $\mathrm{d}P/\mathrm{d}\omega$, obtained in practice by taking the Fourier transform of whatever EM signal was being tracked. Since a power spectrum is one dimensional, only a distribution of speeds will be measurable in one device. A non-directional device would have access to $f(v)$ whereas a single directionally sensitive one would in addition have access to projections of the velocity distribution that would rotate with the Earth. As with the form factor we can write the power spectrum as the sum of a non-directional part and a directional part given by a geometry-weighted speed distribution $f_\mathcal{G}(v;t)$,
\begin{eqnarray}\label{eq:dirpowerspectrum}
\dbd{P}{\omega}(t) &=& P_0\dbd{v}{\omega} \left( \int {\rm d}\Omega_v \, v^2 \, f(\mathbf{v}; t)-\int {\rm d}\Omega_v \, v^2 \mathcal{G}(\mathbf{v})\, f(\mathbf{v}; t)\right)\bigg|_{|\mathbf{v}| = v(\omega)} \\
&=& P_0\dbd{v}{\omega} \bigg(f[v(\omega); t]- f_\mathcal{G}[v(\omega);t]\bigg) \, ,
\end{eqnarray}
where $P_0$ is the total signal power in the 0-velocity limit (i.e. eq.~\eqref{eq:totpower} calculated using $C_0$ instead of $C({\bf v})$ or with $f(\mathbf{v})=\delta^3({\mathbf v})$). Since we can focus an experiment to have sensitivity to frequencies only within the axion bandwidth we can ignore any frequency dependence in $P_0$ and simply pick a benchmark value based on some experimental configuration as we discuss shortly. The derivative $\textrm{d}v/\textrm{d}\omega$ is simply introduced to write the differential power spectrum with frequency in terms of speed distributions which have dimensions of inverse speed. Optimal sensitivity to the three dimensions of $f(\mathbf{v})$ would be achieved if one constructs an experiment consisting of three devices each with a non-zero $\gl^i$ or $\gq^{ii}$ in individual linearly independent directions. Depending on how much information on the velocity distribution one wanted, not all of these devices would be necessary, but the signal in each is analogous so we can treat the setup generally to begin with. The power spectrum in each experiment will be influenced by $v_i$ or $v_i^2$ for $\ell$ and $q$-type experiments, so we rearrange the measured directional and non-directional powers by writing down functions of $\omega$ and $t$ which describe only the directional \emph{corrections} to the power spectra
\begin{equation}\label{eq:ltype_correction}
\Delta^i_\ell(\omega,t) \equiv\frac{p^i_{\rm dir.}-p_{\rm non-dir.}}{p_{\rm non-dir.}} = -\frac{f_\mathcal{G}[v(\omega);t]}{f[v(\omega);t]}= -\frac{\gl\int {\rm d}\Omega_v \,  v_i\, f({\bf v}; t)}{\int {\rm d}\Omega_v \, \, f({\bf v}; t)}\bigg|_{|\mathbf{v}| = v(\omega)} = -\gl\langle v_i(\omega,t)\rangle_{\Omega_v}\, ,
\end{equation}
for linearly dependent experiments or
\begin{equation}\label{eq:qtype_correction}
\Delta^i_q(\omega,t) \equiv \frac{p^i_{\rm dir.}-p_{\rm non-dir.}}{p_{\rm non-dir.}} = -\frac{f_\mathcal{G}[v(\omega);t]}{f[v(\omega);t]}= -\frac{\gq\int {\rm d}\Omega_v \,  v^2_i\, f({\bf v}; t)}{\int {\rm d}\Omega_v \, \, f({\bf v}; t)}\bigg|_{|\mathbf{v}| = v(\omega)} = -\gq \langle v^2_i(\omega,t)\rangle_{\Omega_v}\, ,
\end{equation}
for quadratic dependence. For notational convenience we use the labelling $p = \textrm{d}P/\textrm{d}\omega$ here but in our statistical analysis this will be replaced by the power in one frequency bin. Reducing the power spectrum to a directional correction we can see that they essentially amount to the angular average of $v_i$ or $v_i^2$ over a shell of radius $v$. We can evaluate these integrals for the Maxwellian $f(\mathbf{v})$ by first performing a rotation such that $\vlab$ points along the axis of the experiment. This will introduce a dependence on the angle between $\vlab$ and the axis, $\cos{\theta^i_{\rm lab}(t)}$ (refer to eq.~\eqref{eq:costhlabs} for the full time dependence). For each direction and for linear and quadratic experiments we have,
\begin{eqnarray}\label{eq:power-modification}
		\ell-{\rm type}:  \Delta^i_\ell(\omega,t) &=& -\zeta_\ell(\omega) ~ \cos(\theta^i_{\rm lab}(t))~ \gl \\
		q-{\rm type}: 	 \Delta^i_q(\omega,t)&=&-\left[\zeta_{q1}(\omega) + \zeta_{q2}(\omega) ~ \cos^2(\theta^i_{\rm lab}(t)) \right] \gq, \, ,
\end{eqnarray}
where the functions $\zeta_{\ell,q1,q2}$ are determined only by quantities of the velocity distribution. Explicitly,
\begin{eqnarray}
		\zeta_\ell(\omega) &=& v(\omega) \coth \left(\frac{v(\omega) v_{\rm lab}}{\sigma_v ^2}\right)-\frac{\sigma_v ^2}{v_{\rm lab}}, \\
		\zeta_{q1}(\omega)  &=& \frac{\sigma_v ^2}{v_{\rm lab}} ~\zeta_\ell(\omega) , \label{eq:QuadraticVSens-Constant-Summand} \\		
		\zeta_{q2}(\omega)  &=&  v(\omega)^2 - 3 \zeta_{q1}(\omega) .
\end{eqnarray}
These will in principle be functions of time also, but as long as one does not exceed experimental durations longer than a few tens of days one can treat $v_\textrm{lab}$ as constant in time and hence the $\zeta$'s as functions of only $\omega$. We reiterate here again that the case for a stream is identical after replacing $\vlab \rightarrow \vlab-\vstr$ and $\sigma_v \rightarrow \sigma_{\rm str}$. Now with this general formalism in hand, we can turn our attention to specific cases exhibiting a directional sensitivity. In figure~\ref{fig:powerspectra} we show the shapes of the directionally corrected differential power spectra $p^{\rm dir.}$ as well as the directional corrections isolated, $\Delta_{\ell,\,q}$. The distribution corresponds to a Maxwellian halo that contains a 10\% contribution from the S1 stream. The quantity being shown in each is a ratio of powers, since we have multiplied each differential power value by a binwidth in frequency $\Delta \omega$ and then divided by the total power $P_0$, this way we can clearly illustrate the shape of the effects in frequency and time. As with all our examples we assume the observation begins on Jan 1 at Munich. We assume a linear geometry factor of $\gl c = 71$ and a quadratic one of $\gq c^2 = 10^5$ (see section~\ref{sec:benchmarks} for how we settle on these values). The quadratic correction is only negative but the linear correction can be both positive or negative depending on the orientation of the DM/stream wind with respect to the experimental axis. Importantly for making measurements, the phases and amplitudes of the modulating part of the power spectrum in each experiment are distinct from one another. We now describe the specific experimental designs that can achieve these directional effects in practice.

\subsection{Resonant cavities}
\label{sec:cavity}

We define a rectangular cavity in an $(x,y,z)$ coordinate system with dimensions $(L_x,L_y,L_z)$ to have a homogeneous magnetic field ${\bf B}_{\rm e} = (0,B_e,0)$. The electric fields of the modes that are permitted in the cavity can be separated into time and spatial dependent parts as ${\bf E}(t,{\bf x})= \sum_i E_i(t){\bf e}_i$. Of interest to us are the transverse electric modes TE$_{l0n}$ which have for the spatial part\footnote{The factor of 2 in this formula comes from the normalisation condition $\int {\rm d}V {\bf e}_i\cdot{\bf e}_j = V\delta_{ij}$ that the modes must satisfy.},
\begin{equation}
{\bf e}_{l0n} = \left(0, 2  \sin{\left(\frac{\pi l x}{L_x}\right)} \sin{\left(\frac{\pi n z}{L_z}\right)} , 0\right) \, .
\end{equation}
The resonant frequency of this mode is given by,
\begin{equation}
\omega^2=\left(\frac{l\pi}{L_x}\right)^2+\left(\frac{n\pi}{L_z}\right)^2,
\end{equation}
where we assume that $L_y\lesssim L_x,L_z$ to isolate the fundamental mode. Plugging this mode geometry into the form factor we get,
\begin{align}
C_{l0n} &= \left|\frac{2}{V} \int_V {\rm d}x \,{\rm d} y\, {\rm d}z \, \left[\sin{\left(\frac{\pi l x}{L_x}\right)} \sin{\left(\frac{\pi n z}{L_z}\right)} \, e^{i{\bf p}\cdot {\bf x}}\right] \right |^2 \, , \label{eq:cavint} \\
 &=\left |2 i\pi^2 l n \frac{((-1)^le^{i q_x}+1)(e^{i q_y}-1)((-1)^ne^{i q_z} +1)}{q_y(l^2\pi^2 - q_x^2)(n^2\pi^2-q_z^2)} \right|^2 \, ,
\end{align}
where $q_x = m_a L_x v_x$ etc. Next, we can evaluate the absolute value signs. Expanding and keeping only terms up to $v^2$ we get something that can be written as,
\begin{eqnarray}
C_{l0n}&\simeq&\frac{16 \left((-1)^l-1\right) \left((-1)^n-1\right)}{\pi ^4 l^2 n^2}-\frac{4}{3 \pi ^6 l^4 n^4}{\bigg [}6q_x^2n^2\left((-1)^n-1\right) 
   \left((-1)^l\pi ^2  l^2+ 4+(-1)^{l+1}4 \right) 
   \nonumber \\ &&+6 q_z^2l^2((-1)^l-1)\left((-1)^n\pi ^2  n^2+4+(-1)^{n+1}4 \right)
   \nonumber \\ &&+q_y^2\pi^2  
   l^2 n^2 ((-1)^l-1) \left((-1)^n-1\right) \bigg ] .
\end{eqnarray}
If $l,n$ are even this vanishes, but if $l,n$ are odd then it reduces to
\begin{eqnarray}
C_{l0n}  &\simeq& \frac{64}{\pi^4l^2n^2}\left(1 -  \left(\frac{1}{4}-\frac{2}{\pi ^2 l^2}\right) q_x^2 -   
   \left(\frac{1}{4}-\frac{2}{\pi ^2 n^2}\right)q_z^2 - \frac{q_y^2}{12}\right)\nonumber \\
					&=& C_0(1-\sum_{i=1}^{3} \gq^iv_i^2) \, , \label{eq:cavC}
\end{eqnarray}
where
\begin{equation}
(\gq^x,\gq^y,\gq^z)=m_a^2\left ( L_x^2\left(\frac{1}{4}-\frac{2}{\pi ^2 l^2}\right),  \frac{L_y^2}{12},L_z^2 \left(\frac{1}{4}-\frac{2}{\pi ^2 n^2}\right) \right ) \, .
\end{equation}
As suggested by how we set up our formalism, we see here that the corrections turn out to be negative, however if we assume we have sufficient signal to noise to detect the axion we need only focus on its modulation.

With the expression for $C_{l0n}$ written in this way we can see that it contains the usual zero velocity form factor for the TE$_{l0n}$ mode in a rectangular cavity, and a second term expressed as a velocity dependent geometry factor.  To get directional sensitivity, we desire our device to be elongated in one direction. We will leave $L_z$ small with $n=1$. 
We foresee two options for making $L_x\gg L_{z,y}$. One would be to use the fundamental mode, leaving $l=1$. In this case,
\be\label{eq:gqcavl1}
\gq^x\simeq \frac{\pi^2-8}{4}\left(\frac{L_x}{L_z}\right)^2\gg\gq^{y,z} \, ,
\ee
with
\begin{equation}
\omega\sim \frac{\pi}{L_z}\left(1+\frac{L_z^2}{2L_x^2}\right) \, .
\end{equation}
This gives a potentially excellent velocity dependence for large $L_x$, however one could eventually run into problems of mode crowding. For very large $L_x$ the frequency difference between different low $l$ values becomes extremely small, actually being less than the axion line width at $L_x/L_z={\cal O} (10^3)$.

An alternative approach to extend one dimension would be to use a higher order mode ($l\gg1$). Ensuring the resonant frequency remains at $m_a$ by setting $L_x=\sqrt{2}\pi l /m_a$, we see that
\be
\gq^x\simeq \frac{L_x^2m_a^2}{4}=\frac{\pi^2l^2}{2} \label{eq:gcav2}\, .
\ee
Since $C_0\propto l^{-2}$, the total form factor for the velocity dependent terms is constant with increasing $l$. One might worry from this line of thinking that one gains nothing by moving to higher and higher modes, however such a concern only arises because of the overly complicated way in which the power from resonant cavities is usually expressed. Remembering eq.~\eqref{eq:power} we see that there is also a factor of $QV/m_a$. The quality factor of a cavity is defined to be 
\begin{equation}
Q=\omega\frac{\int {\rm d}V\epsilon({\bf x})|{\bf E_{\bf k}}({\bf x})|^2}{P_{\text{loss}}},
\end{equation}
where $P_{\text{loss}}$ is the power lost from the cavity. Assuming that $P_{\text{loss}}$ is constant, then as one goes to higher mode numbers the quality factor also increases by a factor $l$, i.e., higher order modes have narrower resonances\footnote{Of course, one is still limited by the axion line width so for $Q\gtrsim 10^6$ only part of the axion spectrum is measured.}. Overall we see
\begin{equation}
P_{\bf p}\propto \frac{1}{m_a^2}\left(1-\frac{\pi^2l^2v_x^2}{2}\right )=\frac{1}{m_a^2}-\frac{L_x^2v_x^2}{4}.\label{eq:powerreccav}
\end{equation}
Thus if we keep the size of the cavity fixed, and go to higher masses, one loses out by a factor of $m_a^2$ when going to higher order modes. However, for the case we are interested in, keeping $m_a$ fixed and looking at high mode cavities, the velocity effects indeed increase with $l^2$. To get a large velocity effect from a cavity in this way, one must go to very high modes. However such an experiment may not be so impractical. For high mode numbers in a cavity, the vast majority of the empty volume does not need to be magnetised. Since the integral of the waves in the middle of the device cancels on average, the central region of the cavity plays no role in signal generation aside from allowing the axion to undergo a change of phase. One would only need to have a magnetic field within approximately a half wavelength of the ends of the device, obtaining the same $\gq$ as in eq.~\eqref{eq:gcav2}. However, the form factor is actually enhanced,
\begin{equation}
C_0=\frac{256}{l^2\pi^4}\, .
\end{equation}
This enhancement is because each magnetised half wavelength adds constructively to the produced power. When fully magnetised, the field everywhere in the cavity cancels aside from only one half wavelength. Partial magnetisation would result in a dramatic reduction in magnet costs, though still requires operation at high cavity modes. Thus higher order modes may be a good way to achieve strong velocity dependence without prohibitive magnet requirements, albeit at the cost of signal power. Unfortunately, the issue of mode crowding is not avoided, with typical spacing between modes being $\sim \omega/2l$. A similar concept would be to have two cavities separated by some large distance, and add the signal from each of them together. Such an idea has some advantages in avoiding higher modes, however one would have to very precisely match the phases of each cavity in real time over long distances.

From the above discussion we have seen that there are competing problems of mode crowding at low-$l$ and redundant cavity volume at high-$l$. Fortunately it may be possible to mitigate both issues by loading the cavity with dielectrics to modify the modes~\cite{Morris:1984nu} or by combining multiple coupled cavities~\cite{Goryachev:2017wpw,Melcon:2018dba}. One could also use wires with currents to modify $B_e$~\cite{Rybka:2014cya}, however in this case the issue of mode crowding remains. Each of these methods has the advantage of increasing the form factor significantly, as well as potentially increasing the quality factor. 

To estimate what we could gain such setups, we can take the simplest case, placing a series of transparent (phase thickness $\pi$) dielectrics a half wavelength apart. Such a description captures the essential behaviour of all three possible setups, though depending on the realisation the details might differ. We approximate the dielectric loaded cavity mode by simply using $|\sin(\frac{\pi l x}{L_x})|$ instead of $\sin(\frac{\pi l x}{L_x})$ in the integrand of eq.~\eqref{eq:cavint}, giving
\begin{equation}
C_{l01}\sim\frac{64}{\pi^4}-\frac{16l^2v_x^2}{3\pi^2}\, ,
\end{equation}
where we again take the mode where $n=1\ll l$ and $L_x=l /m_a$. Now not only does the velocity independent part of the form factor not decrease with increasing $l$, the velocity dependent part actually increases. Note that in this case $l$ is not a mode number, rather $l-1$ gives the number of inserted dielectrics. Thus a dielectric loaded resonator would win over an empty cavity at high modes by a factor of $l^2\propto (L_x/m_a)^2$. Such a setup would be ideal for gaining a strong directional sensitivity, while avoiding mode crossings.

Lastly, we remark that unlike the zero velocity limit, it is possible for one of $l,n$ to be even. In this case, it is well known that the velocity independent term vanishes, but the velocity dependent terms do not. If only one of $l,n$ is odd then
\be
C_{l0n} =\frac{4 L_{z,x}^2 m_a^2 v_{z,x}^2}{\pi ^4 l^2 n^2}.
\ee
In this case the form factor actually increases for higher values; if the signal-to-noise was high enough one would have an excellent way of studying the tail of the velocity distribution. 

By modifying the aspect ratios of cavities, there are several ways to gain a strong sensitivity to the axion velocity. In this paper we remain agnostic to the various practical issues as each different realisation has potential advantages and pitfalls, which can only be illuminated by in depth design studies. While any of these devices would be challenging, no frequency scanning is required so one could devote considerable time and resources into perfecting the performance for a single known frequency.

\subsection{Dielectric haloscope}\label{sec:dielectrics}
\begin{figure}[t]
\centering
\mbox{}\hfill
    \begin{subfigure}[t]{0.49\textwidth}
      \includegraphics[width=\textwidth]{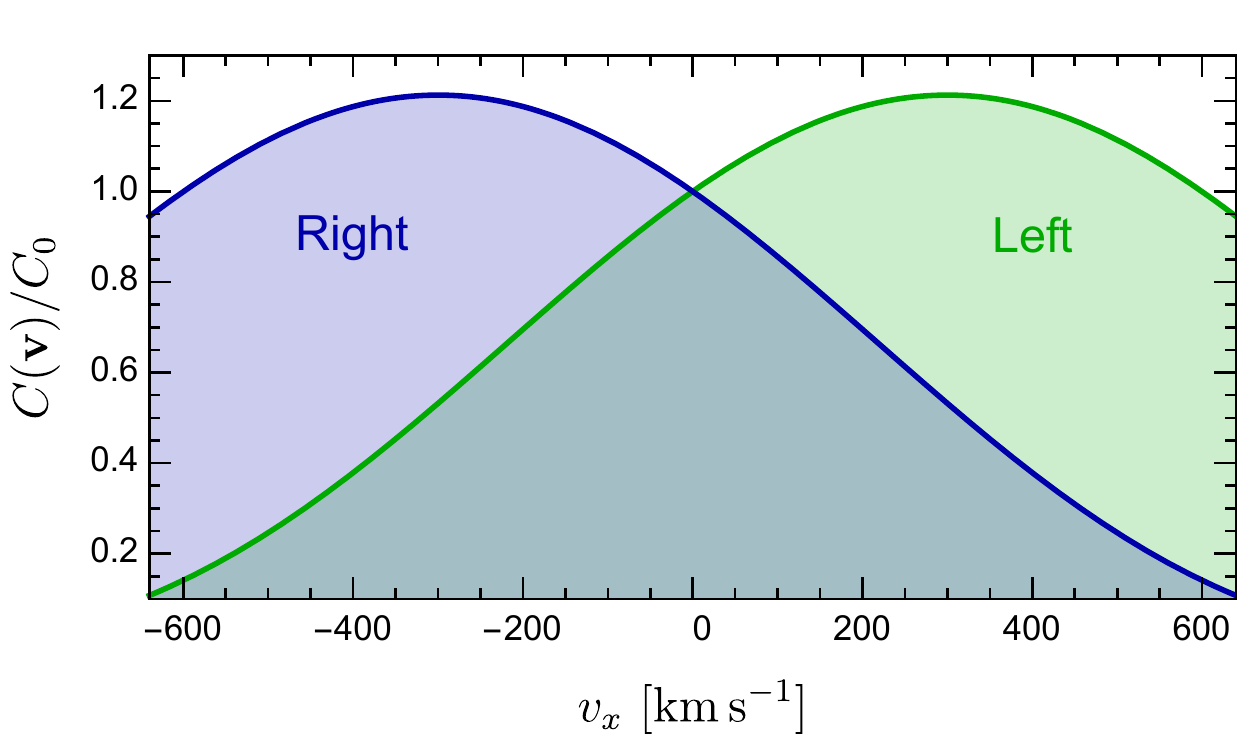}
      \caption{400 transparent disks with $\delta_{\rm t}=\pi$.}\label{fig:transvelocity}
    \end{subfigure}  \hfill
        \begin{subfigure}[t]{0.49\textwidth}
      \includegraphics[width=\textwidth]{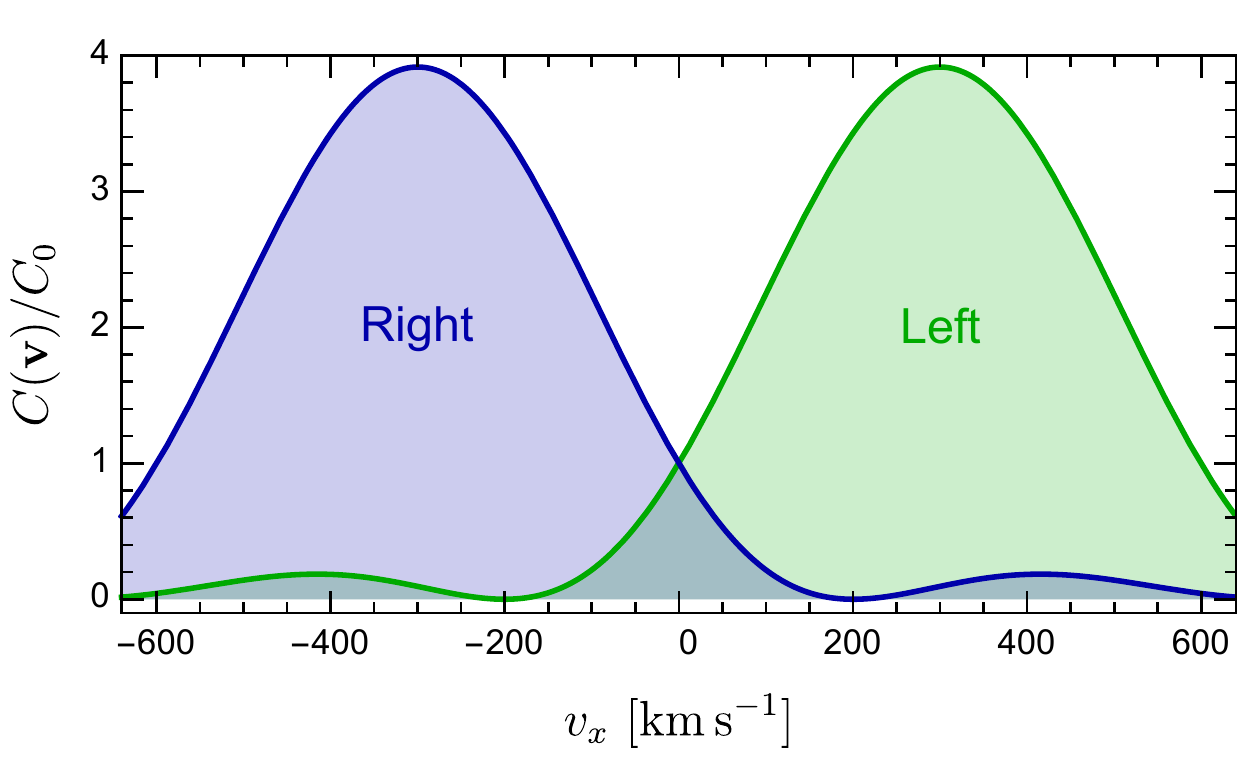}
      \caption{1000 transparent disks with $\delta_{\rm t}=\pi$.}\label{fig:bothsides}
    \end{subfigure}   
    \hfill\mbox{}
\mbox{}\hfill
        \begin{subfigure}[t]{0.49\textwidth}
      \includegraphics[width=\textwidth]{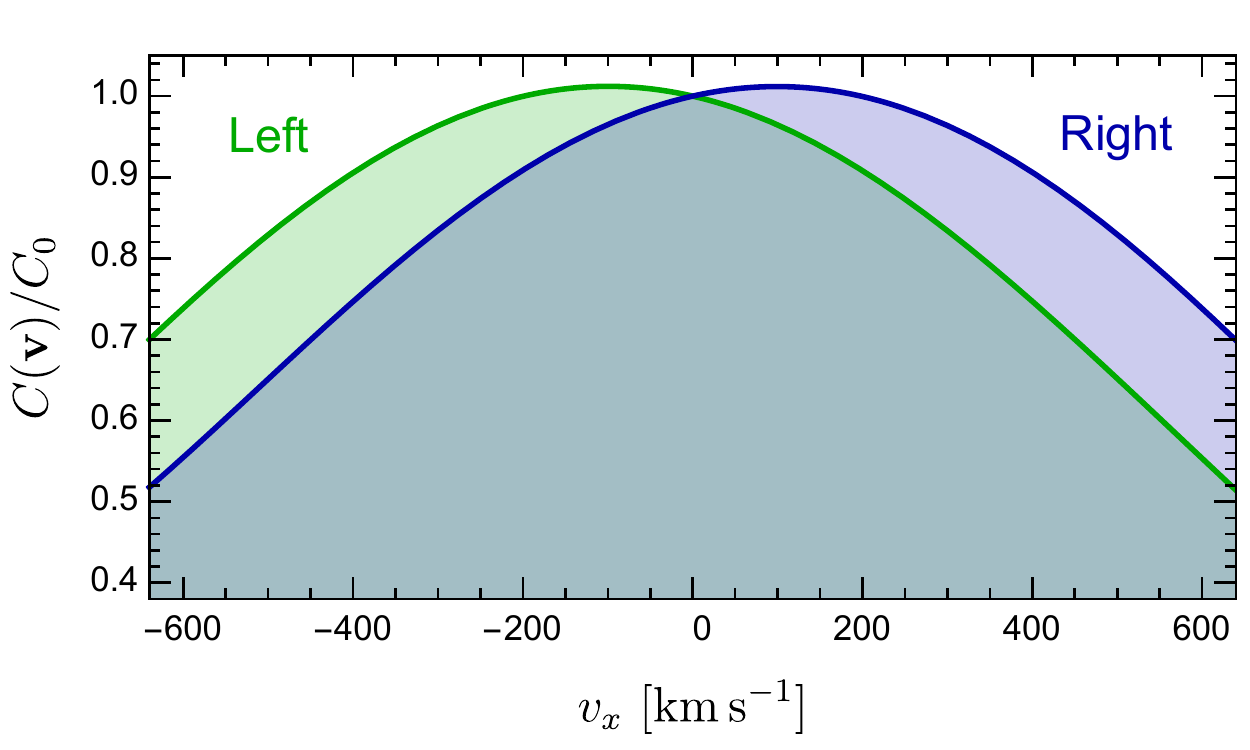}
      \caption{400 partially reflective disks with $\delta_{\rm t}=3.163$.}\label{fig:bothsidesR}
    \end{subfigure}   
     \begin{subfigure}[t]{0.49\textwidth}
      \includegraphics[width=\textwidth]{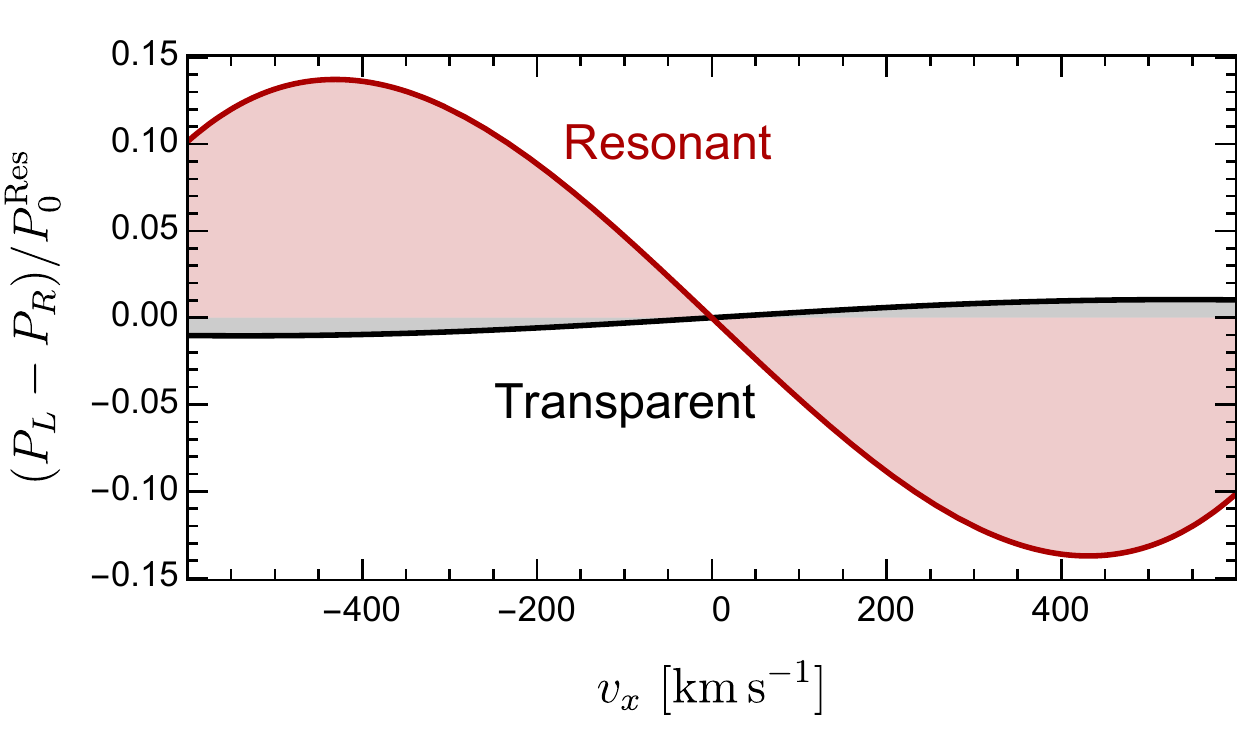}
      \caption{400 transparent/reflective disks}
\label{fig:LvsR}
    \end{subfigure}
       \hfill\mbox{} 
\caption{(a)-(c) display the relative form factor $C({\bf v})/C_0$ corresponding to power emitted from the left (green) and right (blue) hand sides of the dielectric haloscope as a function of the axion velocity in the $x$ direction. All have disks separated by a phase separation of $\delta_{\rm s}=3.138$ with refractive index $n=5$, but varying values of phase thickness $\delta_{\rm t}$. In (d) we show the difference between left and right handed powers for transparent (black) and partially reflecting (red) disks. For illustrative purposes we have assumed a Dirac delta velocity distribution.}
\end{figure}

We now turn our attention to dielectric haloscopes, which consist of a series of dielectric disks placed parallel to a magnetic field, illustrated in figure~\ref{fig:haloscope_designs}. As discussed in ref.~\cite{Millar:2017eoc}, it is possible to either increase the separations of the disks, or add disks, to increase the length of the device to a decent fraction of the axion coherence length. The subsequent velocity effects considered there were described using a classical transfer matrix formalism. To make the connection with our universal notation, we must instead extend the formalism developed in ref.~\cite{Ioannisian:2017srr} to include the axion velocity. 
As discussed in section~\ref{generalformalism} the produced power is given by an overlap of the axion and photon wave functions. While the axion has a trivial plane wave function, the photon's wave function is distorted by the presence of dielectric media. We will assume that the transverse area of the dielectric disks is large, so that the momenta in the transverse directions are approximately conserved. Conservation of momentum requires that these momenta are the same as that of the axion. Thus the photon wave function is given by
\begin{equation}
{\bf E}_{\bf k}({\bf x})= {\bf E}_{\bf k}(x)e^{i{\bf p_{||}}\cdot {\bf x}}\, .
\end{equation}
The ${\bf E}_{\bf k}(x)$ is simply the ``Garibian" wave function considered in~\cite{Ioannisian:2017srr}, which consists of
an incoming plane wave of unit amplitude which is then split by the haloscope into a transmitted and a reflected component. The space is spanned by two such wave functions, depending on the side of the haloscope on which they
are incident. The axion velocity only induces a small shift in frequency of $m_av^2/2$.
 We thus see that 
\begin{equation}\label{eq:matrix-element-dh}
{\cal M}=\frac{g_{a\gamma}}{2\omega L}\int {\rm d} x\,
\,{\bf B}_{\rm e}({\bf x})\cdot{\bf E}^*_{\bf k}(x)e^{i p_x x}\, ,
\end{equation}
where $L$ is the length of the haloscope in the $x$ direction. One can define an effective quality factor $Q_{\rm dh}$ by 
\begin{equation}
Q_{\rm dh} =m_a\frac{1}{4}\frac{\int {\rm d}V\left[\epsilon({\bf x})|{\bf E_{\bf k}}({\bf x})|^2+|{\bf B_{\bf k}}({\bf x})|^2\right]}{2 A}\, , \label{eq:dielectricQ}
\end{equation}
where ${\bf E}_{\bf k}, {\bf B}_{\bf k}$ are the $E$ and $B$-fields of the Garibian wave function. The overall power in EM waves $P_{\bf p}$ is still given by eq.~\eqref{eq:power}, with the coupling efficiency $\kappa=1$. In general $Q_{\rm dh}$ and $C({\mathbf v})$ (and thus $P_{\bf p}$) can be different for photons emerging from either side of the device. 

We saw in section~\ref{generalformalism} that a cavity haloscope always has quadratic dependence on the velocity of the axion. However, a linear dependence on $v$ would potentially provide the full directional information of the velocity distribution. To achieve linear dependence, from eq.~\eqref{eq:cexpand} we know that the free photon wave function must have some travelling wave behaviour (i.e. a spatial variance in the phase). If a dielectric haloscope is strongly resonant, or there is a metallic mirror in the haloscope, the Garibian wave functions will form a standing wave. 

A simple example which can obtain linear dependence would be a series of transparent dielectric disks. If the phase thickness of the disk is $\delta_{\rm t}=\pi$, then each disk is transparent to radiation but still emits photons in the presence of axions~\cite{Jaeckel:2013eha}. This transparent setup does not use resonances to increase signal power, only constructive interference. To calculate the produced power we must solve for the Garibian wave functions of the system. Photons may be emitted from either side of the device, however due to the symmetry of the system the only differences in produced power could come from the direction of the axion velocity itself. Thus we will only consider photons emitted from a single direction, and simply use $v\to-v$ to obtain the other direction.

In general a dielectric haloscope consists of $m-1$ dielectric regions, between some positions $x_1$ and $x_m$, with interfaces at distances $x_r$. For simplicity we consider the case where the $E$-field is inserted as a right moving wave on the left hand side of the haloscope, which allows us to compute the power being emitted from the left side of the haloscope.  In order to evaluate the integral in eq.~\eqref{eq:dielectricQ}, we can break it up into regions of different dielectric material. In each region $r$ we can break up the $E$-field into left and right moving parts,
\begin{equation}
E_{\bf k}^r(x)=R^r e^{in_r\omega\Delta x_r}+L^r e^{-in_r\omega\Delta x_r}\, ,
\end{equation}
where $R^r$ is the amplitude of the right-moving component, $L^r$ the left-moving component, with $\Delta x_r=x-x_r$. We take $\Delta x_0=\Delta x_1$ and follow the same convention as ref.~\cite{Millar:2016cjp}, so that the field amplitudes $R^r$ and $L^r$ of the right and left moving EM waves are defined at the left boundary of every region, except for $R^0$ and $L^0$ which are defined at $x_1$, i.e., the leftmost interface. The $E$-fields for different regions are connected by the boundary conditions, with $E_{||}$ and $H_{||}$ being conserved. 

Consider a series of $N$ transparent disks of refractive index $n$ (thickness $\pi/n\omega$) with a distance $\delta_{\rm s}/\omega$ between each disk, where we will call $\delta_{\rm s}$ the phase separation. For the $j$th dielectric disk, $R$ and $L$ are given by
\begin{equation}
\vv{R^{2j-1}}{L^{2j-1}}=\frac{e^{i(j-1)(\delta_{\rm s}+\pi)}}{2n}\vv{1+n}{n-1}\, , \label{eq:diskE}
\end{equation}
and in the $j$th vacuum region $R$ and $L$ are given by
\begin{equation}
\vv{R^{2j}}{L^    {2j}}=e^{i[(j-1)\delta_{\rm s}+j\pi]}\vv{1}{0}\, .\label{eq:vacE}
\end{equation}
Using these expressions, we can evaluate eq.~\eqref{eq:dielectricQ} to get
\begin{equation}
Q_{\rm dh}=\frac{m_a}{4}\sum_{s=1}^{m-1}{\rm d}x \,n_s^2(x_{s+1}-x_s)(|R_s|^2+|L_s|^2)=\frac{1}{8}\left(n\pi N\left(1+\frac{1}{n^2}\right)+2\delta_{\rm s}(N-1)\right) \, .
\end{equation}
To find the effects of the axion velocity, we must evaluate the overlap integral in the form factor $C({\bf v})$. In terms of our left and right moving waves, we can write
\begin{align}
\int {\rm d}x  \,E_{\bf k}({ x}){ B}_{\rm e}e^{i{ p_x}{ x}}\sim &  \, \frac{ B_{\rm e}}{i\omega}\Bigg[R_0-L_0-(R_m-L_m)e^{i\omega v_xx_m}\nonumber\\
&+\sum_{s=1}^{m-1}\frac{e^{i\omega v_x x_{s-1}}}{n_s}R_s\left(e^{i\omega d_s(n_s+v_x)}-1\right)-L_s\left(e^{-i\omega d_s(n_s-v_x)}-1\right)\Bigg]\, ,
\end{align}
where we have neglected some subdominant terms in the velocity which enter outside of the argument of a phase. Using eqs.~\eqref{eq:diskE} and \eqref{eq:vacE} we then see that
\begin{align}
\int {\rm d}x\,  E_{\bf k}({ x}){ B}_{\rm e}e^{i{ p_x}{ x}}=&\frac{ B_{\rm e}}{2i\omega}\Bigg(2+\frac{\left(1+e^{\frac{i \pi  v_x}{n}}\right) \left(-1+e^{\frac{i N (\delta_{\rm s} n (v_x+1)+\pi  (n+v_x))}{n}}\right)}{n^2 \left(1+e^{i \left(\delta_{\rm s}(1+v_x)+\frac{\pi  v_x}{n}\right)}\right)}\nonumber \\ &-\frac{\left(-1+e^{i \delta_{\rm s} (v_x+1)}\right) \left((-1)^N e^{i \left(\delta_{\rm s} (N-1) (v_x+1)+\frac{\pi  N v_x}{n}\right)}+e^{\frac{i \pi  v_x}{n}}\right)}{1+e^{i \left(\delta_{\rm s} (1+v_x)+\frac{\pi  v_x}{n}\right)}}\Bigg) \, .
\end{align}
Unfortunately this expression is too complicated in general to be reduced to something like eq.~\eqref{eq:cavC}. Depending on the choice of $\delta_{\rm s}$, the velocity dependence can enter either quadratically or linearly. 

If the disks are arranged for maximal constructive interference at zero velocity ($\delta_{\rm s}=\pi$) the system becomes symmetric with respect to the axion velocity, and so must have a quadratic dependence. Specifically, in this case one can show that~\cite{Millar:2017eoc}
\begin{equation}
c\gl=0,\,\,\,
\,c^2\gq\simeq\frac{N^2\pi^2}{12} \, .
\end{equation}
An experiment using this transparent setup (with the addition of a mirror) has been proposed to operate at the optical range, using photon counting rather than linear amplification as we consider here~\cite{Baryakhtar:2018doz}. Photon counting experiments generally have insufficient energy resolution to measure the axion lineshape. However, the second phase of this experiment would use a huge number of dielectric layers, up to $N=1000$. In this case the axion velocity has a major impact on the signal power, leading to significant systematic uncertainties due to the unknown velocity distribution. Of course, if such an experiment had discovered ALPs or hidden photons previously the power modulation could be used for axion astronomy in a similar way as described in this paper, albeit without resolving the line width. Due to the different nature of the statistics for a photon counting experiment, we will not consider this scenario further.

To achieve a linear velocity dependence we can imagine placing each disk slightly out of phase with respect to case where maximal constructive interference occurs at $v_x=0$. Then a velocity in one direction will increase the constructive interference and a velocity in the other decrease it, giving us a discrimination between the two directions. In figure~\ref{fig:transvelocity} we show the relative form factor corresponding to the power produced from the left and right side of a device consisting of 400 dielectric disks with $n=5$ and $\delta_{\rm s}=3.138$. 
The power at zero velocity is given by
\begin{equation}
P_0^{\rm Transp}=3\times 10^{-21}\, {\rm W}\left(\frac{B_{\rm e}}{15\,{\rm T}}\right)^2\left(\frac{A}{1\, {\rm m}^2} \right )\left(\frac{\bar\rho_a}{0.4\, {\rm GeV}\,{\rm cm}^{-3}}\right )\left(\frac{C_{a\gamma}}{1.92}\right)^2\, ,
\end{equation}
where $A$ is the transverse area of disks. At lowest order the velocity effects are linear, with $c\gl=383$. 

Beyond this setup, if we were to use more dielectric disks, the disparity between power produced in each direction can grow until there is an almost complete discrimination between the two. We show such a case in figure~\ref{fig:bothsides}, which shows the relative form factor corresponding to the power emitted from each side of a device consisting of 1000 dielectric disks with $n=5$ and $\delta_{\rm s}=3.138$.  Despite each side being only sensitive velocities in one direction, together the combination covers the full range of realistic galactic velocities. One would then see a modulation of the signal as a transference between the power being emitted from each side. The correlation  between the two measurements would also serve as a useful systematic check, though one would need to take care with possible reflections between the two detector and antenna setups. We will not consider such a case in full detail, as the approximation of the velocity dependence as being linear or quadratic clearly breaks down, which would make the analytic calculations in section~\ref{sec:stats} significantly more complicated.

So far in this discussion, we have completely neglected possible resonant enhancements of the signal strength. While as argued above a strongly resonant behaviour precludes the possibility of a linear velocity dependence, it is actually possible to achieve a stronger absolute linear shift in the power as function of velocity at the expense of the relative size of the effect compared to the total signal power. For resonant behaviour to occur, the disks must be partially reflecting. To show that such a situation can indeed occur, in figure~\ref{fig:bothsidesR} we show a dielectric haloscope consisting of 400 dielectric disks with $n=5$ and a phase thickness of $\delta_{\rm t}=3.163$, with $\delta_{\rm s}=3.138$. 
While the relative size of linear effects is smaller than the previous examples, with $c\gl=71$ and $c^2\gq=105000$, the total power is much greater with
\begin{equation}
P_0^{\rm Res}=2.6\times 10^{-19}\, {\rm W}\left(\frac{B_{\rm e}}{15\,{\rm T}}\right)^2\left(\frac{A}{1\, {\rm m}^2} \right )\left(\frac{\bar\rho_a}{0.4\, {\rm GeV}\,{\rm cm}^{-3}}\right )\left(\frac{C_{a\gamma}}{1.92}\right)^2\, ,
\end{equation}
which we label ``resonant'', though the resonant behaviour is relatively mild. To see whether a given setup is more sensitive to direction of the axion, we can calculate $P_L(v_x)-P_R(v_x)$. Normalised to $P_0^{\rm Res}$ we show the comparison between 400 transparent and 400 mildly resonant disks in figure~\ref{fig:LvsR}. In these terms, the more resonant setup is an order of magnitude more sensitive to the sign of the velocity. Further, one can measure both linear and quadratic behaviour, giving essentially the first and second moment of the velocity distribution. Thus we will use the more resonant example as our benchmark dielectric haloscope.

Dielectric haloscopes have the unique ability to discern the sign of the axion velocity in a specific direction. They further have immense flexibility to enhance the power generated at a specific axion velocity, being tuneable to almost any situation. While only a few examples were shown here, this flexibility would be very beneficial when it comes to the practical design of an experiment, allowing, for example, one to design a device that focused exclusively on the tail of $f(\mathbf{v})$ or on the velocity of a stream. However, this flexibility and ability to obtain a linear velocity dependence can only be achieved if one sacrifices strongly resonant behaviour, which will limit the achievable signal power.  

We assume that the dielectrics are arranged in a left-right symmetric manner as above such that difference between the power emitted from each side is simply given by $v_i\to-v_i$. In terms of our general formalism this means we can isolate either linear or quadratic effects simply with how the left and right hand side powers are combined. This is a clear advantage offered by the dielectric haloscope setup. In terms of our general formalism we can recover the directional corrections introduced in eqs.~\eqref{eq:ltype_correction} and \eqref{eq:qtype_correction} only with a slight modification to the formula,
\begin{equation}
\Delta^i_\ell = \frac{(p^i_{\rm R, dir.}-p^i_{\rm L, dir.})}{2p_{\rm non-dir.}} \, ,
\end{equation}
and
\begin{equation}
\Delta^i_q= \frac{(p^i_{\rm L, dir.}+p^i_{\rm R, dir.})/2 - p_{\rm non-dir.}}{p_{\rm non-dir.}} \, .
\end{equation}

\subsection{Benchmark experimental parameters}\label{sec:benchmarks}
Given our three model haloscope designs --- a low and high-$l$ cavity (with quadratic-$v$ directionality) as well as a dielectric disk haloscope (quadratic and/or linear-$v$ directionality) --- we now summarise the size requirements for each experiment to achieve some benchmark geometry factors, $\gl$ and $\gq$, and signal power $P_0$. Although the design parameters are challenging, we emphasise that this experiment would only have to be built and calibrated once. For instance, one would only need to design the cavity for a single resonant frequency, or for a single set of disk spacings in the case of the dielectric haloscope\footnote{We also refer the reader to the excellent prospects for quantum limited noise in higher mass experiments ($m_a>40\,\mu$eV) with use of single photon detectors~\cite{Lamoreaux:2013koa}.}. As displayed in figure~\ref{fig:haloscope_designs}, we focus on three axion masses in the range $10$--$100\, \mu$eV. For the lower end we require the cavities, fixing masses of 10 $\mu$eV and 40 $\mu$eV, and assigning them a partially magnetised setup with a high-$l$ and a fully magnetised setup with $l=1$ respectively. For larger masses dielectric haloscopes are preferable; we assign it a 100 $\mu$eV axion in this case. We emphasise that our later sections give scaling relations that can be used to reapply our results to any value of $P_0$. This section is merely to highlight that the specific values of $P_0$ that are used for certain figures are experimentally reasonable.

For a reasonable dielectric haloscope setup we showed earlier that we can achieve (now in units of$\kms$),
\begin{subequations}
\begin{eqnarray}
\gl &\simeq & 2.4 \times 10^{-4} \,\,{\rm km}^{-1} \,{\rm s} \, ,\\
\gq &\simeq &  1.1 \times 10^{-6} \, \,{\rm km}^{-2}\, {\rm s}^2 \, ,
\end{eqnarray}
\end{subequations}
with the total power from the $v=0$ calculation,
\begin{equation}\label{eq:benchdhpowers}
m_a = 100\, \mu {\rm eV} \,:\quad P_0^{\rm dh}=2.6\times 10^{-19}\, {\rm W}\left(\frac{B_{\rm e}}{15\,{\rm T}}\right)^2\left(\frac{A}{1\, {\rm m}^2} \right )\left(\frac{\bar\rho_a}{0.4\, {\rm GeV}\,{\rm cm}^{-3}}\right )\left(\frac{C_{a\gamma}}{1.92}\right)^2\, .
\end{equation}
Since the dielectric haloscope observes both a linear and a quadratic effect we use its values of $\gl$ and $\gq$ as benchmarks. They also give conveniently similar sized directional effects,
\begin{equation}
\mathcal{G}_\ell = 7.1 \% \left( \frac{v}{300 \kms}\right) \, , \quad
\mathcal{G}_q = 10.0 \% \left( \frac{v}{300 \kms}\right)^2 \, .
\end{equation}
Now we only need to reproduce these values in our cavities. 

Firstly for a partially magnetised high-$l$ cavity to achieve the same $\gq$ we require $l = 142$ by eq.~\eqref{eq:gqcavl1}. Then since we need to fix the frequency to $m_a = 10\,\mu$eV inside a cavity with one long dimension $L_x$ and two shorter dimensions $L_z = L_y$ this means we need to have $L_{z,y} = \sqrt{2}\pi/m_a = 8.7$ cm and $L_x = l L_z = 12.5$ m. As already discussed this cavity suffers a reduction in $C_0$ by a factor of $l^2/4$ with respect to the fully magnetised $l=1$ case.

Secondly, for the fully magnetised cavity resonating at $l=1$ with $m_a = 40\,\mu$eV, we need to set the aspect ratio using $\gq = (\pi^2-8)/2 (L_x/L_z)^2$. After enforcing the resonant frequency this gives values of $L_x = 7.16$ m and $L_{y,z} = 2.20$ cm.  

\noindent Using eq.~\eqref{eq:power} for the total power in the $v=0$ limit, we get for each of these cavities
\begin{eqnarray}\label{eq:benchcavitypowers}
m_a = 10\, \mu {\rm eV} \,:\quad P^{\rm cav}_0 &=& 1.6\times 10^{-23}\,{\rm W}\,\left(\frac{B_{\rm e}}{15\,{\rm T}}\right)^2 \left( \frac{\bar\rho_a}{0.4 \, {\rm GeV\, cm}^{-3}} \right)  \left( \frac{C_{a\gamma}}{1.92} \right)^2 \\
m_a = 40 \,\mu{\rm eV} \, : \quad P^{\rm cav}_0 &=& 1.2 \times 10^{-20}\,{\rm W}\, \left(\frac{B_{\rm e}}{15\,{\rm T}}\right)^2\left( \frac{\bar\rho_a}{0.4 \, {\rm GeV\, cm}^{-3}} \right)  \left( \frac{C_{a\gamma}}{1.92} \right)^2 \, .
\end{eqnarray}
We summarise the inputs to these calculations in table~\ref{tab:benchmarks}. Whilst our dielectric haloscope benchmark produces more power than the cavities, this is due in part to the very large magnetised volume of the experiment, which would come at high cost. Thus the various benchmarks provide the reader examples of less and more ambitious experiments, ranging from the relatively budget oriented partially magnetised cavity to the more complex and large volume dielectric haloscope.

\begin{table}[h!]\centering
\ra{1.3}
\begin{tabularx}{\textwidth}{l|XXl}
\hline\hline
{\bf Partially magnetised} & Magnetic field & $B_e$ & 15 T \\
{\bf cavity}& Quality factor & $Q$ & $ 10^6$ \\
$m_a = 10\,\mu$eV & Widths & $L_{y,z}$ & 8.7 cm\\
$g_{a\gamma} = 3.84\times 10^{-15}$ GeV$^{-1}$ & Length & $L_x$ & 12.5 m  \\
\quad & Mode number & $l$ & 142 \\
\quad & Form factor & $C_0$ & $256/(142\pi^2)^2$  \\
\cline{2-4}
\quad & \cellcolor{lightgray}Total power &\cellcolor{lightgray} $P_0 $ & \cellcolor{lightgray}$1.6\times 10^{-23} \, {\rm W}$  \\
\quad & \cellcolor{lightgray}Geometry factor & \cellcolor{lightgray}$\gq$ & \cellcolor{lightgray}$1.1 \times 10^{-6} \, \,{\rm km}^{-2}\, {\rm s}^2$ \\

\hline\hline
{\bf Thin cavity} & Magnetic field & $B_e$ & 15 T \\
$m_a = 40\,\mu$eV & Quality factor & $Q$ & $10^6$ \\
$g_{a\gamma} = 1.54\times 10^{-14}$ GeV$^{-1}$ & Widths & $L_{y,z}$ & 2.20 cm\\
\quad & Length & $L$ & 7.16 m  \\
\quad & Mode number & $l$ & 1 \\
\quad & Form factor & $C_0$ & $64/\pi^4$  \\
\cline{2-4}
\quad &\cellcolor{lightgray} Total power & \cellcolor{lightgray}$P_0 $ & \cellcolor{lightgray}$1.2 \times 10^{-20} \, {\rm W}$  \\
\quad & \cellcolor{lightgray}Geometry factor & \cellcolor{lightgray}$\gq$ &\cellcolor{lightgray}$1.1 \times 10^{-6} \, \,{\rm km}^{-2}\, {\rm s}^2$ \\

\hline \hline
{\bf Dielectric disks} & Magnetic field & $B_e$ & 15 T \\
$m_a = 100\,\mu$eV & Number of disks & $N$ & 400 \\
$g_{a\gamma} = 3.84\times 10^{-14}$ GeV$^{-1}$ & Disk area & $A$ & 1 $\text{m}^2$ \\
\quad & Refractive index & $n$ & 5 \\
\quad & Phase separation & $\delta_{\rm s}$ & $3.138$ \\
\quad & Phase thickness & $\delta_{\rm t}$ & $3.163$ \\
\cline{2-4}
\quad & \cellcolor{lightgray} Total power & \cellcolor{lightgray}$P_0 $ &\cellcolor{lightgray}$2.6\times 10^{-19}$ W \\
\quad & \cellcolor{lightgray}Geometry factors& \cellcolor{lightgray}$\gl$ & \cellcolor{lightgray}$2.4 \times 10^{-4} \,\,{\rm km}^{-1} \,{\rm s}$ \\
\quad & \cellcolor{lightgray}\quad &\cellcolor{lightgray} $\gq$ & \cellcolor{lightgray}$1.1 \times 10^{-6} \, \,{\rm km}^{-2}\, {\rm s}^2$\\
\hline\hline
\end{tabularx}
\caption{Summary of the experimental parameters that are required to achieve our baseline geometry factors, as well as the total signal power received in each.}
\label{tab:benchmarks}
\end{table}

\section{Statistical analysis}\label{sec:stats}
To estimate the sensitivity required to do axion astronomy with a directional experiment we utilise a statistical methodology based on the popular profile likelihood ratio test. A related method was used in ref.~\cite{OHare:2017yze}, who performed parameter estimation by first generating mock data using a certain set of input axion and astrophysical parameters and then using the maximum likelihood to reconstruct those parameters. A similar but extended approach was taken in ref.~\cite{Foster:2017hbq}, who also made use of mock data but in addition provided analytic relations using the Asimov data set, see ref.~\cite{Cowan:2010js}. To more straightforwardly and efficiently compare our two different classes of directional experiment the Asimov method is attractive here as well.

\subsection{Profile likelihood ratio test}
To build a likelihood we must decide on the format that our signal will take, and parameterise the noise level that the measurement of such a signal would suffer. We follow a similar procedure to ref.~\cite{Brubaker:2016ktl}. Ultimately we desire that our experiments measure a power spectrum, which can be obtained by taking the Fourier transform of some timestream. The frequency resolution of the subsequent power spectrum will be given by the inverse of the duration of the timestream sample, $\Delta \nu = 1/\delta t$. The power spectrum would have an extent in frequency equal to the bandwidth of the experiment which we label $\Delta \Omega$. For a single power spectrum taken in this way, thermal and quantum noise defined by a system temperature $T_{\rm sys}$ would be white and exponentially distributed across many realisations. The expected power in each frequency bin of the resulting spectrum is $P_N = k_B T_{\rm sys} \Delta \nu$, and since it is exponentially distributed the standard deviation has the same numerical value. Then we imagine that some large number $\mathcal{N}$ of these power spectra are taken and averaged over a time $\Delta t = \mathcal{N} \delta t$ so that in accordance with the central limit theorem the noise approaches a Gaussian distribution with the same expectation value $P_N$ in each bin but with an uncertainty suppressed by $\sqrt{\mathcal{N}}$,
\begin{equation}\label{eq:noise}
\sigma_N = kT_{\rm sys} \sqrt{\frac{\Delta \nu}{\Delta t}} \, .
\end{equation}
The argument is precisely the same for the statistics of the fluctuations in the signal which are similarly suppressed by this stacking, except that the mean value in each bin is given by the axion power which is a function of frequency. In specific examples later on we assume a noise temperature of $T_{\rm sys} = 4$ K, but explicitly quote how one would scale the results for other temperatures. This value is realistic for a dielectric haloscope which needs a large magnetised volume. For the cavities $T_{\rm sys} = 4$~K could be argued is slightly pessimistic, since there may also be the option of quantum limited noise, however the volumes we require here are also larger than is currently used. Additionally for this temperature, the thermal fluctuations will always have the dominant effect on our signal-to-noise relative to the size of the random fluctuations in the signal. At much lower temperatures they will begin to compete but since the statistics of both the noise and the signal are the same the arguments we make here still hold.  

Now that we can assume we have a Gaussian noise spectrum over some time $\Delta t$, we then iterate this entire process of stacking over an even longer time $t_{\rm obs}$ so that we have a total of $N_t = t_{\rm obs}/\Delta t$ grand power spectra all with Gaussian noise. If $t_{\rm obs}$ is longer than $\mathcal{O}({\rm hours})$ then we also expect our signal to have modulated in this time as well.

There are some restrictions on the lengths of the various times at play here. First we must have our smallest interval of time $\delta t$ long enough to resolve the signal lineshape. For example the minimum duration required to achieve a speed resolution of $\Delta v$ or smaller,
\begin{equation}
\delta t > \frac{2 \pi}{m_a v \Delta v} = 0.03\, {\rm s}\, \left( \frac{40\,\mu{\rm eV}}{m_a}\right) \left( \frac{300\,{\rm km\,s}^{-1}}{v}\right) \left( \frac{1\,{\rm km\,s}^{-1}}{\Delta v}\right) \, .
\end{equation}

The next longest interval $\Delta t$ must be long enough such that we have enough $\mathcal{N}$ power spectra to stack to make the assumption of Gaussian noise e.g. $\Delta t/\delta t \sim 1000$~\cite{Duffy:2006aa}. Then we also must make the assumption that our $\Delta t$ is short enough to assume that the signal does not modulate too much within this time, i.e. that we can approximate the signal within this bin to be signal obtained at the time of the centre of the bin. Since most signals will modulate with a period of a day, as long as we have $\Delta t \lesssim 1$ hour this approximation would be suitable. Then we must also require that our longest time $t_{\rm obs}$ is long enough to see whatever property of the signal we desire, e.g. a day in the case of a daily modulation. We make these arguments simply to demonstrate the timescales that would be required by a real experiment for the steps taken to derive our analytic formulae to be valid, in particular in approximating our later sums over frequency and time bins as integrals. Fortunately these three durations of time are sufficiently distinct from one another for all three mass benchmarks that we believe the signal modelling assumptions to be quite safe.

We can now write down the likelihood for such a dataset, given some model to describe the signal and noise it contains. To summarise, we have a total of $N_t$ power spectra which each have a total of $N_\omega = \Delta \Omega/2\pi \Delta \nu$ frequency bins across the bandwidth. In each of these bins the noise is normally distributed with standard deviation $\sigma_N$ so we can construct a likelihood from the products of the probabilities of seeing measured powers $P_{ij}\simeq \Delta  \omega (\textrm{d}P(\omega_i,t_j)/\textrm{d}\omega)$ in each bin, given the expectation $P^{\rm exp}_{ij}(\Theta)$. This will be dependent on some set of model parameters $\Theta$ that are free in the model $\mathcal{M}$. We write the log likelihood as,
\begin{equation}
\ln \mathcal{L}(P|\mathcal{M},\Theta) = -\frac{1}{2} \sum_{i = 1}^{N_t} \sum_{j = 1}^{N_\omega} \left(\frac{P_{ij} - P^{\rm exp}_{ij}(\Theta)}{\sigma_N}\right)^2 \, ,
\end{equation}
where we have left out the constants from the normalisation of each individual probability that will cancel when we take ratios of likelihoods. Here in assuming a flat standard deviation $\sigma_N$ we have assumed that the dominant statistical fluctuation in the value of the binned power is from thermal noise, neglecting the random fluctuations in the signal. Finally if we wish to build our observatory by combining the signal from multiple experiments, this essentially constitutes an additional sum over each one. For ease of reading we neglect this sum for now, but reintroduce implicitly later in our final results once we are armed with our final analytic formulae.

The profile likelihood ratio test comprises a hypothesis test of some model $\mathcal{M}_1(\Theta)$ (named the alternative hypothesis) against the null hypothesis $\mathcal{M}_0(\Theta)$. One organises $\mathcal{M}_0$ to be a subset of the alternative model, usually by setting some parameter in $\Theta$ to zero.  First we define the maximum likelihood ratio $\Lambda$ which is the ratio between the values of the likelihood that are maximised when $\Theta = \hat{\Theta}$ under model $\mathcal{M}_1$ and $\Theta = \hat{\hat{\Theta}}$ under model $\mathcal{M}_0$,
\begin{equation}\label{eq:likelihood-ratio}
\Lambda = \frac{ \mathcal{L} (P | \mathcal{M}_1,\hat{\Theta} ) }{\mathcal{L} (P | \mathcal{M}_0,\hat{\hat{\Theta}} )  }\, .
\end{equation}
If our null model $\mathcal{M}_0$ is recovered after the application of a constraint on the more general $\mathcal{M}_1$ then we can define a profile likelihood ratio test statistic $D = -2 \ln \Lambda$. According to Wilks' theorem the test statistic is  $\chi^2_{\mu_1-\mu_0} \text{-distributed}$, where the degree of freedom for the $\chi^2$-distribution is given by the difference of free parameters $\mu_1 - \mu_0$ between the two models~\cite{Wilks:1938dza}. So for example if we have some data set and we are trying to test for the presence of one parameter that separates the null and alternative hypotheses. Then we would observe some value of $D = D_{\rm obs}$ and calculate the $\chi_1^2$ cumulative distribution function from this value, which would give the probability of measuring at most $D_{\rm obs}$ if the null hypothesis is indeed false, usually called the significance of the result e.g.
\begin{equation}\label{eq:p-val}
\mathcal{S} = 1-\int_{D_{\rm obs}}^\infty \chi^2_1(D) \,{\rm d} D = \sqrt{D_{\rm obs}} \, ,
\end{equation}
(note that this equates to $\sqrt{D_{\rm obs}}$ only in the case of one parameter).

One way to determine how sensitive an experiment must be to test for some property of a model (e.g. daily modulation or a tidal stream) would be to Monte Carlo generate many sets of mock data and compute the test statistic on each one thus building a distribution of values of $D$. This way one can account for the look elsewhere effect by quoting the required sensitivity in terms of a statistical power $\mathcal{P}$, defined as the probability of obtaining a given result if the alternative hypothesis is true. In other words, the significance is a measure of rejecting the null hypothesis but the power is a measure of accepting the alternative hypothesis. So we could require that our generated distribution of $D$ was such that a fraction $\mathcal{P}$ of them had a significance greater than $\mathcal{S}$. Say if we required $\mathcal{P} = 0.9$ and $\mathcal{S}=0.95$ then an experiment that generated a distribution of $D$ under $\mathcal{M}_1$ that passed these criteria would be able to successfully measure the effect in question to a 95\% significance, 90\% of the time. 

However we will not do this. In fact we can use a much simpler method that does not require us to expensively Monte Carlo many mock datasets, while simultaneously allowing us to obtain analytic relationships between experimental requirements and astronomical goals across wide parameter spaces. First we must define the Asimov data set, i.e. that in which the data in each bin exactly matches the expectation for that bin given model $\mathcal{M}_1$, $P_{ij} = P^{\rm exp}_{ij}$. The likelihood under $\mathcal{M}_1$ will then be correctly maximised and thus equal to 0, but the likelihood under $\mathcal{M}_0$ will be left with a piece that corresponds to the difference between the two models. As long as the number of observations (bins in this case) is high then the Asimov data set will give an excellent estimate to the median value of $D$ one would expect if one were to Monte Carlo the problem. This method is advantageous as it saves significant computational expense and in our case allows more enlightening analytic formulae to be obtained.

\subsection{Measuring modulations}
We search for modulations in the directional correction to the power spectrum, defined in terms of our general formalism in eqs.~\eqref{eq:qtype_correction} and \eqref{eq:ltype_correction}. Under a discretisation in frequency and time we construct the likelihood function from the data set, 
\begin{equation}
P_{ij} = P_0\Delta \omega \bigg[ f(\omega_i) - f_\mathcal{G}(\omega_i,\,t_j)\bigg]\, ,
\end{equation}
where we have written the distributions as functions of $\omega$ using $f(\omega) = \frac{\textrm{d}v}{\textrm{d}\omega} f(v)$ etc. For daily modulations we can treat the function $f(\omega)$ as the sole contribution under the null hypothesis and assume that only the directional correction modulates sinusoidally with time. This adds three parameters, the mean value, amplitude and phase of the modulation: $\{c_0,\,c_1,\, \phi\}$ respectively. Recalling eq.~\eqref{eq:power-modification} we have,
\begin{eqnarray}
		\ell-{\rm type}:  f_\mathcal{G}(\omega_i,t_j) &=& \gl f(\omega_i) \, \zeta_\ell(\omega_i) ~\left[c_0 + c_1 \cos(\omega_d t_j + \phi)\right] \\
		q-{\rm type}: 	f_\mathcal{G}(\omega_i,t_j)&=& \gq f(\omega_i) \left[\zeta_{q1}(\omega_i) + \zeta_{q2}(\omega_i) ~ \left(c_0 + c_1 \cos(\omega_d t_j + \phi)\right)^2 \right] \, ,
\end{eqnarray}
for linear and quadratic experiments respectively. We then want to compare the unmodulated and modulated powers,
\begin{align}
		\mathcal{M}_0 &: \quad P_{ij} = P_0 \Delta \omega f(\omega_i), \quad  \Theta = \{P_0,  v_\lab, \sigma_v\} \, , \\
		\mathcal{M}_1 &: \quad P_{ij} = P_0 \Delta \omega [f(\omega_i) - f_\mathcal{G}(\omega_i, t_j)], \quad \Theta = \{P_0,  v_\lab, \sigma_v, c_0, c_1, \phi\} \, .
\end{align}
Computing $D$ using Asimov data just ends with needing to sum over the directional corrections,
\begin{equation}
		D = \sum_{i,j} \left( \frac{\int_{\rm bin} {\rm d} \omega\, P_0 f_\mathcal{G}(\omega_i, t_j)}{\sigma_N}\right)^2\, .
\end{equation}
Assuming that the bin size is small enough and our data contain most of the signal, we can approximate the sum with an integral
\begin{align}
	D \approx& \frac{\Delta \omega}{\Delta t} \int_{0}^{t_{\rm obs}} {\rm d}t \int_{m_a}^\infty {\rm d} \omega ~ \frac{\left(P_0 f_\mathcal{G}(\omega, t)\right)^2}{\sigma_N^2} \, .
\end{align}
Notice that while in the linear-type case the correction only modulates, in the quadratic case we have a modulation as well as an overall offset that persists over time. We in fact only have to calculate parts of the test statistic for these 3 cases individually:
\begin{eqnarray}
	\ell{\rm -type:} \quad &D_\ell& \approx \, \frac{\Delta \omega}{\Delta t} \int_{0}^{t_{\rm obs}} {\rm d}t \int_{m_a}^\infty {\rm d} \omega ~ \left[\frac{ \gl\,P_0 f(\omega) \, \zeta_\ell(\omega) \, \cos\theta_{\rm lab}(t)}{\sigma_N}\right]^2 \nonumber \\
q{\rm-type\,(offset):} \quad		&D_{q1}& \approx \frac{\Delta \omega}{\Delta t} \int_{0}^{t_{\rm obs}} {\rm d}t \int_{m_a}^\infty {\rm d} \omega ~ \left[\frac{\gq\, P_0 f(\omega) \, \zeta_{q1}(\omega) }{\sigma_N}\right]^2 	\nonumber \\
q{\rm -type\, (modulation):} \quad &D_{q2}& \approx \frac{\Delta \omega}{\Delta t} \int_{0}^{t_{\rm obs}} {\rm d}t \int_{m_a}^\infty {\rm d} \omega ~ \left[\frac{\gq\,P_0 f(\omega) \, \zeta_{q2}(\omega) \, \cos^2\theta_{\rm lab}(t)}{\sigma_N}\right]^2 \, . \nonumber
\end{eqnarray}
The integrals over time and frequency can be separated in each case. After replacing $\sigma_N$ using eq.~\eqref{eq:noise} we find we can write the $\ell$-type test statistic as,
\begin{equation}\label{eq:teststatgeneral}
D_{\ell} = 2 \pi \left (\frac{P_0}{k_B T_{\rm sys}}\right)^2 \mathpzc{g}_{\ell}^2 \, \mathcal{I}^\ell_\omega \,\mathcal{I}^\ell_t \, ,
\end{equation}
where $\mathcal{I}^\ell_\omega$ and $\mathcal{I}^\ell_t$ encode the integrals over $\omega$ and $t$ respectively. The quadratic experiments need to include the offset and modulation term,
	\begin{align}
	D_q = 2 \pi \left (\frac{P_0}{k_B T_{\rm sys}}\right)^2 \gq^2 \, \left(\mathcal{I}^{q1}_\omega \,\mathcal{I}^{q1}_t +\mathcal{I}^{q2}_\omega \,\mathcal{I}^{q2}_t  +\mathcal{I}^{q12}_\omega \,\mathcal{I}^{q12}_t  \right) \, .
	\end{align}	
 We use the label `$q1$' for the integrals of the offset term and `$q2$' for the integrals of the modulation term. Since we integrate over the square of the directional correction (which contains both) we need to include the mixing term which we label `$q12$'.  All the integrals in the above formulae can be written analytically for both the SHM and a stream. The integrals over $\omega$ contain the dependence on the shape of the linewidth, i.e. $\sigma_v$, $v_{\rm lab}$, and therefore scale $\propto {m_a}^{-1}$. The integrals over $t$ encode the information gained from the modulation of the signal, i.e. $c_0, c_1, \phi$, and thus scale $\propto t_{\rm obs}$. We list these in full in Appendix~\ref{sec:analyticformulae}.

\subsection{Parameter constraints}
To estimate the uncertainty on some parameter measurement, we look towards the unmaximised likelihood ratio
\begin{eqnarray}
	d(\Theta) = 2 \ln \frac{\mathcal{L} (P | \mathcal{M}_1,\, \Theta )}{ \mathcal{L} (P | \mathcal{M}_{\rm 0},\, \Theta)} \, ,
\end{eqnarray}
where if $\Theta$ only contains the parameters of interest, $d(\hat{\Theta}) = D$. The uncertainty on a model parameter $\vartheta \in \Theta$ can be estimated from the curvature of $d$ around the value $\hat{\vartheta}$ that maximises the likelihood under $\mathcal{M}_1$
\begin{align}
	\sigma_{\vartheta}^{-2} = - \frac{1}{2} \frac{\partial^2}{\partial \vartheta^2} d \big|_{\vartheta=\hat{\vartheta}} \, \,.
\end{align}
For a daily modulation we are interested in the set $\Theta = \left\{c_0, c_1, \phi\right\}$. If the true values are at $\Theta_{\rm true}$, we may compute $d$ for the Asimov data set giving,
\begin{equation}\label{eq:dTheta}
	d(\Theta) = - \sum_{i,j} \chi_{ij,\,1}^2(\Theta) + \sum_{ij} \chi_{ij,\,0}^2 \, , 
\end{equation}
where,
	\begin{eqnarray}
\chi_{ij,\, 1}^2(\Theta) &\equiv& \frac{1}{\sigma^2_N} \left[\int_{\rm bin} \textrm{d} \omega\, P_0 \big(f_\mathcal{G}(\omega_i,\, t_j \, | \, \Theta) - f_\mathcal{G}(\omega_i,\, t_j \, | \, \Theta_{\rm true})\big) \right]^2\\ 
 \chi_{ij,\, 0}^2 &\equiv& \frac{1}{\sigma^2_N} \left[\int_{\rm bin}\textrm{d} \omega\,  P_0 f_\mathcal{G}(\omega_i,\, t_j\, | \, \Theta_{\rm true}) \right]^2 \, .
\end{eqnarray}
Here only the first sum depends on $\Theta$, so taking the derivative with respect to any $\vartheta \in \Theta$ sees the second one vanish. Taking the $\ell$-type case first we see that we in fact just have the expression,
	\begin{align}
	d(\Theta) \approx& \, 2 \pi \left (\frac{P_0}{k_B T_{\rm sys}}\right)^2 \gl^2 \int_{0}^{t_{\rm obs}} {\rm d}t \int_0^\infty {\rm d} \omega ~ \left[f(\omega) \, \zeta_\ell(\omega)   \, ( \cos(\theta_{\rm lab})|_\Theta - \cos(\theta_{\rm lab})|_{\Theta_{\rm true}}) \, \right]^2 \, ,
	\end{align}
which involve the same integrals that have already been introduced in computing $D$. For the $\ell$-type case this leads to,
	\begin{align}
	\sigma_{\vartheta}^{-2} =& \,  -\pi \left (\frac{P_0}{k_B T_{\rm sys}}\right)^2  \gl^2 \,  \mathcal{I}^\ell_\omega \, \frac{\partial^2 d_t}{\partial \vartheta^2}\bigg|_{\Theta={\Theta}_{\rm true}} \, ,
	\end{align}
where the derivatives of the time integral $d_t$ for each parameter are,
	\begin{align}
	-\frac{\partial^2 d_t}{\partial c_0^2} \bigg|_{{\Theta}_{\rm true}} &=  2t_{\rm obs} \label{eq:sens-c0-linear} \\
	-\frac{\partial^2 d_t}{\partial c_1^2}\bigg|_{{\Theta}_{\rm true}} &=  t_{\rm obs}-\frac{\sin (2 \phi_{\rm true} )}{2 \omega_d }+\frac{\sin (2 (t_{\rm obs} \omega_d +\phi_{\rm true} ))}{2 \omega_d }  \label{eq:sens-c1-linear}\\
	-\frac{\partial^2 d_t}{\partial \phi^2} \bigg|_{{\Theta}_{\rm true}} &=  {c_{1,{\rm true}}}^2 \left(t_{\rm obs}
	+\frac{\sin (2 \phi_{\rm true})}{2 \omega_d }
	-\frac{\sin (2 (t_{\rm obs} \omega_d +\phi_{\rm true}))}{\omega_d }
	+\frac{\sin (2 t_{\rm obs} \omega_d +2 \phi_{\rm true})}{2 \omega_d }	\right) \, .\label{eq:sens-phi-linear}
	\end{align}
The $q$-type offset is insensitive to any $\{c_0,c_1,\phi\}$ but the $q$-type modulation leads to more lengthy terms, analogous to eqs.~\eqref{eq:sens-c0-linear}-- \eqref{eq:sens-phi-linear} so we make an approximation for large times $t_{\rm obs}$ so that the $\sin$ terms are negligible and get
	\begin{align}
	-\frac{\partial^2 d_t}{\partial c_0^2} \bigg|_{{\Theta}_{\rm true}} &\approx (8 c_{0, {\rm true}}^2 + 4 c_{1, {\rm true}}^2) t_{\rm obs}   \, ,\label{eq:sens-c0-quadratic} \\
	-\frac{\partial^2 d_t}{\partial c_1^2} \bigg|_{{\Theta}_{\rm true}} &\approx (4 c_{0, {\rm true}}^2 + 3 c_{1, {\rm true}}^2) t_{\rm obs}    \, ,\label{eq:sens-c1-quadratic}\\
	-\frac{\partial^2 d_t}{\partial \phi^2} \bigg|_{{\Theta}_{\rm true}} &\approx (4 c_{0, {\rm true}}^2 c_{1, {\rm true}}^2 + c_{1, {\rm true}}^4) t_{\rm obs}   \, . \label{eq:sens-phi-quadratic}
	\end{align}
 Including the offset term ends exactly as above since it does not depend on $\Theta$. This is as expected, since a known offset should not alter the level at which the parameters of an oscillation can be inferred. As one would expect from a Gaussian likelihood, the uncertainty on each modulation parameter scales with $t_{\rm obs}^{-1/2}$ and inversely with signal-to-noise $(P_0/k_B T_{\rm sys})^{-1}$. We also notice that the uncertainty on the phase $\phi$ scales with the amplitude of the modulation as $c_1^{-1}$, which is also to be expected since if a signal does not modulate ($c_1=0$) then the phase is undefined and unmeasurable.

\section{Results}\label{sec:results}
\subsection{Measuring the daily modulation}\label{sec:dailymod}
\begin{figure}[t]
\centering
\includegraphics[width=\textwidth,trim={0cm 0cm 0cm 0cm},clip]{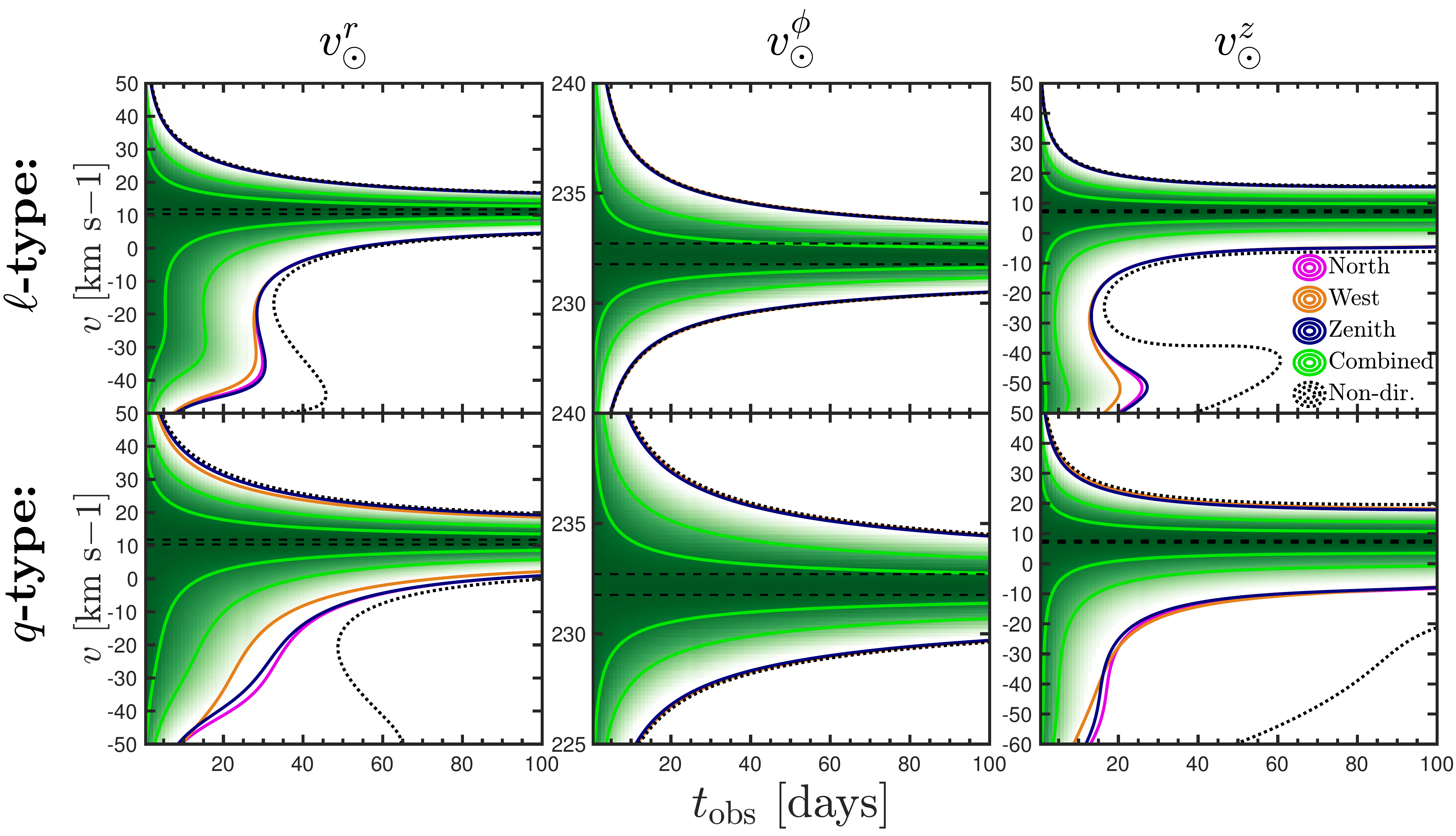} 
\caption{One dimensional constraints on the Solar velocity components as a function of experimental duration $t_{\rm obs}$. There are five sets of constraints corresponding to using the signal from each experimental orientation separately (magenta, orange and blue lines for north, west and zenith directions) and then one (green) for the three experiments combined. We show the non-directional constraint for an equivalent experiment with $\gl=0$ or $\gq = 0$ as a black dotted line. For all experiments except for the combined we only show the 2$\sigma$ upper and lower uncertainties. For the combined constraint we show both 1 and 2$\sigma$ uncertainties and shade according to the value of the profile likelihood. The horizontal black lines indicate the size of the statistical uncertainty on the Solar peculiar velocity (the part of $v^\phi_\odot$ from the galactic rotation speed also has sizable model dependent systematic uncertainties which we do not show).}
\label{fig:dailymod_deltat}
\end{figure}
Employing the machinery described in the previous section we now estimate the general scale of signal required in a directional experiment to measure the daily modulation. For measuring a modulation controlled by three parameters to a 3$\sigma$ discrimination against an unmodulated hypothesis, we need a test statistic of $D = 13.93$. The power required in $\ell$ and $q$-type experiments following the formulae detailed in the previous section lead us to the general sizes,
\begin{equation}
P_\ell > 1.3 \times 10^{-21}\, {\rm W}\, \left(\frac{T_{\rm sys}}{4\,{\rm K}} \right) \left(\frac{\gl}{2.4\times 10^{-4}\,{\rm km}^{-1} \,{\rm s}} \right)^{-1} \left(\frac{t_{\rm obs}}{4\,{\rm days}} \right)^{-\frac{1}{2}}  \left( \frac{m_a}{100 \, \mu{\rm eV}}\right)^\frac{1}{2} \, ,
\end{equation}
\begin{equation}
P_q > 8.6 \times 10^{-22}\, {\rm W}\, \left(\frac{T_{\rm sys}}{4\,{\rm K}} \right) \left(\frac{\gq}{1.1\times 10^{-6} \,{\rm km}^{-2} \,{\rm s}^2} \right)^{-1} \left(\frac{t_{\rm obs}}{4\,{\rm days}} \right)^{-\frac{1}{2}}  \left( \frac{m_a}{100 \, \mu{\rm eV}}\right)^\frac{1}{2} \, ,
\end{equation}
where we assume that three identical experiments pointing along the north, west and zenith axes have been combined. These values of power are in line with the three benchmarks of section~\ref{sec:benchmarks} so we expect to be able to measure the daily modulation in the experiments as they are described.

The implication of detecting the daily modulation will be the ability to infer the 3-dimensional components of the Solar velocity in the galactic rest frame. We expressed the constraint one can place on the modulation parameters in analytic form in the previous section but now we can translate it into the astrophysical language by computing the constraints on the Solar velocity, $\mathbf{v}_\odot$, from the likelihood directly. This way we can account for both the daily and annual modulations. We write the velocity in galactocentric coordinates now as $\mathbf{v}_\odot = (v_\odot^r,\,v_\odot^\phi,\,v_\odot^z)$. The second component of this velocity (which includes the rotation speed of the local standard of rest) still possesses sizable systematic uncertainties and is very sensitive to the modelling of the Milky Way rotation curve~\cite{McMillan:2009yr}. We show the constraints on the components of the Solar velocity as a function of total experimental duration $t_{\rm obs}$ in figure~\ref{fig:dailymod_deltat}. In this result and in subsequent results we will be comparing a benchmark quadratic velocity experiment with a linear velocity one. We will also be comparing experiments pointing along each of our three laboratory axes, as well as an experiment using signals from all three experiments combined. Here we assume that the experiments have a 4~K noise temperature and total power of $P_\ell = 2.6\times 10^{-21}$ W and $P_q = 1.7\times10^{-21}$ W, which are chosen so that the both combined $\ell$ and $q$-type experiments measure the modulation to the same significance in 4 days.

As anticipated when we wrote down our analytic formulae, the constraint on our modulation parameters and subsequently the components of the Solar velocity decrease with total time $\propto t_{\rm obs}^{-1/2}$. The uncertainty on $v^\phi_\odot$ and the upper boundary of the constraints on $v^r_\odot$ and $v^z_\odot$ in fact exhibit two scaling regimes both $\propto t_{\rm obs}^{-1/2}$ but with different gradients for short and long times. We associate these with the daily modulation at short times and the annual modulation for longer times. One very noticeable feature for the lower limits of the $r$ and $z$ components is the multiple solutions for short durations. This is most pronounced in the non-directional limit. Even though the full annual modulation signal is sufficient to discriminate between these solutions --- the lower solution for these velocity components does eventually disappear --- this requires $t_{\rm obs} \gtrsim 40$--$60$ days. In particular for the power used in the $q$-type experiment we require even longer times before the uncertainty on $v^z_\odot$ reaches below 10 $\kms$. The impact of the multiple solutions for $\mathbf{v}_\odot$ is dampened significantly with the inclusion of directional information. Since we have normalised the values of power so that the \emph{daily} modulation is detected to the same significance, the evolution for small $t_{\rm obs}$ is very similar. However towards larger durations the uncertainty bands decrease slightly faster for the $l$-type experiments when the dominant influence is the fact that the $\ell$-type power is slightly higher. In the transition between these two regimes, the incorrect solution for $\mathbf{v}_\odot$ vanishes slightly faster in the $\ell$-type experiment since it cannot reproduce the modulation signals as well when sign information is present. Ultimately the prospects for measuring the Solar velocity are very good in axion experiments generally. Additionally here we are beginning to see that the directional information is making marked improvements to the discovery reach especially for short duration experiments. In all experiments we are able to get good constraints on $v_\odot^\phi$ since this parameter is also largely involved in setting the shape of the power spectrum, which our likelihood function is integrating over as well as the modulation.

\subsection{Measuring the anisotropy of the DM halo}\label{sec:anisotropy}
\begin{figure}[t]
\centering
\includegraphics[width=\textwidth]{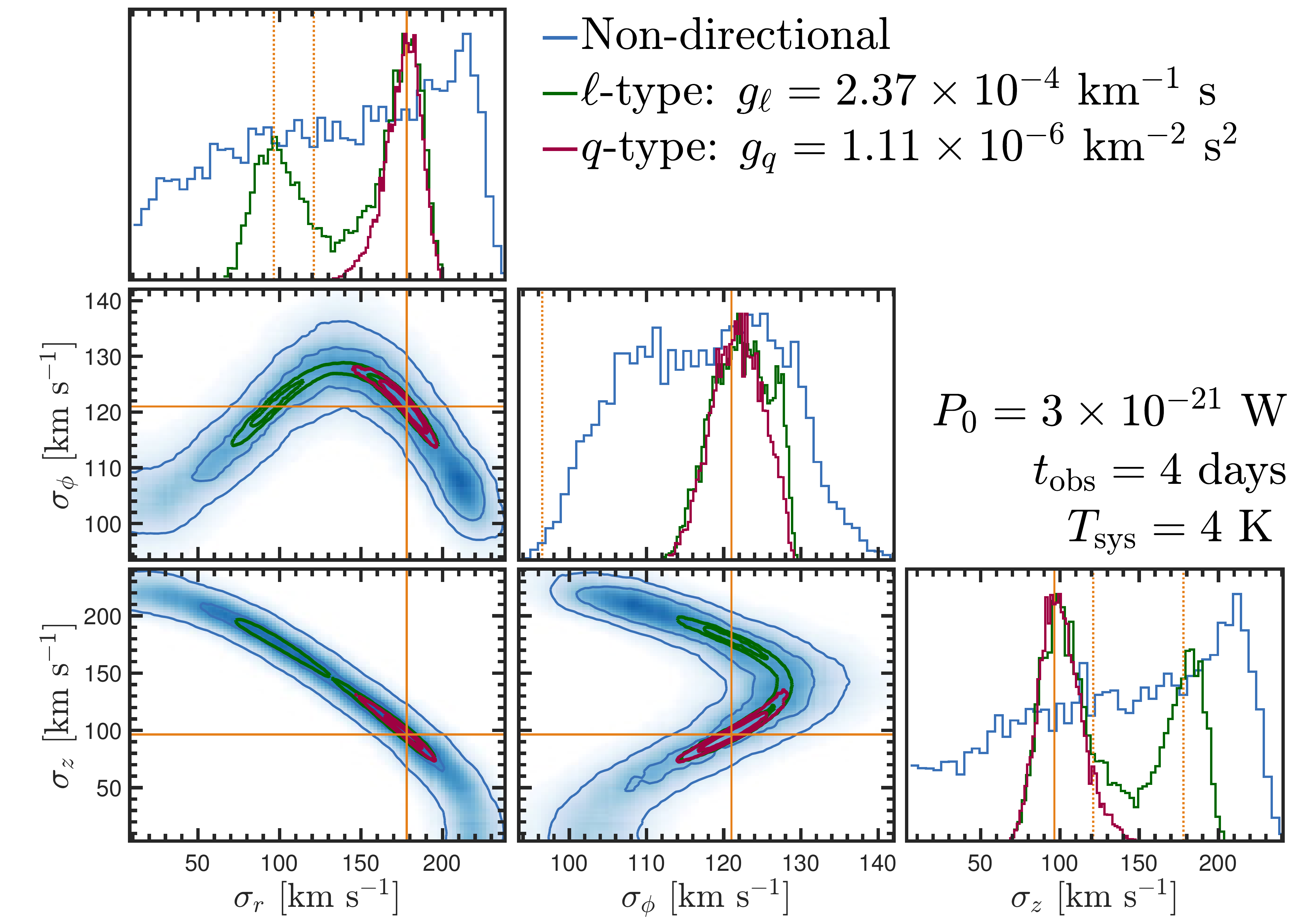} 
\caption{One and two-dimensional marginalised posterior distributions for the reconstruction of the anisotropy of the velocity distribution from its three dispersion components $\sigma_{r,\,\phi,\,z}$. There are three sets of constraints corresponding to the signals from non-directional (blue), linear-directional (green) and quadratic-directional (red) experiments. For the directional experiments we have combined signals from all three experimental axes. The input values of the dispersion velocities in each direction are indicated with orange lines. In each one-dimensional panel we also show the true values of the other two dispersion components as dotted lines (the value of $\sigma_r$ is beyond the limits shown for the $\sigma_\phi$ panel).}
\label{fig:anisotropy_params}
\end{figure}
It is predicted that the smooth component of the velocity distribution of a dark matter halo cannot be perfectly isotropic. It may be possible for an axion experiment with directional sensitivity to detect some anisotropy in the velocity ellipsoid of our own halo, even if it was present at a low level. Milky Way analogues in N-body simulations generically observe halos with some level of anisotropy, see e.g. refs.~\cite{Wojtak:2008mg,Ludlow:2011cs,Lemze:2011ud}, and indeed models for the real MW halo share this prediction~\cite{Hunter:2013vua,Fornasa:2013iaa,Bozorgnia:2013pua}. For galaxies forming from radial infall this usually results in a larger velocity dispersion in the radial direction. In our own Milky Way indeed a significantly larger velocity dispersion in the radial direction was observed in the kinematics of halo stars~\cite{Herzog-Arbeitman:2017zbm}. Such an anisotropy would likely be difficult to observe with the frequency dependence of the power spectrum alone. But one would expect a velocity distribution that was slightly hotter in one direction to alter the phases and amplitudes of daily modulations in a more complicated way than simply being controlled by $\cos{\theta_{\rm lab}}$. Detecting this anisotropy will be one of the key benefits of a directional experiment so it is a useful exercise.

The degree of anisotropy in the velocity ellipsoid of some component of a galactic halo is usually parameterised\footnote{If the halo model is allowed to possess triaxiality the anisotropy parameter can depend on other galactic coordinates as well as radius.} with $\beta(r)$,
\begin{equation}
\beta(r) = 1 - \frac{\sigma^2_t}{2\sigma^2_r} \, ,
\end{equation} 
where $\sigma_{r,t}$ are velocity dispersions in the radial and tangential directions. If at a given radius $\sigma^2_t = 2\sigma^2_r$ then $\beta = 0$ and the distribution is isotropic. N-body halos typically have anisotropy parameters that are zero for $r\rightarrow 0$ which then grow to values $\beta(r > 8\,{\rm kpc}) \sim 0.2$--$0.4$~\cite{Wojtak:2008mg,Ludlow:2011cs,Lemze:2011ud}, although it has been suggested that the inclusion of baryons may make the local distribution less anisotropic~\cite{Kelso:2016qqj}. We can model a velocity distribution with some anisotropy by generalising the isotropic Maxwellian introduced earlier,
\begin{equation}
f(\mathbf{v}) = \frac{1}{(8 \pi^3 \det{\boldsymbol{\sigma}^2})^{1/2}} \exp\left(-\frac{1}{2}(\mathbf{v} + \vlab)^T \boldsymbol{\sigma}^{-2} (\mathbf{v} + \vlab) \right) \, ,
\end{equation} 
where $\sigma^2_t = \sigma^2_\phi + \sigma^2_z$ at our position. If we assume that the dispersion tensor is diagonal $\boldsymbol{\sigma}^2 = \textrm{diag}(\sigma^2_r, \sigma^2_\phi, \sigma^2_z)$ then this is,
\begin{equation}
f(\mathbf{v}) = \frac{1}{(2 \pi)^{3/2} \sigma_r \sigma_\phi \sigma_z} \exp\left( - \frac{(v_r+v_\textrm{lab}^r)^2}{2\sigma^2_r}- \frac{(v_\phi+v_\textrm{lab}^\phi)^2}{2\sigma^2_\phi}- \frac{(v_z+v_\textrm{lab}^z)^2}{2\sigma^2_z}\right) \, .
\end{equation}
One can allow for correlations between the dispersions in different directions with off-diagonal elements, however for simplicity we neglect this possibility. Reference~\cite{Herzog-Arbeitman:2017fte} does observe a slight tilt in the velocity ellipsoid of their halo stars, but mostly only due to one correlation (the $\sigma_r \sigma_z$ element).

Starting from the isotropic case, we increase the dispersion velocity slightly in the radial direction and decrease it in the tangential directions. We then attempt to measure the resulting anisotropy by placing the above velocity distribution into our statistical analysis as before\footnote{Our directional integrals do not yield analytic results here so the analysis in this section is purely numerical.}. For this example we choose as a benchmark the best fit values of the dispersion components of the distribution of metal-poor halo stars from Ref.~\cite{Herzog-Arbeitman:2017fte}; we use the set with metallicities [Fe/H]$<-1.8$. These values are $\sigma_r = 178$ km s$^{-1}$, $\sigma_\phi=121\kms$ and $\sigma_z = 96.5 \kms$, giving an anisotropy parameter of $\beta = 0.62$ which is a relatively high value. We sample the posterior distribution generated by our Asimov likelihood over linear priors in all parameters\footnote{We use the {\sc MultiNest} nested sampling algorithm~\cite{Feroz:2013hea,Feroz:2008xx,Feroz:2007kg} with 5000 live points to do this.}. 

In figure~\ref{fig:anisotropy_params} we display the one and two-dimensional posterior distributions. We also set the components of $v_\odot$ as free parameters but marginalise over them since their resulting uncertainties are essentially the same as the results of the previous section. In addition to our two directionally sensitive experiments we include the result from the same analysis in an equivalent experiment but with no directional effect, i.e. setting $\gl$ or $\gq$ to zero. The total power $P_0$ is assumed to be identical in all experiments. As expected the non-directional signal has poor sensitivity to the anisotropy with large bands of viable values of the three dispersion components able to reproduce the shape of $f(v)$ (in this case the experimental duration is insufficient to measure any modulation in frequency). In particular the measurement of $\sigma_r$ and $\sigma_z$ is very poor, with the constraints consistent with 0 at the 95\% level for both parameters. This is because the value of $\sigma_\phi$ is in the direction that $f(v)$ is primarily boosted, so ends up having the greatest impact on its shape.

With directional sensitivity on the other hand we gain major sensitivity to the velocity anisotropy. Peculiarly though, the $q$-type case performs much better here. In fact the $\ell$-type experiment exhibits a multimodal solution for the values of $\sigma_r$ and $\sigma_z$ where it seems to struggle to distinguish between the numerical values of the two parameters. This is perhaps at first counter-intuitive since one would expect that the $\ell$-type experiment would always be more sensitive by not discarding the sign information on $\mathbf{v}$. However this is only true when searching for individual directions, for instance the direction of $\vlab$. Here we are trying to constrain parameters which control the \emph{shape} of the distribution. Note that the dispersion parameters only ever enter the signal as the square, there is no sign information there to measure. The $q$-type experiments turn out to be more sensitive because they receive larger (albeit negative) directional corrections over the span of frequencies where the dispersion values are playing the greatest role. Moreover the directional correction has a persistent offset, analogous to the $\zeta_{q1}(\omega)$ term in Eq.~\eqref{eq:power-modification}. Whereas in the $\ell$-type experiments, since the correction can be both positive and negative over one day, there are times when it disappears (or can become very small) and the signal becomes essentially non-directional. Again for measuring individual velocities, the vanishing of the directional correction can only happen when they line up with the axis of the experiment correctly. However here this effect is a hindrance since without any directional correction at all the signal cannot distinguish between shape parameters controlling the widths of the distribution in different directions. However we should emphasise the excellent reconstruction shown in the $q$-type experiment, which is able to distinguish each dispersion component from each other at over the 2$\sigma$ level.

\subsection{Measuring a stream}\label{sec:stream}
\begin{figure}[t]
\centering
\includegraphics[width=\textwidth,trim={1.5cm 0 17.7cm 0},clip]{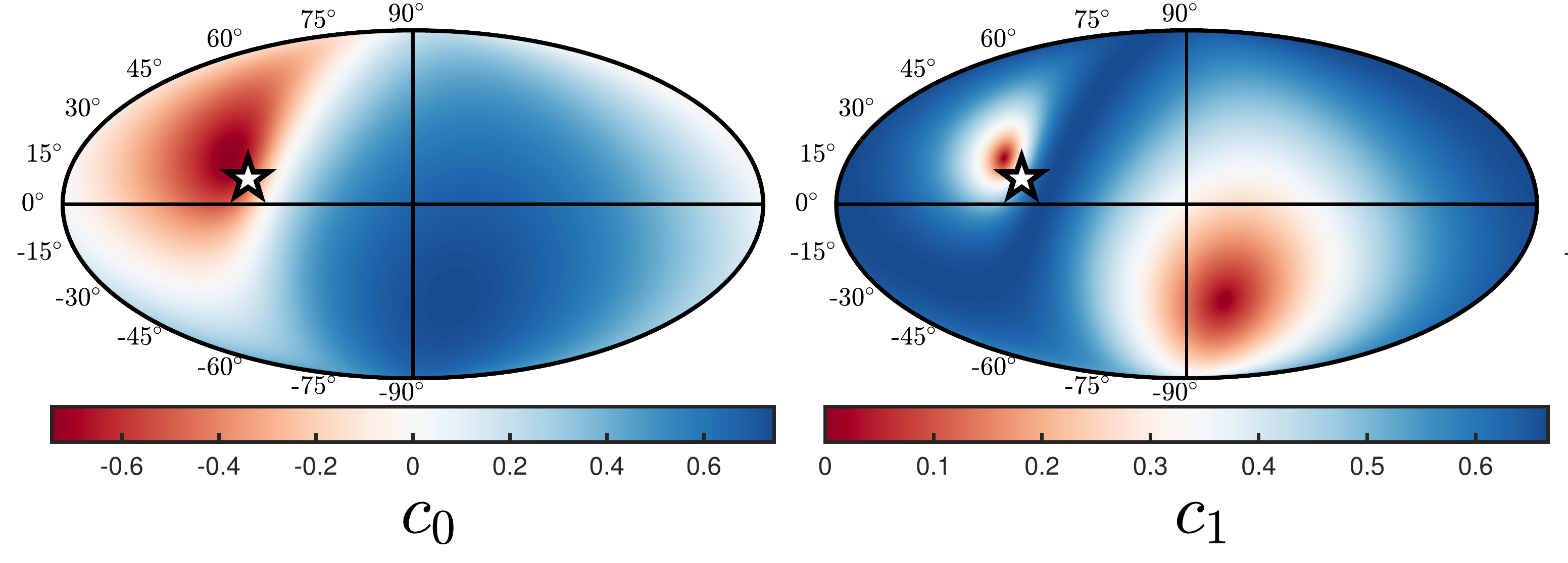} 
\includegraphics[width=0.55\textwidth,trim={32.7cm 0 1.5cm 0},clip]{Figures/streammodulation-eps-converted-to.pdf} 
\caption{Mollweide projections of the parameters $c_0$, $c_1$ and $\phi$ for the daily modulation of the stream-experiment angle $\cos{\theta_{\rm str}}$, as a function of the stream's galactic longitude and latitude $(l_{\rm str},\,b_{\rm str})$. The modulation parameter values correspond to a measurement on January 1 in a north-pointing experiment. We also fix the stream with a galactic frame speed of 300 km s$^{-1}$. The white star indicates the direction of the lab velocity. There are two points at which $\phi$ is undefined and $c_1 = 0$, which are where $|\vlab - \vstr|$ coincides with the rotation axis of the Earth. No daily modulation would be present for streams observed to be pointing in these directions.}
\label{fig:streammodulation}
\end{figure}

\begin{figure}[t]
\centering
\includegraphics[width=\textwidth]{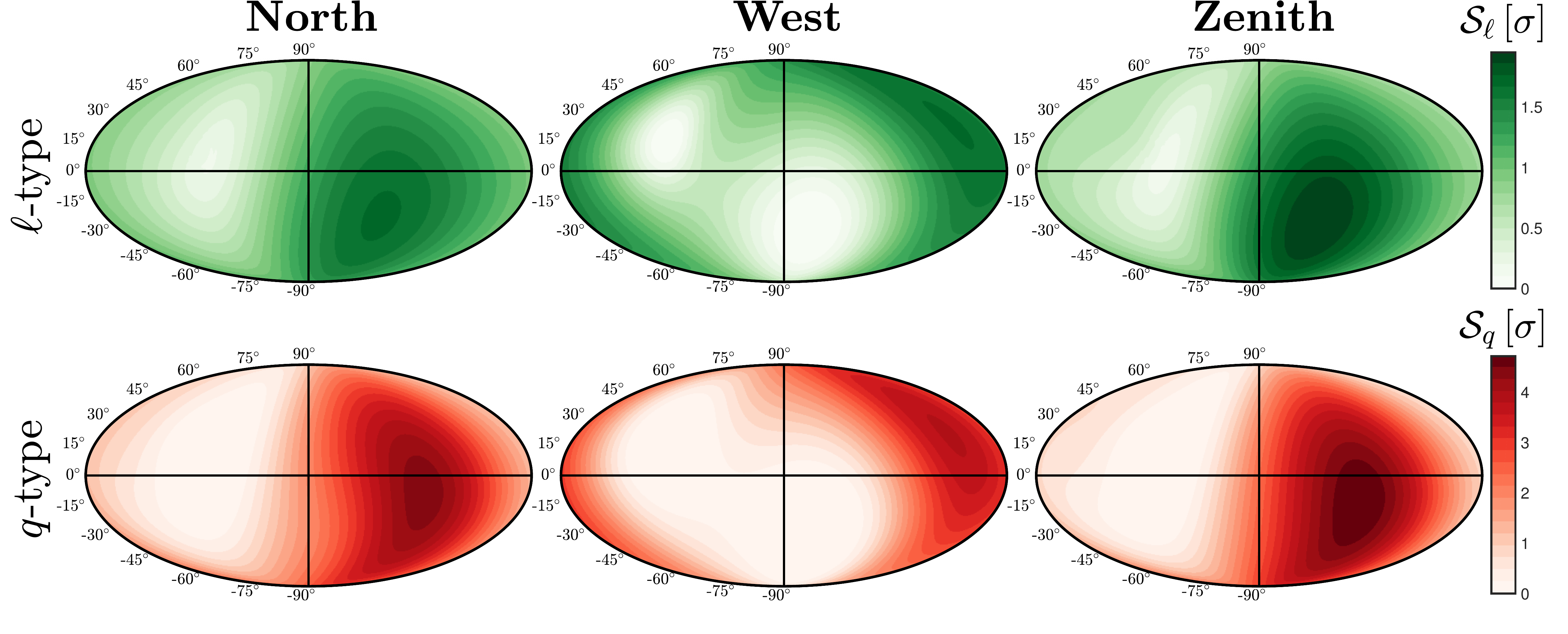} 
\caption{Significance for measuring the daily modulation of a stream as a function of stream longitude and latitude in each experiment and for linear and quadratic types. The measurement time is assumed to be over 4 days beginning on January 1 and the stream speed is fixed at 300 km s$^{-1}$.}
\label{fig:streamD_indiv}
\end{figure}

\begin{figure}[t]
\centering
\includegraphics[width=\textwidth,trim={1.2cm 0 2.8cm 0cm},clip]{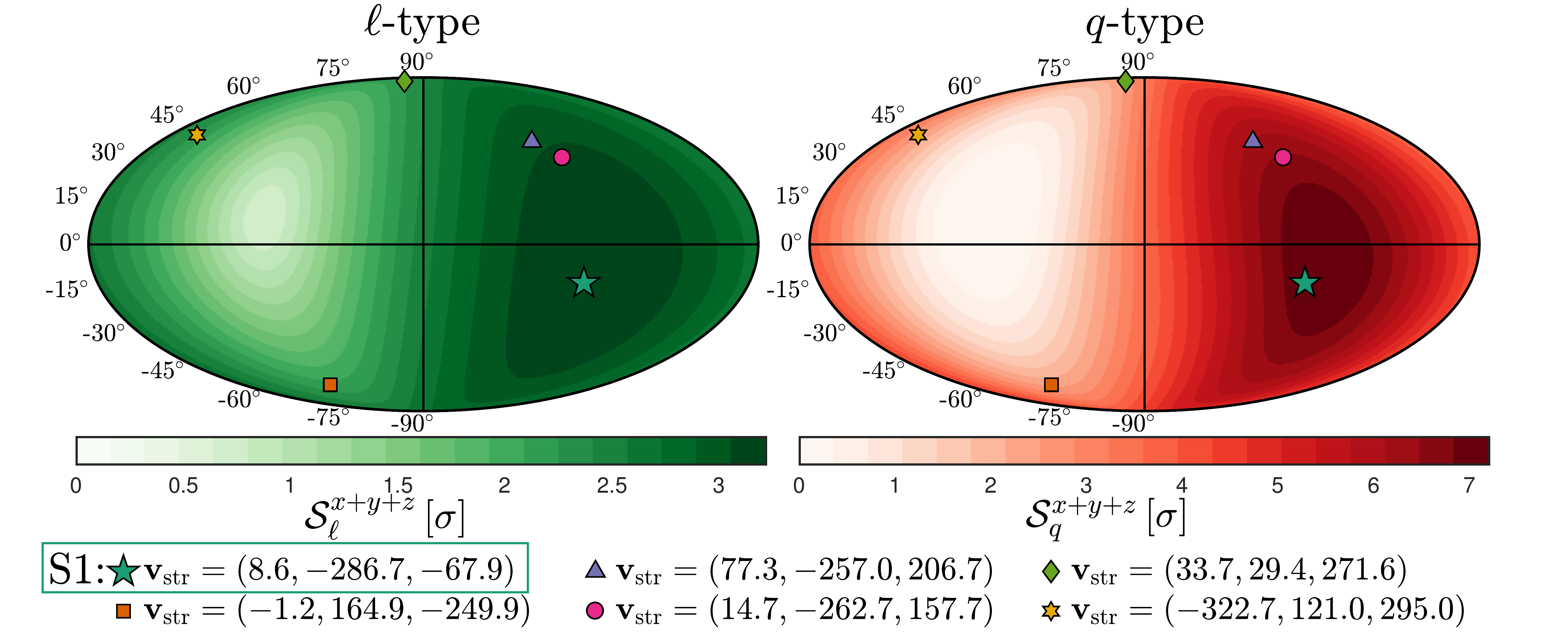} 
\caption{Significance for measuring the daily modulation of a stream after combining all three experimental axes, shown above in figure~\ref{fig:streamD_indiv}. We also overlay the galactic longitude and latitude of the directions of the six nearby objects reported in refs.~\cite{Myeong:2017skt,Myeong:Preprint}. As before the stream speed is assumed to be 300 km s$^{-1}$, which is approximately the speed of most of these features. We highlight the S1 stream which can be most confidently claimed to intersect the Solar position.}
\label{fig:streamD_total}
\end{figure}

\begin{figure}[t]
\centering
\includegraphics[width=\textwidth,trim={1cm 0cm 0cm 0cm},clip]{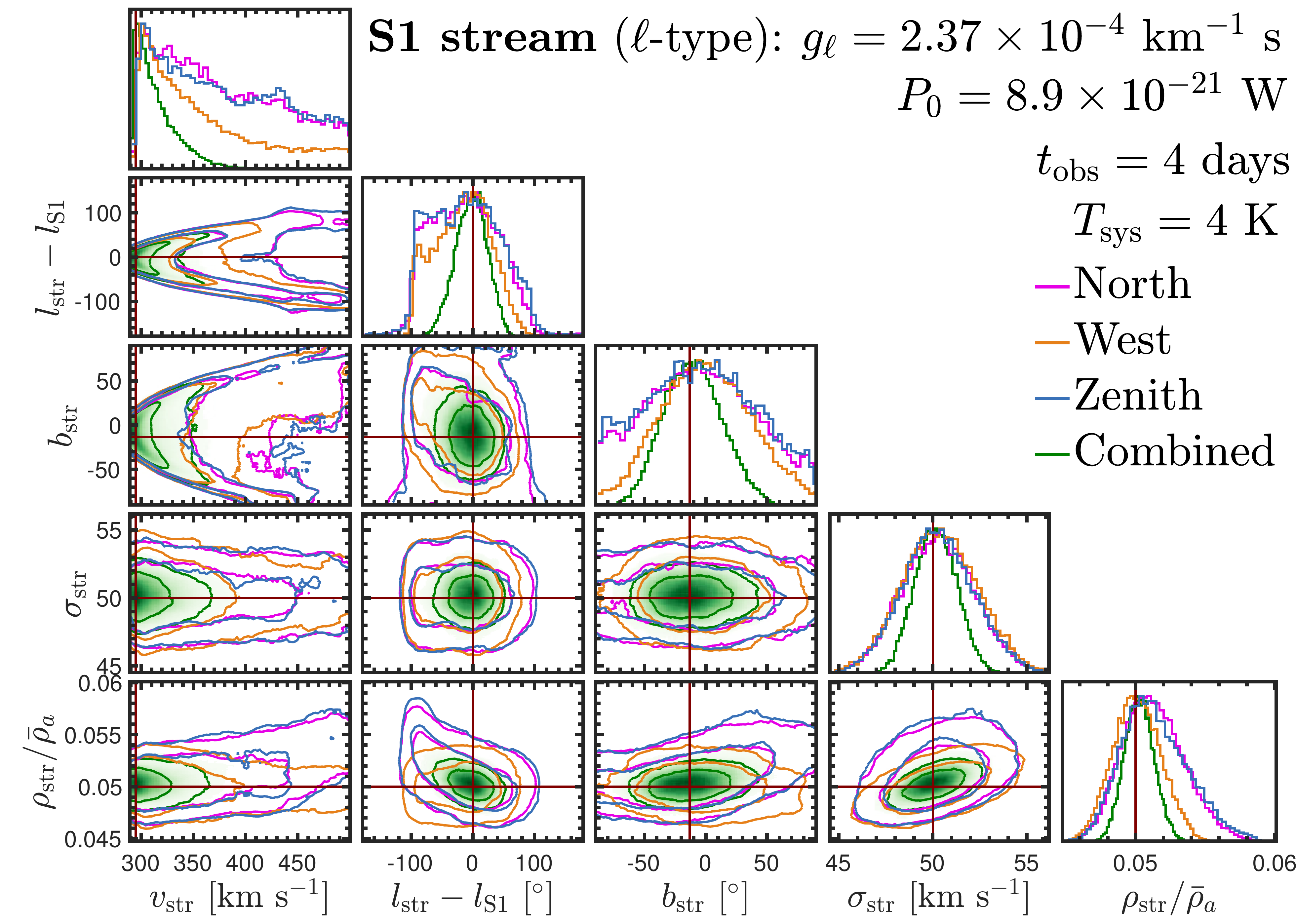} 
\caption{One and two-dimensional marginalised posterior distributions for the reconstruction of the five parameters of the S1 stream: its velocity, dispersion and density (from left to right horizontally). There are four sets of constraints corresponding to using data from each $\ell$-type experimental orientation separately (magenta, orange and blue contours for north, west and zenith directions) and then one (green) for the three experiments combined. The straight red lines mark the true parameter values.}
\label{fig:streammodulation-ltype}
\end{figure}

\begin{figure}[t]
\centering
\includegraphics[width=\textwidth,trim={1cm 0cm 0cm 0cm},clip]{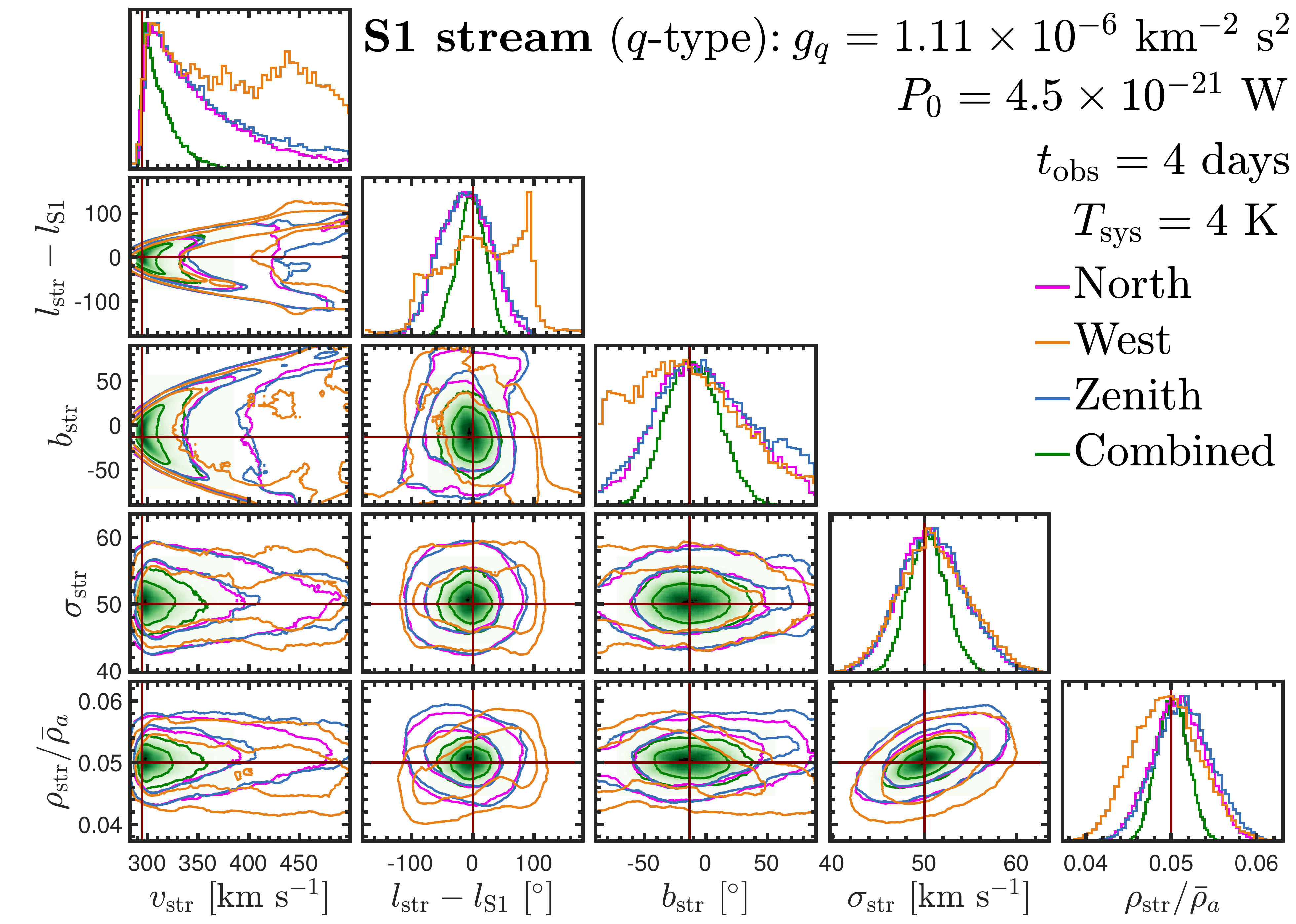} 
\caption{Posterior distributions for the reconstruction of the S1 stream as in figure~\ref{fig:streammodulation-ltype} but here for $q$-type experiments.}
\label{fig:streammodulation-qtype}
\end{figure}

The treatment of the daily modulation due to $\vlab$ is entirely analogous to a treatment one can make of the daily modulation induced by a stream. As mentioned previously one only needs to make the substitutions $\vlab \rightarrow \vlab - \vstr$, $\sigma_v \rightarrow \sigma_{\rm str}$ and $\bar{\rho}_a \rightarrow \rho_{\rm str}$. Streams could in principle appear to originate from any galactocentric velocity, so our prior on the stream direction can only be the whole sky. We write the stream direction using the usual galactic longitude and latitude $(l_{\rm str},b_{\rm str})$\footnote{The conversion to galactocentric cylindrical coordinates is defined as $(v^r,\,v^\phi,\,v^z) = v(\cos{b}\cos{l},\,\cos{b}\sin{l},\,\sin{b})$.}. We show how the values of the daily modulation parameters for a stream vary with this direction in figure~\ref{fig:streammodulation}. For clarity we show the parameters $\{c_0,\,c_1,\,\phi\}$ for a north-pointing experiment at Munich, measuring a stream with a galactic frame speed of $300\,\kms$ on January 1. The modulation parameters are defined precisely as before, 
\begin{equation}
\cos{\theta^i_{\rm str}}(t) =  \frac{\hat{\mathbf{x}}^i\cdot(\vlab -\vstr)}{|\vlab - \vstr|} = c_0+c_1 \cos(\omega_d t + \phi)
\end{equation}
where $\hat{\mathbf{x}}^i$ is one of our axes (north, west, zenith). Here and for several figures to come we display functions of the full sky by mapping the the galactic $(l,\,b)$ with a Mollweide projection. As is convention we put the galactic centre at the origin. The longitude $l$ is read horizontally and the latitude (which is also labelled numerically) is read vertically. One should interpret a position on this projection as the direction that a given stream \emph{points towards}. So at the position of the white star is a stream that is co-rotating with us. Symmetrically opposite would be a stream that is counter-rotating, e.g. S1.

The most noticeable feature in figure~\ref{fig:streammodulation} is that there appear to be two regions of the sky where the modulation amplitude $c_1$ vanishes and subsequently the modulation phase becomes undefined. We associate these two points with the rotation axis of the Earth, where naturally if the Earth frame stream direction happens to coincide with our axis of rotation, no daily modulation will occur. The location of these ``blind spots'' in $\hat{\mathbf{v}}_{\rm str}$ varies with $v_{\rm str}$ and over the course of the year as the Earth's rotation axis moves relative to the halo. But at any given day there will always be certain streams that will not induce a modulation. This fact is of course conspicuous and would not prevent the stream from being observed using the frequency of the feature. However in the case of quadratic-type experiments that can only measure $|c_0|$, we remark that roughly half of the stream `sky' is degenerate with the other half. The location of the poles also varies over the year as well as with the value of $v_{\rm str}$, however the skymaps are qualitatively similar.

Next we show how the measurability of a stream via its daily modulation is dependent on the direction of the stream. In figure~\ref{fig:streamD_indiv} we show the significance achievable in measuring the daily modulation of a stream comparing against the model in which the stream is unmodulated (i.e. a non-directional experiment). The modulation adds three parameters so we compute the significance from the value of the test statistic and the $\chi^2_3$ distribution, then converting to a ``Gaussian $\sigma$'' i.e. $68\% \rightarrow 1\sigma$ etc. We again display the result for both linear and quadratic experiments along each axis separately. The significance is displayed as a function of stream direction $(l_{\rm str},\,b_{\rm str})$ projected using the same Mollweide mapping as in the previous figure. Following this we also show the total test statistic (all three experiments combined) in figure~\ref{fig:streamD_total}. We see that especially in the west-pointing experiment the significance vanishes along the same directions as highlighted earlier: those that align with the rotation axis of the Earth. A stream is undetectable via its daily modulation in this particular experiment, however comparing the same point in the north and zenith-pointing experiments shows that it is indeed measurable in those. Whilst the quadratic experiments observe a greater maximum significance value for head-on streams, the linear experiments observe a consistently large significance over the full sky (keeping in mind that the dielectric haloscope can be both a linear and a quadratic experiment when using different combinations of signals from the left and right hand sides of the device). The head-on streams are the most well-measured when looking for \emph{modulations} because faster features give greater deviations away from the non-directional power, cf. $\mathcal{G}_\ell \propto v$ and $\mathcal{G}_q \propto v^2$. This is also the reason why quadratic experiments require smaller overall powers to measure the faster features to the same significance. So a stream originating from the opposite direction to the one we are moving (e.g. S1) will always be the most well-measured directionally.

In figure~\ref{fig:streamD_total} we also mark the directions of the six nearby substructures reported in refs.~\cite{Myeong:2017skt,Myeong:Preprint}. As mentioned in section~\ref{sec:fv} the first of these objects labelled `S1' has been confidently claimed to be a stream that intersects our position. S1 arrives head-on with respect to our galactic orbit, placing it in prime orientation for detection. If present this should be easily picked up by an axion search, and subsequently fully measurable in a directional experiment. Computing the test statistic for the S1 stream we find that measuring the velocity components of the stream from its daily modulation requires powers in $\ell$ and $q$-type experiments of,
\begin{equation}\label{eq:S1powerltype}
P_\ell \gtrsim  8.9\times 10^{-21}\, {\rm W}\, \left( \frac{\rho_{\rm str}}{0.05\bar\rho_a}\right)^{-1} \left(\frac{T_{\rm sys}}{4\,{\rm K}} \right) \left(\frac{\gl}{2.4\times 10^{-4}\,{\rm km}^{-1} \,{\rm s}} \right)^{-1} \left(\frac{t_{\rm obs}}{4\,{\rm days}} \right)^{-\frac{1}{2}}  \left( \frac{m_a}{100 \, \mu{\rm eV}}\right)^\frac{1}{2} \, ,
\end{equation}
\begin{equation}\label{eq:S1powerqtype}
P_q \gtrsim  4.5\times 10^{-21}\, {\rm W}\, \left( \frac{\rho_{\rm str}}{0.05\bar\rho_a}\right)^{-1} \left(\frac{T_{\rm sys}}{4\,{\rm K}} \right) \left(\frac{\gq}{1.1\times 10^{-6} \,{\rm km}^{-2} \,{\rm s}^2} \right)^{-1} \left(\frac{t_{\rm obs}}{4\,{\rm days}} \right)^{-\frac{1}{2}}  \left( \frac{m_a}{100 \, \mu{\rm eV}}\right)^\frac{1}{2} \, ,
\end{equation}
As with our daily modulation these are essentially within the scope of the benchmark experiments detailed in table~\ref{tab:benchmarks}.

As a final word on the topic of tidal streams we would like to display how well a directional experiment can make measurements of all the properties of the stream in conjunction. We take the aforementioned case of the S1 stream and perform a maximum likelihood fit, taking the threshold required powers to measure the modulation eqs.~\eqref{eq:S1powerltype} and~\eqref{eq:S1powerqtype}. We explore the posterior distribution generated by sampling our likelihood function over linear priors in the five parameters defining the stream: the velocity, dispersion (which is $\sim$ 50 km s$^{-1}$ when written as a single variate Maxwellian~\cite{Myeong:Preprint}) and the density. We make the assumption that the density of dark matter in S1 comprises 5\% of $\bar{\rho}_a$, although we stress that this parameter is completely unknown. 

The marginalised posterior distributions are shown in figures~\ref{fig:streammodulation-ltype} and~\ref{fig:streammodulation-qtype} for $\ell$ and $q$-type experiments respectively, again performing four separate analyses in each case. The first three use data from each (north, west and zenith) experiments separately and then a fourth with all three combined. As expected with all three experiments combined the stream is very well-measured. Individually we can see that in the $\ell$-type case the west-pointing experiment appears to constrain the stream most successfully, but in the $q$-type experiment it is the worst. An intuition for this result can be gleaned by looking back at figure~\ref{fig:powerspectra} which shows the actual signal for this stream. In the $\ell$-type spectra the west-pointing experiment has a very large modulation amplitude since it has both the sensitivity to the sign of $\cos{\theta_{\rm lab}^\mathcal{W}}$ and a value of $c_1$ slightly larger than the other two directions (which have their amplitudes suppressed by factors of $\cos{\lambda_{\rm lab}}$ or $\sin{\lambda_{\rm lab}}$). On the other hand in the $q$-type experiment this large modulation gets folded into purely negative values. So the west-pointing experiment becomes much less useful for measuring S1 when quadratic effects are considered. However as already discussed the effect in all three $q$-type experiments is enhanced due to a large extra factor of $|\vlab - \vstr|$ meaning they need less power to reach an equal significance. Our measurements are mostly nicely Gaussian with the exception of $v_{\rm str}$ which looks to be approximately one-sided for speeds larger than the true speed of S1 $\approx 300\,\kms$. This is due to the fact that S1 is incoming head-on, so for all directions other than the true direction a faster stream is needed to reproduce the correct peak frequency. As before, we reiterate that our signal powers are reasonable based on the experimental setups summarised in table~\ref{tab:benchmarks}. The signal requirements can be rescaled according to eqs.~\eqref{eq:S1powerltype} and~\eqref{eq:S1powerqtype} whereas the experimental requirements to reach those signals can be rescaled using eqs.~\eqref{eq:benchdhpowers} and~\eqref{eq:benchcavitypowers}

\subsection{Prospects for minicluster streams}
\begin{figure}[t]
\centering
\includegraphics[width=0.8\textwidth,trim={0cm 0 0cm 5.5cm},clip]{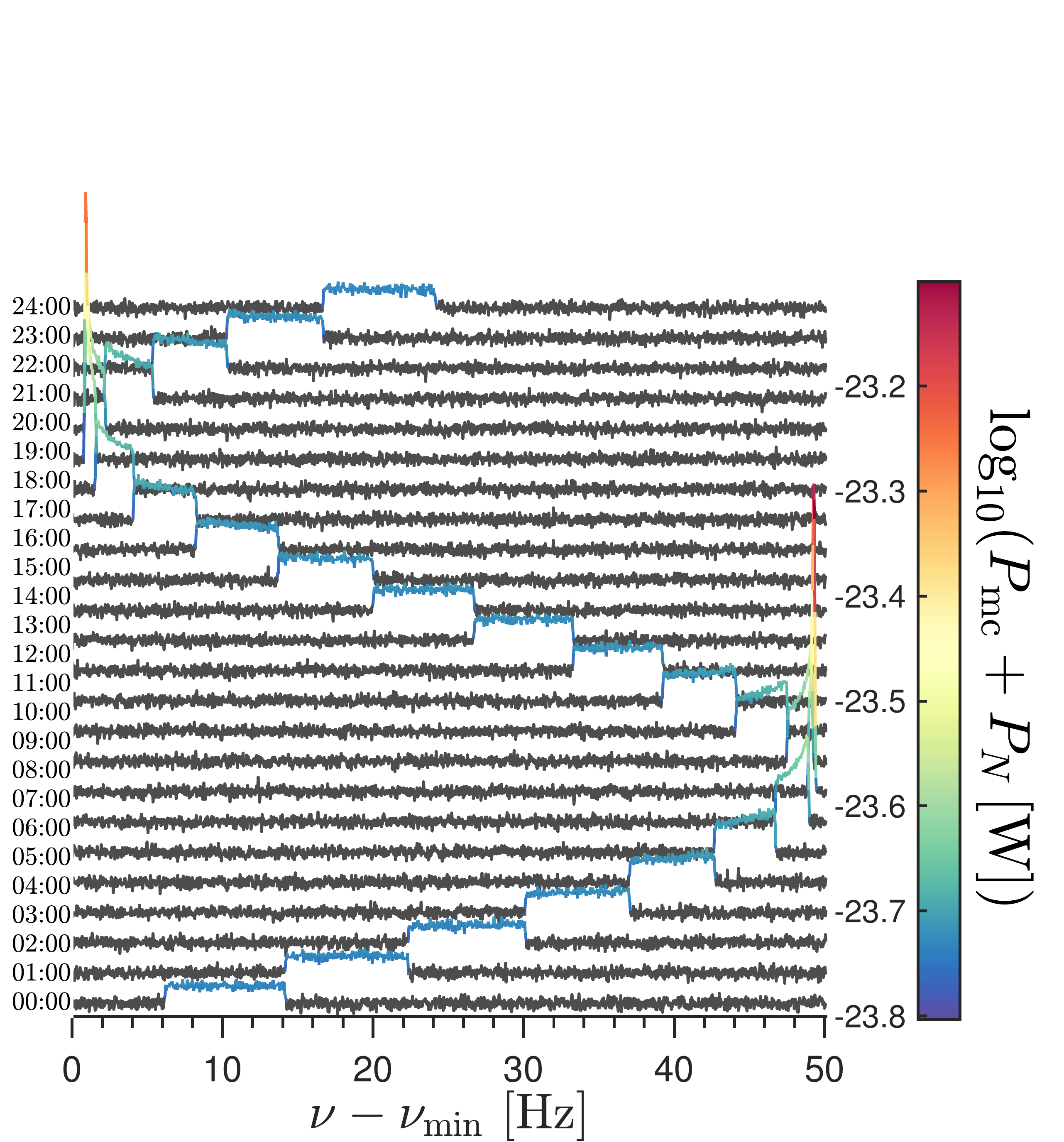} 
\caption{Simulated power spectrum for the observation of a minicluster stream as a function of time binned with durations of one hour. Since the minicluster stream linewidth is so much smaller than the change in $v_{\rm lab}$ over one hour due to the rotation of the Earth, the signal modulates roughly sinusoidally over the day with an amplitude and phase related to $|\vlab - \mathbf{v}_\textrm{mstr}|$.}
\label{fig:miniclusterspectrum}
\end{figure}
\begin{figure}[t]
\centering
\includegraphics[width=1\textwidth,trim={0cm 0cm 0cm 0cm},clip]{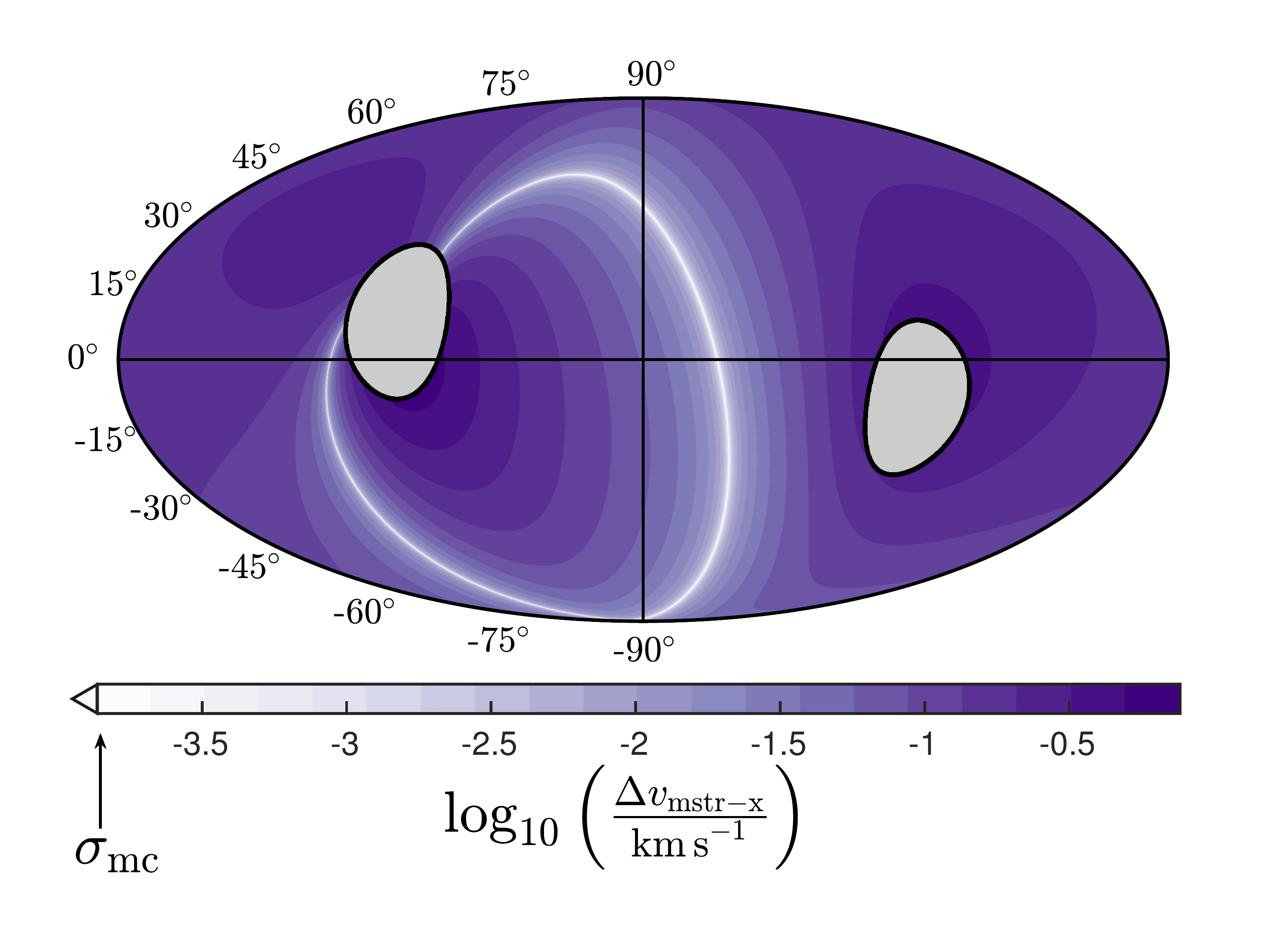} 
\caption{Shift in the velocity of a ministream within a time of $T_{\rm mstr-x}$, as a function of the stream direction, again using a Mollweide projection of its galactic longitude and latitude. We assume a stream speed of 300 km s$^{-1}$, an overdensity of $\delta = 10$ and an initial minicluster mass of $M_{\rm mc} = 10^{-13}\,M_{\odot}$. The grey regions show stream directions for which $T_{\rm mstr-x}>0.25$ days, which will be approximately those which can clearly observe a daily modulation signal. The lower limit of the colour scale corresponds to the dispersion of the minicluster stream $\sigma_{\rm mc} = 1.4 \times 10^{-4} \kms$.}
\label{fig:ministream_deltav}
\end{figure}
A scenario that has been gaining interest in the last couple of years is the possibility that a decent chunk of an axionic dark matter halo could be bound up in miniclusters (see refs.~\cite{Berezinsky:2013fxa,Tkachev:2014dpa,Fairbairn:2017sil,Fairbairn:2017dmf,Enander:2017ogx,Davidson:2016uok,Levkov:2018kau} for the most recent progress on the topic). Miniclusters have intriguing signatures for indirect detection, but a punishingly small direct encounter rate on Earth. It was suggested however in refs.~\cite{Tinyakov:2015cgg, Dokuchaev:2017psd} that over time and many passages through the disk and bulge that miniclusters may become appreciably tidally disrupted by stellar interactions. Crossings through their trailing ministreams would likely be more frequent. Even if the density of a given stream was diluted over many Gyr, the initial density of a minicluster is so high that the detection prospects are not completely unfathomable. Remaining agnostic with regards to how often such a passage could occur\footnote{This requires much more in depth numerical analysis accounting for the initial mass function and abundance of miniclusters, in turn needing a full simulation of the axion field through the QCD phase transition.}, we can nevertheless describe how the signal from the crossing of a ministream could be used to measure the properties of its progenitor.

We assume the simplest model for a minicluster~\cite{Kolb:1994fi}, that of a sphere with density $\rho_{\rm mc}$ and mass $M_{\rm mc}$. The densities are very large, typically labelled by some contrast $\delta$,
\begin{equation}
\rho_{\rm mc} = 7\times 10^6\, {\rm GeV \, cm}^{-3}\, \delta^3(1+\delta) \, .
\end{equation}
Miniclusters have a characteristic mass given by the horizon enclosure at matter-radiation equality, around $M_{\rm mc} \simeq 10^{-12}\,M_\odot$. The precise spectrum and mass function of miniclusters is the subject of much ongoing work. Here we focus only on the heuristic arguments regarding their detection and suggest that for now one resorts to the scaling relations detailed below if concerned about the specificities some minicluster.

Without directional sensitivity one can extract the density $\rho_{\rm mstr}$, dispersion $\sigma_{\rm mc}$ and the lab frame {\rm speed} $|\vlab - \mathbf{v}_{\rm mstr}|$ from the power spectrum. These are related to the properties of the minicluster as well as the age of the stream $t_{\rm mstr}$. We have,
\begin{eqnarray}\label{eq:ministreamdensity}
\rho_{\rm str} &\simeq & \rho_{\rm mc} \frac{R_{\rm mc}}{\sigma_{\rm mc} t_{\rm mstr}} \\
&\simeq & 19.8 \, {\rm GeV \, cm}^{-3}\,\,\delta^{3/2} (1+\delta)^{1/2}  \left( \frac{1\, {\rm  Gyr}}{t_{\rm mstr}} \right) \, , \nonumber
\end{eqnarray}
for the stream density (which is diluted linearly since the instance of disruption) and for the virial velocity dispersion,
\begin{equation}\label{eq:miniclusterdispersion}
\sigma_{\rm mc} = \sqrt{\frac{G M_{\rm mc}}{R_{\rm mc}}} = 6.28\times 10^{-5} \, {\rm km\,s}^{-1}\, \delta^{1/2} (1+\delta)^{1/6} \left(\frac{M_{\rm mc}}{10^{-12} \, M_\odot}\right)^{1/3} \, .
\end{equation}
We assume that the stream retains the original temperature of its progenitor. In principle this would only be a lower limit on the dispersion since tidal effects will likely heat the minicluster by some amount proportional to the timescale of disruption, see e.g.~ref.~\cite{Schneider:2010jr}. We also have an additional observable, the minicluster stream crossing time, dependent on the radius of the stream and its orientation relative to our trajectory,
\begin{equation}\label{eq:streamcrossingtime}
T_\textrm{mstr-x}= \frac{2 R_{\rm mc}}{v_\textrm{lab}\sin{(\vartheta_{\rm mstr})}} \approx \frac{4\, {\rm days}}{\delta (1+\delta)^{1/3}}  \left( \frac{M_\textrm{mc}}{10^{-12} M_\odot} \right)^{1/3} \left( \frac{ \sin{(60^\circ)}}{\sin{(\vartheta_{\rm mstr})}}\right) \, .
\end{equation}
We denote the angle between the ministream velocity and the lab velocity by
\begin{equation}
\sin{(\vartheta_{\rm mstr})}=\sqrt{1-\left(\frac{\textbf{v}_\textrm{lab}\cdot\mathbf{v}_{\rm mstr}}{v_\textrm{lab}v_{\rm mstr}}\right)^2} \, .
\end{equation}
Notice we have six unknown parameters $\{\delta,\,M_{\rm mc}, \, t_{\rm mstr},\,\mathbf{v}_{\rm mstr} \}$ but only four equations with which to determine them (eqs.~\eqref{eq:ministreamdensity},~\eqref{eq:miniclusterdispersion},~\eqref{eq:streamcrossingtime}, and the frequency of the stream which provides $|\vlab - \mathbf{v}_{\rm mstr}|$). In a more sophisticated model we may also wish to describe the density profile of the minicluster. So we are going to require additional information, or so it would seem. In fact, the situation is slightly more complicated than the cases considered before. So far we have ignored the daily modulation in \emph{frequency} due to the rotation {\rm speed} of the Earth $\sim$0.47 km s$^{-1}$ which is negligible when considering the full axion power spectrum with a width of $\sim$ 300$\kms$. But here we are dealing with features that have characteristic linewidths 4 orders of magnitude smaller than even this smallest correction. So this means that one could in fact extract two additional pieces of information from a non-directional signal --- the phase and amplitude of the daily modulation in frequency. We illustrate a  signal in figure~\ref{fig:miniclusterspectrum}, showing a single day's worth of modulation in the power spectrum. Since the minicluster linewidth is much smaller than the variation in $\vlab$, integrating the spectrum over one hour produces a signal that has swept out a segment of frequencies. At times when the Earth rotates along the direction of the stream the modulation turns over, leading to a very large enhancement in power (note that the signal is plotted with a log scale). The modulation in frequency is not a perfect sinusoid with a period of one day because the revolution speed of the Earth is slightly different at the beginning of the day compared with the start. This is only now visible when at such fine resolution. The full six parameters of this very simple model would be measurable to high accuracy, as long as the experiment could achieve this spectral resolution (which only requires an increase in timestream sample duration, see section~\ref{sec:miniclusterstats} below).

So it seems here that we have no need of directionality. But consider those minicluster streams that would give rise to signals with no daily modulation in frequency. This can happen in two ways. Firstly if the stream crossing time is much smaller than 1 day then the modulation parameters of the signal and thus the three components of the stream velocity will not be measurable (as in figure~\ref{fig:dailymod_deltat} for very small durations). Complementary to this, if the stream dispersion is wider than the frequency shift the feature undergoes during the crossing time, then this too would mean the modulation is poorly measured. Miniclusters that produce wide frequency band streams with very short radii are those with very high values of $\delta$, or an $M_{\rm mc}$ much smaller than $10^{-12}$ $M_\odot$. In figure~\ref{fig:ministream_deltav} we show the shift in speed over the crossing time (labelled $\Delta v _{\rm mstr-x}$) as a function of the stream direction for a particular minicluster input with $\delta = 10$ and $M_{\rm mc} = 10^{-13}\,M_\odot$. For this minicluster the velocity dispersion is $\sigma_{\rm mc} = 1.3\times10^{-4}$ km s$^{-1}$ which is used as the lower limit of the colour scale. Ministream directions in the light band that stretches across the image give frequency shifts during the crossing time that are smaller than the linewidth. We also exclude stream directions which have crossing times above 1/4 of a day, which as one would expect are roughly collinear with our trajectory $\pm \vlab$. Here we assume that each ministream crossing began on January 1. The skymaps for other times look qualitatively similar however the light band of stream directions will move across the sky with the rotation axis of the Earth. Clearly this is a rather fine tuned region of minicluster parameter space, but we should mention again that the estimate for the stream dispersion used here can only be a lower limit. One could expect hotter ministreams to be possible if heating during tidal disruption was accounted for. This issue along with the many other mysteries surrounding miniclusters we leave for future work.

\subsubsection*{A note on noise statistics for miniclusters}\label{sec:miniclusterstats}
Miniclusters have extremely small velocity dispersions which means one would need to make some modifications to the binning as described in section~\ref{sec:stats}. To gain a sufficient frequency resolution to measure a feature with a typical minicluster width centred around $v\sim 300 \kms$ we need to have a single power spectrum constructed from $\delta t > 180$ seconds of integration time (for $m_a = 100 \,\mu$eV).  For our analytic treatment this is not a problem; our formulae are independent of the choice of $\delta t$ since we make the assumption that the sum over power spectra bins can be approximated by an integral. It does mean though that we are forced to consider the case where the frequency binning is small enough to pick up the shape of the feature in the first place. This leads to a different problem regarding the statistics of the noise and the randomness of the signal. Since we wish our larger daily modulation bins to have durations of $\lesssim$ 1 hour (so that our \emph{time} sum can be approximated by an integral), we will only be able to construct them from at most $\mathcal{N}\sim 20$ power spectra. This is potentially worrying since we made the assumption that the central limit theorem was making our noise and signal fluctuations Gaussian. The average of $\mathcal{N}$ exponentially distributed numbers with an expectation value of $P_N$ is a gamma distribution with a shape parameter of $\mathcal{N}$ and a scale parameter of $P_N$. For $\mathcal{N}\sim 20$ the discrepancy should not be important (given our other approximations). For our lower mass benchmarks however we would need to consider values $\mathcal{N}\sim 2 - 7$ for the typical minicluster\footnote{For more information on this particular statistical issue we refer the reader to ref.~\cite{Duffy:2006aa} searching for cold flows of axions in ADMX.}. This would cause the noise to be noticeably non-Gaussian and the observed signal much more influenced by random correlations in the phases of the axion field.  Moreover it may be that the assumption of completely uncorrelated phases is not the ideal description for a disrupted minicluster. They could in fact retain some of the highly correlated nature that is characteristic of a minicluster. Though saying much more on this issue would require an in depth study. 

\section{Summary}\label{sec:summary}
In developing a general formalism to describe directional effects in axion detection we have settled on three designs that would be able to implement them in reality. The first two we discussed consist of modifications to the conventional resonant cavity. In cavities -- and any experiment with electric fields that have standing wave behaviour --- there is only the possibility to gain sensitivity to the projection of the \emph{square} of the axion velocity along the elongated axis of the device. We have described how one could construct such cavities that are large enough to approach the de Broglie wavelength of the axion. For masses between $10\,\mu$eV  and $40\,\mu$eV we can set up cavities at high and low mode numbers respectively. The low mass end with high mode numbers requires a rather lengthy cavity but it turns out that only the ends of the cavity need to be magnetised to achieve a usable directional effect. For an effect of the same magnitude at the higher end of this mass range the cavity needs to be fully magnetised, but can have a very thin aspect ratio while using only the lowest mode. 

At higher masses still we have developed a way to extend the dielectric haloscope concept employed by MADMAX to exploit phase differences across the device as suggested in ref.~\cite{Millar:2017eoc}. In this latter case we have devised a setup where the disks are spaced symmetrically just out of phase with respect to perfect constructive interference at $v=0$. Adding or subtracting the signals from either side of the experiment can then give quadratic or linear dependence on the axion velocity. For all the experiments discussed, the various real world requirements for measuring a $\sim 10$\% directional effect are summarised in table~\ref{tab:benchmarks}. These benchmarks informed the feasibility of doing axion astronomy but one need not be more optimistic than we have. Even if parameters such as our benchmark magnetic field or noise temperature are not achievable, much of the astronomy can be done but for slightly longer times than the (already very brief) benchmark duration of $t_{\rm obs}~=~4$~days.

Directional experiments pose excellent prospects for the post-discovery era. Signals that exhibit pronounced daily modulations give us access to the full three dimensional velocity distribution in a much shorter time than is required for non-directional experiments. We find that these experiments can straightforwardly pick up this daily modulation and use it to infer the components of the Solar velocity. Directional experiments are particularly novel since they are able to measure the anisotropy of the velocity ellipsoid of the Milky Way dark matter halo. Due to higher order alterations to the daily modulation, the signals can distinguish increased or decreased velocity dispersions in different galactocentric directions. Such a fine sensitivity to the multidimensional structure of the velocity distribution is not possible in non-directional experiments. We also find that substructure in the form of streams are measurable in very short periods of time for almost any orientation across the sky for our linearly sensitive experiments. The local S1 stream which has been shown to directly pass through the Solar position is in fact in prime position for detection since it is incoming almost head-on with a very fast laboratory frame speed. This would lead to a large directional correction in quadratic and linear experiments even if the dark matter content was scarce, at 5\% of $\rho_0$ or less. We also showed that all the properties of the S1 stream can be reconstructed with the daily modulation signal alone. This is especially interesting if one considers the possibility of very small scale substructure due to the disruption of miniclusters by stars in the Milky Way that would give rise to enhancements in the signal over timescales of a day or less. We have found a small range of hot dense miniclusters with streams that are roughly orthogonal to our trajectory through the galaxy that would \emph{require} a directional experiment to measure. For all other miniclusters the full set of properties could be reconstructed thanks to the non-directional daily modulation due to the rotation \emph{speed} of the Earth.

In addition to the concepts detailed here we expect that there will be many more extensions one could devise to cover the remaining axion windows. For instance in a cavity resonating at lower frequencies (e.g. 2 $\mu$eV), we have checked that under quantum limited noise it would be possible to directly measure the electric field with $S/N >1$ in a single coherence time. Tracking the phase of the electric field in several of these experiments (which would need to be separated by $\sim$km) and combining them in real time would allow this array of cavities to measure the instantaneous axion velocity and populate $f(\mathbf{v})$ ``mode by mode''. At larger masses it may be possible to extend the dish antenna method to gain directional sensitivity to axion dark matter~\cite{Jaeckel:2015kea}. Beyond even these masses, at the upper end of the unexcluded axion window, a dielectric haloscope for optical frequencies has been suggested~\cite{Baryakhtar:2018doz}. Since it is analogous to the dielectric haloscope concept used by MADMAX it could be extended in precisely the same way as we have described. The only difference would be present in the statistical treatment of the background necessary when using bolometers to do photon counting as proposed by the authors. Whatever new experiments enter the stage, we hope that the general formalism we have developed here will be of use.

\section*{Acknowledgements}
We thank N. W. Evans for discussion and further information on the local streams. We also thank E. Vitagliano for enlightening discussions. CAJO is very grateful for the benevolent hospitality of the Max Planck Institute for Physics in Munich, where much of this work took place. CAJO is spoilt by the grant FPA2015-65745-P from the Spanish MINECO and European FEDER. AJM acknowledges partial support by the Deutsche Forschungsgemeinschaft through Grant No. EXC 153 (Excellence Cluster “Universe”) and Grant No. SFB 1258 (Collaborative Research Center ``Neutrinos, Dark Matter, Messengers''), as well as the European Union through Grant No. H2020-MSCA-RISE-2015/690575 (Research and Innovation Staff Exchange project ``Invisibles Plus''). JR is supported by the Ramon y Cajal Fellowship 2012-10597, the grant FPA2015-65745-P (MINECO/FEDER), the EU through the ITN “Elusives” H2020-MSCA-ITN-2015/674896 and the Deutsche Forschungsgemeinschaft under grant SFB-1258 as a Mercator Fellow.

\appendix

\section{Lab velocity and modulation parameters}\label{sec:labvelocity}
This appendix deals with the computation of the lab velocity, in particular its three dimensional components in our laboratory coordinate system and the derivation of our definition of the daily modulation parameters $\{c_0,\,c_1,\, \phi\}$. 

The lab velocity $\vlab(t)$ is annually and diurnally modulated by the revolution and rotation of the Earth.  To compute $\vlab(t)$ we need to first define the galactic coordinate system $(\hat{\textbf{x}}_g,\hat{\textbf{y}}_g,\hat{\textbf{z}}_g)$ with axes in directions pointing to the galactic centre, galactic rotation (at the position of the Solar system), and the galactic north pole. We can transform vectors from the galactic to the laboratory system with the following transformation,
\begin{equation}
 \begin{pmatrix}\hat{\mathcal{N}}\\\hat{\mathcal{W}}\\\hat{\mathcal{Z}}\end{pmatrix} = R_{\rm lab}(t)\left(R_{\rm gal} \begin{pmatrix}\hat{\textbf{x}}_g\\\hat{\textbf{y}}_g\\\hat{\textbf{z}}_g\end{pmatrix} \right) \, ,
\end{equation}
where the transformation from the galactic to the intermediate equatorial system is given by the matrix,
\begin{equation}
R_{\rm gal} =
\begin{pmatrix}
-0.05487556 & +0.49410943 & -0.86766615 \\
-0.87343709 & -0.44482963 & -0.19807637 \\
-0.48383502 & +0.74698225 & +0.45598378
\end{pmatrix}  \, ,
\end{equation}
with values assuming the International Celestial Reference System convention for the right ascension and declination of the North Galactic Pole, $(\alpha_{\rm GP},\delta_{\rm GP}) = (192^\circ.85948,\,+27^\circ.12825)$ as well as the longitude of the North Celestial Pole $l_{\rm CP} = 122^\circ.932$~\cite{BinneyGalacticAstronomy}. Then, from the equatorial to the laboratory system at latitude $\lambda_{\rm lab}$ we use the matrix,
\begin{equation}\label{eq:eqt2lab}
 R_{\rm lab}(t) = 
\begin{pmatrix}
 -\sin(\lambda_\textrm{lab})\cos(\tau_d ) & -\sin(\lambda_\textrm{lab})\sin(\tau_d ) & \cos(\lambda_\textrm{lab}) \\
 \sin(\tau_d ) & -\cos(\tau_d ) & 0\\
 \cos(\lambda_\textrm{lab})\cos(\tau_d ) & \cos(\lambda_\textrm{lab})\sin(\tau_d ) & \sin(\lambda_\textrm{lab})
\end{pmatrix} \, .
\end{equation}
The Local Apparent Sidereal Time, $\tau_d$, is expressed as an angle for convenience, 
\begin{equation}
\tau_d  = \omega_d(t-t_d) + \phi_{\rm lab} \, ,
\end{equation}
where $\omega_d = 2 \pi/(0.9973\,{\rm days})$ and $t_d = 0.721$ days (making sure to measure $t$ in days since January 1). Recall that $\phi_\textrm{lab}$ is the longitude of the laboratory location so naturally sets the phase of the diurnal modulation. The frequency should be one sidereal day, but since we will use the definition of the Solar day when we construct the Earth orbital velocity, the frequency here is slightly faster than once per day day. This distinction is mostly unimportant, but can be used as a useful cross check and ensures that the value of the daily modulation does not drift anomalously over the course of the year.

The lab velocity is in total,
\begin{equation}
\textbf{v}_\textrm{lab} = {\bf v}_{\rm LSR} + {\bf v}_{\rm pec} + {\bf v}_\oplus + {\bf v}_{\rm rot} \, .
\end{equation}
The galactic rotation velocity $\mathbf{v}_\textrm{LSR}$ and Solar peculiar velocity $\mathbf{v}_\textrm{pec}$ are both fixed in galactic coordinates. The velocity of the local standard of rest (LSR) is defined in galactic coordinates as (0, $v_0$, 0) where $v_0$ is the circular rotation speed of the Milky Way. The standard value tends to be $v_0\sim220$~km~s$^{-1}$~\cite{Kerr:1986hz}, but astronomical determinations of this speed are heavily dependent on the model used for the MW rotation curve, e.g. ref.~\cite{McMillan:2009yr} quote values of $v_0$ from $200 \pm 20$~km~s$^{-1}$ to $279 \pm 33$~km~s$^{-1}$. The Solar peculiar velocity can also be measured with kinematic data, we use the value from ref.~\cite{Schoenrich:2009bx} of $\textbf{v}_{\rm pec} = (11.1^{+0.69}_{-0.75},12.24^{+0.47}_{-0.47},7.25^{+0.37}_{-0.30})$~km~s$^{-1}$ with additional $\sim$ 0.5 - 2 $\kms$ sized systematic uncertainties. In a direct detection experiment on Earth only the combination of these first two velocities is measurable,
\begin{equation}
{\bf v}_{\rm LSR} + {\bf v}_{\rm pec} \equiv \mathbf{v}_\odot = v_\odot (0.0477,0.9984,0.0312) \, ,
\end{equation} 
where  $v_\odot = 232.6$ km s$^{-1}$. The Earth revolution velocity is calculable in {\it galactic coordinates} to be \cite{McCabe:2013kea},
\begin{equation}
\mathbf{v}_\oplus = v_\oplus \left( \cos[\omega_y(t-t_y)] \epsa + \sin[\omega_y (t-t_y)] \epsb\right) \, ,
\end{equation}
where $\omega_y = 2\pi/(365\,{\rm days})$, $t_y = $ March 20 and $v_\oplus = 29.79$ km s$^{-1}$. The vectors are,
\begin{eqnarray}
\epsa &=& (0.9940,0.1095,0.0031) \, , \\ 
\epsb &=& (-0.0517, 0.4945, -0.8677) \, .
\end{eqnarray}
Since we have two separate modulations: a daily one with frequency $\omega_d$ and an annual one with frequency $\omega_y$, to compress our formulae we again write both times as angles,
\begin{equation}
\tau_d = \omega_d (t-t_d) + \phi_\textrm{lab} \, , \quad 
\tau_y = \omega_y (t-t_y) \, .
\end{equation}
Finally, we have the rotational velocity of the Earth which always points east\footnote{Apart from at the poles when it is 0.} in laboratory coordinates
\begin{equation}
\textbf{v}_\textrm{rot} = v_{\rm rot} \cos{\lambda_{\rm lab}} \begin{pmatrix}0\\-1\\0\end{pmatrix} \, ,
\end{equation}
where $v_{\rm rot} = 0.47 {\rm \, km \, s}^{-1}$.

\begin{figure}[t]
\centering
\includegraphics[width=\textwidth]{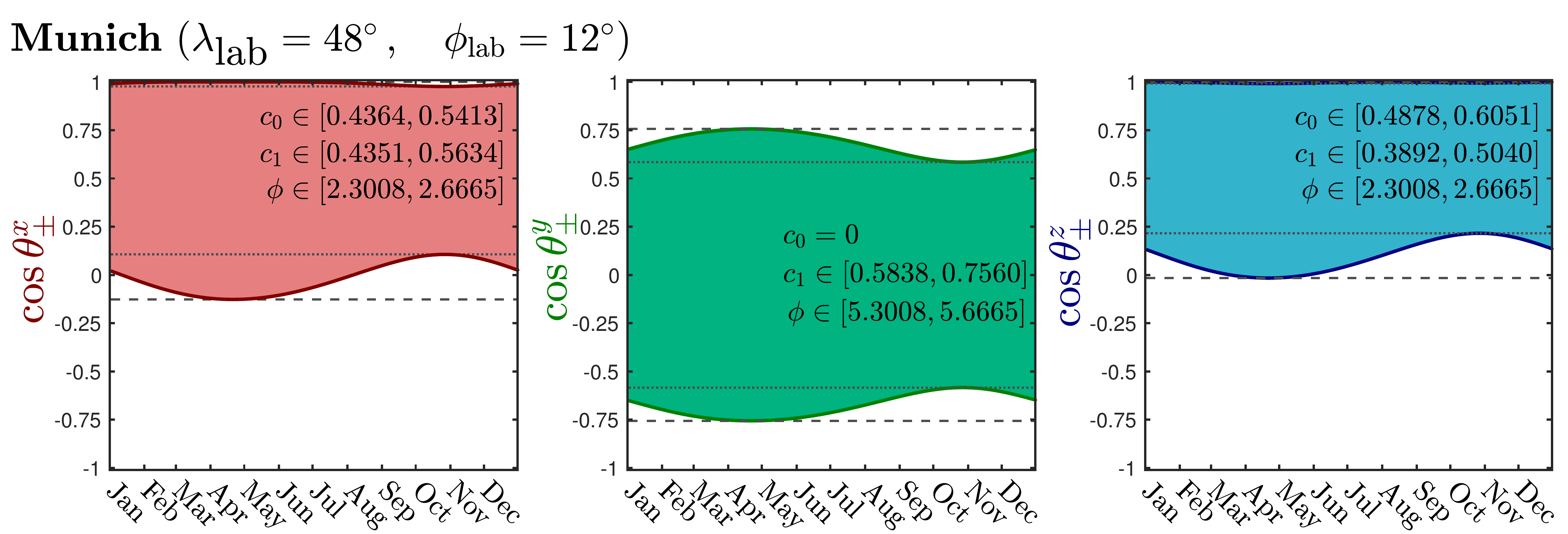} 
\caption{Annual modulation of the amplitude of daily modulation, expressed as $\cos{\theta^i_\pm}$ for north, west and zenith-pointing experiments (from left to right) located at Munich. The shaded region indicates the maximum and minimum value that $\cos{\theta^i_{\rm lab}}$ takes over a single daily modulation as a function of time during the year, as defined in eqs.~\eqref{eq:costhminmaxx}--\eqref{eq:costhminmaxz}. We also display the ranges of the daily modulation parameters $c_0$, $c_1$ and $\phi$ over the year.}
\label{fig:dailymodulation}
\end{figure}

Putting everything together we find that we need to calculate,
\begin{equation}
\vlab(t) = R_{\rm lab}(\tau_d)R_{\rm gal}(\mathbf{v}_\odot + \mathbf{v}_\oplus(\tau_y)) + \textbf{v}_\textrm{rot} \, .
\end{equation}
Focusing on a particular axis in turn we can write down,
\begin{eqnarray}
v_{\rm lab}^\mathcal{N} &\equiv & \vlab \cdot \hat{\mathcal{N}} = \sigma_3 \cos{\lambda_\textrm{lab}} - \sin{\lambda_\textrm{lab}} (\sigma_1\cos{\tau_d} + \sigma_2\sin{\tau_d}) \, ,
\\
v_{\rm lab}^\mathcal{W} &\equiv & \vlab \cdot \hat{\mathcal{W}} = -\sigma_2 \cos{\tau_d} + \sigma_1 \sin{\tau_d} - v_\textrm{rot} \cos{\lambda_\textrm{lab}} \, , \\
v_{\rm lab}^\mathcal{Z} &\equiv & \vlab \cdot \hat{\mathcal{Z}} = \sigma_3 \sin{\lambda_\textrm{lab}} - \cos{\lambda_\textrm{lab}} (\sigma_1\cos{\tau_d} + \sigma_2\sin{\tau_d}) \, ,
\end{eqnarray} 
and we have defined,
\begin{eqnarray}
\sigma_1 (\tau_y) &=& \begin{pmatrix} -0.05487556 \\ +0.49410943 \\ -0.86766615 \end{pmatrix} \cdot \left(\mathbf{v}_\odot + v_\oplus \cos(\tau_y) \epsa + v_\oplus \sin(\tau_y) \epsb\right) \, ,\\ 
\sigma_2 (\tau_y)&=& \begin{pmatrix} -0.87343709 \\ -0.44482963 \\ -0.19807637 \end{pmatrix} \cdot \left(\mathbf{v}_\odot + v_\oplus \cos(\tau_y) \epsa + v_\oplus \sin(\tau_y) \epsb\right)\, ,\\
\sigma_3 (\tau_y)&=& \begin{pmatrix} -0.48383502 \\ +0.74698225 \\ +0.45598378 \end{pmatrix} \cdot \left(\mathbf{v}_\odot + v_\oplus \cos(\tau_y) \epsa + v_\oplus \sin(\tau_y) \epsb\right) \, .
\end{eqnarray}

If we have a cavity that is primarily sensitive to only one direction ($\hat{\mathcal{N}}$, $\hat{\mathcal{W}}$, or $\hat{\mathcal{Z}}$), then the signal correction is dependent on the magnitude of the lab velocity, and the angle between the preferred direction and $\mathbf{v}_\textrm{lab}(t)$ which we write as $\cos{\theta^{\mathcal{N},\mathcal{W},\mathcal{Z}}_{\rm lab}}(t)$. Therefore to estimate the significance of a modulation in these signals we need only know the size of the modulation in $v_\textrm{lab}(t)$ and $\cos{\theta^{\mathcal{N},\mathcal{W},\mathcal{Z}}_{\rm lab}}(t)$ over a day or a year.

Firstly, the speed of the lab we can compute in any coordinate system, so we do this in galactic coordinates with ease (for simplicity we ignore the 0.2\% contribution from the Earth's rotation here). It can be written as,
\begin{equation}
v_{\rm lab}(t) =  \sqrt{v_\odot^2 + v_\oplus^2 + 2 \alpha v_\odot v_\oplus \cos(\tau_y - \omega_y \bar{t}\,)} \, ,
\end{equation}
where $\alpha = 0.491$ and $\bar{t} = 72.4$ days. Then for each axis we have $\cos{\theta^i_{\rm lab}} = v^i/v_{\rm lab}$. The full formulae for these are long winded if including both daily and annual modulation, but we can use the fact that the daily modulation is much faster than the annual to write a simplified description to aid in our analytic treatment of the test statistic.

Looking at the daily modulation first, we take $\sigma_{1,2,3}$ and $v_\textrm{lab}$ as constant, and we reduce each angle down to the form $\cos{\theta} = c_0 + c_1 \cos{\left(\omega_d t + \phi\right)}$ in the following way,
\begin{eqnarray}
\cos{\theta_{\rm lab}^\mathcal{N}} &=& b_0 \cos{\lambda_{\rm lab}} -b_1 \sin{\lambda_{\rm lab}} \cos{\left(\omega_d t + \phi_{\rm lab} + \psi \right)} \, ,\\
\cos{\theta_{\rm lab}^\mathcal{W}} &=& b_1 \cos{\left(\omega_d t + \phi_{\rm lab} + \psi-\pi\right)} \, ,\\
\cos{\theta_{\rm lab}^\mathcal{Z}} &=& b_0 \sin{\lambda_{\rm lab}}  + b_1 \cos{\lambda_{\rm lab}} \cos{\left(\omega_d t + \phi_{\rm lab} + \psi\right)} \, ,
\end{eqnarray}
where for our directional experiments the only unknowns regarding the daily modulation are,
\begin{eqnarray}
b_0 &=& \sigma_3 / v_\textrm{lab} \, ,\\
b_1 &=& \sqrt{\sigma_1^2 + \sigma_2^2}/ v_\textrm{lab}\, , \\
\psi &=& \tan^{-1}\left( \sigma_1/ \sigma_2\right) - \omega_d t_d - \pi /2 \, . \label{eq:psi}
\end{eqnarray}
For example on January 1 we have,  $\{b_0,b_1,\psi\} = \{0.7589,0.6512, -3.5336\}$. Then when we use the definition $\cos{\theta} = c_0 + c_1 \cos{\left(\omega_d t + \phi\right)}$ we just absorb the laboratory location into the experiment specific constants $\{c_0,\,c_1,\,\phi\}$. The ranges for these experiment specific values over the full year are displayed in figure~\ref{fig:dailymodulation}.

Assuming we have knowledge of $v_\textrm{lab}$ on a given day from the frequency dependence of the power spectrum, the daily modulation can be inverted,
\begin{equation}\label{eq:vodot_sol}
\mathbf{v}_\odot = v_\textrm{lab} R_{\textrm{gal}}^{T} \begin{pmatrix}
  b_1 \sin{\left(\psi + \omega_d t_d + \pi/2\right)} \\ b_1 \cos{\left(\psi + \omega_d t_d + \pi/2\right)} \\ b_0
\end{pmatrix}
 - \mathbf{v}_\oplus \, .
\end{equation}
The uncertainties on each component of the velocity will then depend linearly on the uncertainties of the constants. Notice that in fact you only need the daily modulation in one of the cavities ($\mathcal{N}$ or $\mathcal{Z}$) to measure all three constants. The west pointing experiment cannot measure $b_0$, because it always rotates in the same direction the experiment points so the modulation in angle will always be centred around 0\footnote{The west-pointing experiment is still useful since its modulation amplitude is larger than the other two directions.}. For a $q$-type experiment we can only measure the square of the cosine of each angle, so there will be degenerate solutions for $\{b_0,\,b_1\}$ and $\{-b_0,\,-b_1\}$ but given that we know that the the second component of $\mathbf{v}_\odot$ is  $\sim 220$ km s$^{-1}$ there will only be one solution that is consistent with galactic rotation. For streams we do not necessarily have a prior expectation on a velocity so this will not be possible and there will always be multiple solutions. However the procedure is the same,
\begin{equation}
\mathbf{v}_{\rm str} = \mathbf{v}_\odot + \mathbf{v}_\oplus -  |\mathbf{v}_\textrm{lab}-\mathbf{v}_\textrm{str}| R_{\textrm{gal}}^{T} \begin{pmatrix}
  b_1 \sin{\left(\psi + \omega_d t_d + \pi/2\right)} \\ b_1 \cos{\left(\psi + \omega_d t_d + \pi/2\right)} \\ b_0
\end{pmatrix} \, ,
\end{equation}
where again the value of  $|\mathbf{v}_\textrm{lab}-\mathbf{v}_\textrm{str}|$ can be independently inferred from the frequency of the feature.

In this study we focus on daily modulations, but it will still be possible to search for annual modulations and indeed this will always improve statistics. However this does not require directional sensitivity and was covered extensively in previous work~\cite{OHare:2017yze,Foster:2017hbq}. We can show the size of annually modulating features in our directional haloscopes by coarse graining over the daily modulation to only consider its \emph{range} at each day of the year. We do this by defining $\cos{\theta_\pm}$ which is the maximum and minimum value of $\cos{\theta_{\rm lab}}$ within one day, where the parameters that modulate annually are taken to be constant over that day (which they approximately are),
\begin{eqnarray}
v_{\rm lab} \cos{\theta}_\pm^\mathcal{N} &=& \sigma_3 \cos{\lambda_\textrm{lab}} \pm \sqrt{\sigma^2_1 + \sigma^2_2}\sin{\lambda_\textrm{lab}} \, ,\label{eq:costhminmaxx}\\
v_{\rm lab} \cos{\theta}_\pm^\mathcal{W} &=& \pm \sqrt{\sigma^2_1 + \sigma^2_2} \, , \label{eq:costhminmaxy}\\
v_{\rm lab} \cos{\theta}_\pm^\mathcal{Z} &=& \sigma_3 \sin{\lambda_\textrm{lab}} \pm \sqrt{\sigma^2_1 + \sigma^2_2}\cos{\lambda_\textrm{lab}}\, . \label{eq:costhminmaxz}
\end{eqnarray}
When focusing on the daily modulation it is sufficient to say that at a given day during the year, each angle oscillates sinusoidally between $\cos\theta_-$ and $\cos\theta_+$. Each term in the above formulae are time dependent with a frequency of 1 year. We claimed that in our experiments the daily modulation is the more important effect. We display how the size of the daily modulation varies on top of the annual modulation in figure~\ref{fig:dailymodulation}. We see that the amplitude of each daily modulation is larger by up to factor of 4 for each experiment whilst also varying a factor of 365 more quickly. One can also observe that the ranges for the modulation parameters are rather small and would only induce an error of 25\% if left constant over the whole year (and we generally only use times shorter than a few days). For times much longer than this (e.g. figure~\ref{fig:dailymod_deltat}) we account for the full calculation including diurnal and annual modulations.

\section{Analytic formulae for the test statistic}\label{sec:analyticformulae}
In eq.~\eqref{eq:teststatgeneral} we claimed that the integrals over frequency and time in the test statistic can be written analytically for a Maxwellian distribution (describing the SHM or a stream). In the interest of readability we have deposited them here. 
\subsection{Linear experiments} 
First for the $\ell$-type experiments we need to do
	\begin{align}
	D_\ell \approx& \, \frac{\Delta \omega}{\Delta t} \int_{0}^{t_{\rm obs}} {\rm d}t \int_{m_a}^\infty {\rm d} \omega ~ \left(\frac{P_0 f(\omega) \, \zeta_\ell(\omega) \, \cos[\theta_{\rm lab}(t)] \, \gl }{\sigma_N} \right)^2\\
	=& \, 2 \pi \left (\frac{P_0}{k_B T_{\rm sys}}\right)^2 \gl^2 \, \mathcal{I}^\ell_\omega \,\mathcal{I}^\ell_t \, ,
	\end{align}
where we have substituted $\sigma_N = k_B T_{\rm sys} \sqrt{\Delta \omega/2 \pi \Delta t}$. We simplify the notation here (and for similar expressions later) by separating the time and frequency integrals. Recalling that we are writing the daily modulation of the lab velocity angle as $\cos{\theta_{\rm lab}(t)} = c_0 + c_1 \cos{(\omega_d t + \phi)}$ we can integrate over this as well as over a Maxwellian $f(\omega)$ to give the following,
	\begin{align}
	\mathcal{I}^\ell_\omega =& \frac{\sqrt{\pi } \left(2 v_\lab^2-\sigma_v^2\right) \text{erf}\left(\frac{v_\lab}{\sigma_v }\right)+2 \sigma_v v_\lab e^{-\frac{v_\lab^2}{\sigma_v^2}}}{4 \pi v_\lab \sigma_v} \cdot \frac{1}{m_a} \, ,\\
	\mathcal{I}^\ell_t =& \Bigl[{c_0}^2 t+\frac{{c_1}^2 t}{2}+\frac{2 {c_0} {c_1} \sin (\omega_d t +\phi )}{\omega_d }+\frac{{c_1}^2 \sin (\omega_d t +\phi ) \cos (\omega_d t +\phi )}{2 \omega_d } +\frac{2 {c_0}^2 \phi +{c_1}^2 \phi}{2 \omega_d }
	\Bigr]_{0}^{t_{\rm obs}} \, .\nonumber
	\end{align}
	The factor $1/m_a$ comes from the fact that we integrate the square of the distribution $f(\omega)$\footnote{Consider:
		\begin{align*}
			\int_{m_a}^\infty {\rm d} \omega f(\omega)^2 = 
			\int_{m_a}^\infty {\rm d} \omega \left( \frac{{\rm d} v}{{\rm d} \omega} f(v)\right)^2 = 
			\frac{1}{m_a}\int_0^\infty {\rm d} v  \frac{f(v)^2}{v} \, .
		\end{align*}}.
		 In order to write the integral over time starting at 0 we need to ensure that the phase is defined according to the same definition of time. Throughout we assume our origin is January 1 where $\psi = −3.5336$ (see eq.~\eqref{eq:psi}). For daily modulations with a total measurement time $t_{\rm obs}$ over several days we can simplify the time integral to,
	\begin{align}
		\mathcal{I}^\ell_t &\approx \left({c_0}^2 + \frac{1}{2} {c_1}^2\right) t_{\rm obs} \, .
	\end{align}	
	Also if we have a stream, we can approximate further as long as $\sigma_v \equiv \sigma_{\rm str} \ll |\vlab - \vstr|$, giving instead
	\begin{align}
	\mathcal{I}^\ell_\omega &\approx \frac{\left(2  |\vlab - \vstr|^2-\sigma_{\rm str}^2\right)}{4 \sqrt{\pi }  |\vlab - \vstr| \sigma_{\rm str}}  \cdot \frac{1}{m_a} ,
	\end{align}

\subsection{Quadratic experiments}
Next, we consider the quadratic case. We can write the test statistic in the same way except the directional correction has an unmodulated `offset' term and a modulated term which means the full test statistic has to be written as,
	\begin{align}
	D_q = 2 \pi \left (\frac{P_0}{k_B T_{\rm sys}}\right)^2 \gq^2 \, \left(\mathcal{I}^{q1}_\omega \,\mathcal{I}^{q1}_t +\mathcal{I}^{q2}_\omega \,\mathcal{I}^{q2}_t  +\mathcal{I}^{q12}_\omega \,\mathcal{I}^{q12}_t  \right) \, .
	\end{align}
 We use the label `$q1$' for the integrals of the offset term $\zeta_{q1}(\omega)$ and `$q2$' for the integrals of the modulation term $\zeta_{q2}(\omega)\cos{\theta_{\rm lab}}$. Since we integrate over the square of the directional correction we need to include the mixing term which we label `$q12$'.\\
	
	\noindent First for the offset we have,
	\begin{align}
	\mathcal{I}_{\omega}^{q1} &= \left(\frac{\sigma_v ^2}{v_{\rm lab}}\right)^2 \mathcal{I}_{\omega}^\ell \, , \\
	\mathcal{I}_t^{q1} &= t_{\rm obs} \, .
	\end{align} 

\noindent Then for the modulation,
	\begin{align}
	\mathcal{I}^{q2}_\omega =& \frac{\sqrt{\pi } \left(-4 \sigma_v^2 v_\lab^2+4 v_\lab^4+3 \sigma_v^4\right) \text{erf}\left(\frac{v_\lab}{\sigma_v}\right)+e^{-\frac{v_\lab^2}{\sigma_v^2}} \left(4 \sigma_v v_\lab^3-6 \sigma_v^3 v_\lab\right)}{8 \pi  \sigma_v v_\lab}  \cdot \frac{1}{m_a}  \, , \\
	\mathcal{I}^{q2}_t =& 
	\Bigg[  \left(\frac{3 c_1^4}{8} +
	3 c_1^2 c_0^2 +
	c_0^4\right) t \\ &+
	\left(\frac{4 c_1 c_0^3}{\omega_d }+
	\frac{3 c_1^3 c_0}{\omega_d }\right) \sin (t \omega_d +\phi )+
	\left(\frac{3 c_1^2 c_0^2}{2 \omega_d }+
	\frac{c_1^4 }{4 \omega_d }\right) \sin (2 (t \omega_d +\phi )) \\&+
	\frac{c_1^3 c_0 }{3 \omega_d } \sin (3 (t \omega_d +\phi ))+
	\frac{c_1^4 }{32 \omega_d } \sin (4 (t \omega_d +\phi )) \\&+	
	\frac{c_0^4 \phi }{\omega_d }+
	\frac{3 c_1^2 c_0^2 \phi }{\omega_d }+
	\frac{3 c_1^4 \phi }{8 \omega_d } \Bigg]_{0}^{t_{\rm obs}} \, ,
	\end{align}
	
\noindent Then finally the mixing term can be written in terms of integrals already calculated,
	\begin{align}
\mathcal{I}_\omega^{q12}\mathcal{I}_t^{q12}  = & \,2 \int_0^{t_{\rm obs}} \textrm{d} t\,\int_{m_a}^\infty \textrm{d} \omega ~ \zeta_{q1}(\omega) \zeta_{q2}(\omega) \left( f(\omega) \cos(\theta_{\rm lab})\right)^2 \\
	=& ~ 2 \frac{\sigma_v^2}{v_\lab^2} ~ \mathcal{I}_{\omega}^{q2} ~ \mathcal{I}^\ell_t \, .
	\end{align}
	
	Two of these integrals can be simplified in a similar way to before. If we approximate over several days we can write down,
	\begin{align}
	\mathcal{I}^{q2}_t \approx& \left(\frac{3 c_1^4}{8} +
	3 c_1^2 c_0^2 +
	c_0^4\right) t_{\rm obs} \, .
	\end{align}
And if we have a low dispersion stream we can use,
	\begin{align}
	\mathcal{I}^{q2}_\omega \approx& \, \frac{\sqrt{\pi } \left(-4 \sigma_{\rm str}^2 |\vlab-\vstr|^2+4 |\vlab-\vstr|^4+3 \sigma_{\rm str}^4\right)}{8 \pi  \sigma_{\rm str} |\vlab-\vstr|}   \, .
	\end{align}

\bibliographystyle{JHEP}
\bibliography{axiostronomy}

\end{document}